\documentclass[paper]{JHEP3}
\usepackage{amssymb}
\usepackage{amsmath}
\usepackage[english]{babel}
\usepackage[dvipdf]{graphicx}
\usepackage{cite}
\usepackage{wasysym}

\def\as{\alpha_{s}}

\def\asb{\bar{\alpha}_{s}}
\def\Ltau{L}
\def\Lp{\Ltau'}
\def\asl{\alpha_{s}\Ltau}
\def\aslp{\alpha_{s}\Ltau'}

\def\aslh{\alpha_{s}\hat{L}}

\def\d{\hbox{d}}

\def\cf{C_{F}}
\def\ca{C_{A}}
\def\nf{n_{F}}

\newcommand{\lp}{\left}
\newcommand{\rp}{\right}

\title{Two-Loop Soft Corrections and Resummation of the Thrust Distribution in the Dijet Region}

\abstract{The thrust distribution in electron-positron annihilation is a classical precision QCD observable. Using renormalization group (RG) evolution in Laplace space, we perform the resummation of
logarithmically enhanced corrections in the dijet limit, $T\to 1$ to next-to-next-to-leading logarithmic (NNLL) accuracy. We independently derive the two-loop soft function for the thrust distribution
and extract an analytical expression for the NNLL resummation coefficient $g_3$. Our findings confirm earlier NNLL resummation results for the thrust distribution in soft-collinear effective theory.
To combine the resummed expressions with the fixed-order results, we derive the $\log(R)$-matching and $R$-matching of the NNLL approximation to the fixed-order NNLO distribution.} 

\author{Pier~Francesco~Monni, Thomas~Gehrmann\\Institut f\"ur Theoretische Physik, Universit\"at Z\"urich,
Winterthurerstrasse 190,\\CH-8057 Z\"urich, Switzerland }

\author{Gionata~Luisoni\\Institute for Particle Physics Phenomenology, University of Durham,\\
Science Laboratories, South Rd, Durham DH1 3LE, UK}

\keywords{QCD, Thrust, Event-shape, Soft gluons, Resummation}

\preprint{IPPP/11/28; DCPT/11/56; ZU-TH 08/11}

\begin{document}

\section{Introduction}
Event-shape distributions in $e^+e^-$ annihilation are observables which measure the geometrical properties of energy-momentum flow in a hadronic final state. They have been 
measured over a broad range in energies at LEP~\cite{aleph,opal,l3,delphi} as well as at 
earlier electron-positron colliders~\cite{sld,jade}. The event-shape distributions allow 
for  a detailed probe of the dynamics of QCD and especially for a precise determination of the 
strong coupling constant~$\alpha_s$. 
Owing to their infrared and collinear
safety, they can be computed systematically  in perturbation theory. They usually span the range between the kinematical situation of two 
collimated back-to-back jets (dijet limit) and a perfectly spherical final state. 

The fixed-order description, which expands the distribution in 
powers of the strong coupling constant to leading order (LO), next-to-leading order (NLO),
next-to-next-to-leading order (NNLO) and so on, is reliable and convergent over most of the 
kinematical range of the event-shape. In the dijet limit, which is attained for 
the thrust variable~\cite{thrust} as $T\to 1$, the convergence of the fixed-order expansion 
is spoilt by large logarithmic terms $\log(1-T)$ at each order in the strong coupling constant,
which necessitates a resummed description. Resummation of the 
event-shape distribution accounts for the logarithmically enhanced terms to all orders in perturbation theory, and 
ensures a reliable prediction in the dijet region. The resummed cross section is organized in terms of leading logarithms (LL), next-to-leading logarithms (NLL)
and so on. During LEP times, precision studies of 
a standard set of six 
event-shapes were based on the combination of fixed-order NLO 
calculations~\cite{Ellis:1980wv,Ellis:1980nc,Kunszt:1980vt,Vermaseren:1980qz,Fabricius:1981sx,Kunszt:1989km,Giele:1991vf,Catani:1996jh} with NLL 
resummation~\cite{CTTW,broadenings,y3}. To avoid the double counting of terms, both 
expansions need to be matched onto each other and different matching procedures are 
available~\cite{Jones:2003yv}.

In the recent past, substantial progress was made both on the fixed-order and the 
resummed description of event-shapes. Following the development of new 
methods for calculations of QCD jet observables at NNLO~\cite{ourant}, 
the NNLO corrections to $e^+e^-\to 3$ jets and related event-shape observables 
were computed~\cite{GehrmannDeRidder:2007jk,GehrmannDeRidder:2007hr,GehrmannDeRidder:2007bj,Weinzierl:2009nz,Weinzierl:2009ms,weinzierljetnew}. These calculations are based on a numerical integration of the relevant
three-parton, four-parton and five-parton matrix elements, which are combined into a parton-level event generator.
Next-to-leading order electroweak corrections were also computed very
recently~\cite{Denner:2009gx,Denner:2010ia}. Determinations of the strong coupling constant using the newly available NNLO results~\cite{Dissertori:2007xa,asjet,jaquier} 
and the matched~\cite{Gehrmann:2008kh} 
NLL+NNLO~\cite{davisonwebber,jadeas,Dissertori:2009ik,opalas}
predictions led to a big improvement in the scale 
uncertainty and 
showed the need to go beyond NLL in resummation. 

The resummation of large logarithmic corrections is based on a factorization of the 
event-shape cross section in the dijet limit into a convolution of three leading regions (soft, collinear and hard virtual). In 
the conventional approach~\cite{Ster87,CTTW}, the resummation of large logarithms is accomplished in 
Mellin (Laplace) space and the resummed distributions are obtained by an inverse transformation. 
In this approach, the NLL corrections to all standard event-shape variables were 
obtained~\cite{CTTW,broadenings,y3}, as well as NNLL results on the energy-energy 
correlation function~\cite{eec}. By applying soft-collinear effective theory (SCET,~\cite{scet}) to 
event-shape distributions~\cite{hoangtop,schwartzMHnll}, the resummation can be performed
directly in momentum space. SCET offers moreover a systematic framework to compute all 
soft, collinear and hard contributions. In this framework, the N$^3$LL resummation for 
thrust~\cite{Becher:2008cf,Abbate:2010xh} and the heavy jet mass~\cite{Chien:2010kc}
have been performed and applied for a 
precise determination of $\alpha_s$, and the framework for the resummation of the jet broadening 
distributions has been outlined~\cite{Chiu:2011qc, Becher:2011pf}. In these calculations,
the hard subprocess is inferred from the three-loop quark form factor~\cite{3Lform1,3Lform2}, 
 the collinear jet function
is required to the two-loop order, which is known from earlier calculations of 
SCET resummation in heavy meson decays and 
deep inelastic scattering~\cite{SCETJet,bechneubdis},  while the 
soft subprocess to two-loop order
could be computed from the renormalization group invariance of the cross section 
only up to a constant term. Using the  fixed-order NLO 
results~\cite{Catani:1996jh}, this term
was determined numerically by two indepedent groups~\cite{hoangkluth,Becher:2008cf},
obtaining mutually inconsistent results. Motivated by this discrepancy, we perform an
analytical calculation of the full two-loop soft subprocess for thrust 
from first principles in this paper. 
A few days before the completion of this work, the two-loop hemisphere soft function
was derived within SCET~\cite{schwartznew}, and applied to compute the soft 
subprocess for the thrust and the heavy jet mass. Shortly after the release of 
our work, two further SCET-based calculations of the logarithmic 
contributions to the two-loop di-jet soft function~\cite{stewartnew} and 
of the soft function for the Drell-Yan process~\cite{mantrynew} appeared.

The major difference between conventional and SCET-based resummation is the 
handling of intermediate scales in the calculation.  
In the SCET-based resummation, these scales are fixed to their natural 
values directly in momentum space, on the other hand in the conventional approach they are sampled along a complex contour in peforming the Laplace inversion. 
Although both approaches yield identical results at NLL, 
owing to the presence of a Landau pole in the QCD coupling constant, power-suppressed 
differences between them could appear 
in higher order logarithmic contributions. These power-suppressed terms are outside the 
scope of the logarithmic resummation, so both approaches could in 
principle yield different but equally correct results. 
To address the compatibility between the two resummations, 
we perform the NNLL resummation of the thrust distribution in Laplace space
in this paper.

The paper is structured as follows: in Section~\ref{sec:framework}, we review 
the description of the thrust distribution in fixed-order and resummed perturbation theory
and establish the factorization of the cross section in the dijet limit. Using the eikonal 
approximation, we compute the two-loop soft subprocess in Section~\ref{sec:soft}. The 
NNLL resummation of the thrust distribution in Laplace space and its 
inversion to momentum space are derived in Section~\ref{sec:resummation of large logs}, and the 
matching to the fixed-order NNLO distribution is performed in Section~\ref{sec:matching}. 
The numerical impact of these corrections is studied and discussed in Section~\ref{sec:results}.
Finally, Section~\ref{sec:conc} contains our conclusions and an outlook onto 
future applications. An appendix collecting key formulae and technical details of the 
calculation is enclosed.

\section{Thrust distribution in perturbation theory}
\label{sec:framework}
The thrust variable for a hadronic final state in $e^+e^-$ annihilation is 
defined as~\cite{thrust} 
\begin{align}
T=\max_{\vec{n}}
\left(\frac{\sum_i |\vec{p_i}\cdot \vec{n}|}{\sum_i |\vec{p_i}|}\right)\,,
\label{thrust}
\end{align}
where $\vec{p}_i$ denotes the three-momentum of particle $i$, with the sum running 
over all particles. The unit vector $\vec{n}$ is varied to find  the 
thrust direction $\vec{n}_T$ which maximises the expression in parentheses 
on the right hand side. In the present paper we will mostly work with the quantity $\tau\equiv 1-T$.

It can be seen that a two-particle final state has fixed $T=1$,
consequently the thrust distribution receives its first non-trivial 
contribution from three-particle final states, which, at order $\alpha_s$, 
correspond to three-parton final states. Therefore, 
both theoretically and experimentally, the thrust distribution is 
closely related to three-jet production.

\subsection{Fixed-order and resummed calculations}
The differential thrust distribution in perturbation theory is known at NNLO~\cite{GehrmannDeRidder:2007hr,Weinzierl:2009ms}. At a centre-of-mass energy $Q$ and for a renormalization scale $\mu$ takes
the form
\begin{equation}\label{eq:fixedordercs0}
\frac{1}{\sigma_{0}}\, \frac{\d\sigma}{\d \tau}(\tau,Q) = \bar\alpha_s (\mu) \frac{\d A}{\d \tau}(\tau)+ \bar\alpha_s^2 (\mu) \frac{\d B}{\d \tau} (\tau,x_\mu) + \bar\alpha_s^3 (\mu) \frac{\d C}{\d
\tau}(\tau,x_\mu) +
{\cal O}(\bar\alpha_s^4)\;,
\end{equation}
with
\begin{equation}
\asb = \frac{\alpha_s}{2\pi}\;, \qquad x_\mu = \frac{\mu}{Q}\;,
\end{equation}
and where the explicit dependence on the renormalization scale is given by
\begin{align}\label{eq:fixedordermudep}
\frac{\d B}{\d \tau}(\tau,x_\mu) =&\,\frac{\d B}{\d \tau}(\tau)+2\beta_{0}\,\log(x_{\mu}^{2})\frac{\d A}{\d \tau}(\tau),\nonumber\\
\frac{\d C}{\d \tau}(\tau,x_\mu) =&\,\frac{\d C}{\d \tau}(\tau)+2\,\log(x_{\mu}^{2})\left(2\beta_{0}\frac{\d B}{\d \tau}(\tau) +2\beta_{1}\frac{\d A}{\d
\tau}(\tau)\right)+\left(2\beta_{0}\,\log(x_{\mu}^{2})\right)^{2}\frac{\d A}{\d \tau}(\tau).
\end{align}
For the QCD $\beta$-function we follow the convention given in Appendix~\ref{sec:constants}. In theoretical computations it is customary to normalize the distributions to the Born cross
section $\sigma_{0}$ since, for massless quarks, the normalization cancels all electroweak coupling factors. However, experimentally it is easier to normalize the
distributions to the total hadronic cross section $\sigma$. The correction for the normalization can be done by expanding the ratio $\sigma_{0}/\sigma$ in powers of $\asb$, which is
nowadays known at four loops~\cite{Baikov:2010iw}. In the massless case the ratio is given by
\begin{equation}\label{eq:sighadratio}
\begin{split}
\frac{\sigma}{\sigma_{0}}\, =&\,1+\asb K_{1}+\asb^{2}\left[K_{2}+2\beta_{0}\,\log(x_{\mu}^{2})K_{1}\right]+\mathcal{O}(\bar{\alpha}_{s}^{3}),\end{split}
\end{equation}
where
\begin{eqnarray}
K_{1}&=&\frac{3}{2}\cf,\notag\\ 
K_{2}&=&\frac{1}{4}\left[-\frac{3}{2}\cf^{2}+\cf\ca\left(\frac{123}{2}-44\zeta_{3}\right)+\frac{\cf\nf}{2}\left(-22+16\zeta_{3}\right)\right].
\end{eqnarray}

The dependence on the renormalization scale $\mu$ is universal and is the same in (\ref{eq:fixedordermudep}) and (\ref{eq:sighadratio}). Inserting the expansion of $\sigma_{0}/\sigma$ we
obtain the following expression:
\begin{equation}\label{eq:fixedordercshad}
\frac{1}{\sigma}\frac{\d\sigma}{\d \tau}(\tau,Q) = \asb(\mu) \frac{\d \bar{A}}{\d \tau}(\tau)
+ \asb^2 (\mu) \frac{\d \bar{B}}{\d \tau}(\tau,x_\mu) + \asb^3 (\mu) \frac{\d \bar{C}}{\d \tau}(\tau,x_\mu)+{\cal O}(\asb^4)\;,
\end{equation}
where $\bar{A},\bar{B},\bar{C}$ are related to $A,B,C$ by
\begin{align}
\bar{A}(\tau)=&\,A(\tau)\,,\\
\bar{B}(\tau,x_\mu)=&\,B(\tau,x_\mu)-K_{1} A(\tau)\,,\\
\bar{C}(\tau,x_\mu)=&\,C(\tau,x_\mu)-K_{1} B(\tau,x_{\mu})+\lp(K_{1}^{2}-K_{2}\rp)A(\tau)\,.
\end{align}

For later convenience we consider also the integrated distribution
\begin{equation}\label{eq:Rfixed}
R_{T}(\tau)\,\equiv\,\frac{1}{\sigma}\int_{0}^{1}\frac{d\sigma\left(\tau',Q\right)}{d\tau'}\Theta(\tau-\tau^{\prime})d\tau',
\end{equation}
which has the following fixed-order expansion:
\begin{equation}\label{eq:Rfixedorder}
R_{T}\left(\tau\right)\,=\,1+\mathcal{A}\left(\tau\right)\asb(\mu^{2})\,+\,\mathcal{B}\left(\tau,\mu^{2}\right)\asb^{2}(\mu^{2})\,+\,\mathcal{C}
\left(\tau, \mu^ {2} \right)\asb^{3}(\mu^{2}).
\end{equation}
The fixed-order coefficients $\mathcal{A}$, $\mathcal{B}$, $\mathcal{C}$ can be obtained by integrating the distribution~(\ref{eq:fixedordercshad}) and imposing $R_{T}(\tau_{{\rm max}},Q)=1$ to all
orders, where $\tau_{{\rm max}}$ is the maximal kinematically allowed value.

In the two-jet region the fixed-order thrust distribution is enhanced by large infrared logarithms which spoil the convergence of the perturbative series. The convergence can be restored by resumming
the logarithms to all orders in the coupling constant.  The resummed cross section can in general be written as
\begin{align}\label{eq:Rres}
R_{T}(\tau) =
\bigg(1+\sum_{k=1}^{\infty}C_{k}\asb^{k}\bigg)\,\exp\left[\Ltau g_{1}(\asl)+g_{2}(\asl)+\frac{\alpha_{s}}{\pi}\beta_{0}g_{3}(\asl)+\ldots\right]\,,
\end{align}
where $L\equiv\log(1/\tau)$. The function $g_{1}$ encodes all the leading logarithms, the function $g_{2}$ resums all next-to-leading logarithms and so on. 

The last equation gives a better prediction of the thrust distribution in the two-jet region, but fails to describe the multijet region $\tau\rightarrow\tau_{\max}$, where non-singular pieces of the
fixed-order prediction become important. To achieve a reliable description of the observable over a broader kinematical range the two expressions~(\ref{eq:Rfixedorder}) and~(\ref{eq:Rres}) can be
matched, taking care of avoiding double counting of logarithms appearing in both expressions. To this purpose we reexpand the functions $g_{i}$ in powers of $\asl$:
\begin{align}
\Ltau g_{1}(\asl)=&G_{12}\asb \Ltau^{2}\,+\,G_{23}\asb^{2}\Ltau^{3}\,+\,G_{34}\asb^{3}\Ltau^{4}\,+\ldots\,,\notag\\
g_{2}(\asl)=&G_{11}\asb \Ltau\,+\,G_{22}\asb^{2}\Ltau^{2}\,+\,G_{33}\asb^{3}\Ltau^{3}\,+\ldots\,,\notag\\
\frac{\alpha_{s}}{\pi}\beta_{0}g_{3}(\asl)=&G_{21}\asb^{2}\Ltau\,+\,G_{32}\asb^{3}\Ltau^{2}\,+\,G_{43}\asb^{4}\Ltau^{3}\,+\ldots\,.
\end{align}
In a second step we reexpand the exponential function recovering the full logarithmic dependence of the fixed-order series at NNLO
\begin{align}
R^{(1)}_{\rm log}\left(\tau\right)=&C_{1}\,+\,G_{11}\Ltau\,+\,G_{12}\Ltau^{2}\,,\label{eq:R1}\\
R^{(2)}_{\rm log}\left(\tau\right)=&C_{2}\,+\,\left(G_{21}+C_{1}G_{11}\right)\Ltau\,+\,\left(G_{22}+\frac{1}{2}G_{11}^{2}+C_{1}G_{12}\right)\Ltau^{2}\,\nonumber\\
&+\,\left(G_{23}+G_{12}G_{11}\right)\Ltau^{3}+\,\frac{1}{2}G_{12}^{2}\Ltau^{4}\,,\label{eq:R2}\\
R^{(3)}_{\rm log}\left(\tau\right)=&C_{3}+\left(G_{31}+C_{1}G_{21}+C_{2}G_{11}\right)\Ltau\nonumber\\
&+\left(G_{32}+C_{1}G_{22}+\frac{1}{2}C_{1}G_{11}^{2}+C_{2}G_{12}+G_{11}G_{21}\right)\Ltau^{2}\nonumber\\
&+\,\left(G_{33}+G_{11}G_{22}+G_{12}G_{21}+C_{1}G_{11}G_{12}+\frac{1}{6}G_{11}^{3}+C_{1}G_{23}\right)\Ltau^{3}\nonumber\\
&+\,\left(G_{34}+G_{12}G_{22}+\frac{1}{2}C_{1}G_{12}^{2}+G_{11}G_{23}+\frac{1}{2}G_{11}^{2}G_{12}\right)\Ltau^{4}\nonumber\\
&+\,\left(G_{12}G_{23}+\frac{1}{2}G_{12}^{2}G_{11}\right)\Ltau^{5}\,+\,\frac{1}{6}G_{12}^{3}\Ltau^{6}\label{eq:R3}\,.
\end{align}
The difference between the logarithmic part and the full fixed-order series at different orders is given by
\begin{align}
d_{1}(\tau)=&\mathcal{A}(\tau)-R^{(1)}_{\rm log}(\tau)\,,\\
d_{2}(\tau)=&\mathcal{B}(\tau)-R^{(2)}_{\rm log}(\tau)\,,\\
d_{3}(\tau)=&\mathcal{C}(\tau)-R^{(3)}_{\rm log}(\tau)\,.
\end{align}
The functions $d_{i}(\tau)$ contain the non-logarithmic part of the fixed-order contribution and vanish for $\tau\to 0$. We collect them into a function $D(\tau,Q)$ defined as
\begin{equation}
D(\tau,Q)\,=\,\asb d_{1}(\tau)+\asb^{2} d_{2}(\tau)+\asb^{3} d_{3}(\tau)+\mathcal{O}(\bar{\alpha}^{4}_{s}).
\end{equation}

\subsection{Kinematics and factorization of thrust}
\label{KinFac}

\begin{figure}[!htp]
\begin{center}
\includegraphics[width=50mm]{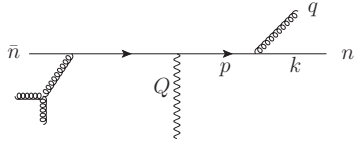}
\caption{First branching kinematics.\label{branching}}
\end{center}
\end{figure}

The definition of thrust (\ref{thrust}) splits the final-state into two hemispheres $S_{\vec{n}_{T}}$ and $\bar{S}_{\vec{n}_{T}}$ separated by the plane $P_{\vec{n}}$ orthogonal to the unit vector $\vec{n}_{T}$. Each final state particle
with momentum $\vec{p}_{i}$ is assigned to either $S_{\vec{n}_{T}}$ or $\bar{S}_{\vec{n}_{T}}$ depending on whether $\vec{p}_{i}\cdot \vec{n}_{T}$ is positive or negative. As it was shown in \cite{CTTW},
no final state momenta can lie in $P_{\vec{n}}$. We denote with $p$ and $\bar{p}$ the total momenta in the hemispheres  $S_{\vec{n}_{T}}$ and $\bar{S}_{\vec{n}_{T}}$ respectively.
We can parametrize the two total momenta as 
\begin{align}
p^{\mu} = \frac{p\cdot n}{2}\bar{n}^{\mu}+p_{\perp}^{\mu}+\frac{p\cdot\bar{n}}{2}n^{\mu}, \qquad
\bar{p}^{\mu} = &\frac{\bar{p}\cdot\bar{n}}{2}n^{\mu}+\bar{p}_{\perp}^{\mu}+\frac{\bar{p}\cdot n}{2}\bar{n}^{\mu},
\end{align}
where $n^{\mu} = (1,\vec{n}_{T})$ and $\bar{n}^{\mu} = (1,-\vec{n}_{T})$. A simple kinematical analysis \cite{CTTW} shows that in the dijet limit  we can recast (\ref{thrust}) as 
\begin{align}
\label{T2jet1}
 \tau \equiv 1-T = \frac{p^{2}}{Q^{2}}+\frac{\bar{p}^{2}}{Q^{2}}+\mathcal{O}((1-T)^{2}),
\end{align}
where we are neglecting terms of relative order $1-T$ which give rise to power-suppressed contributions to the cross section. In our analysis we need to separate soft particle contributions from 
collinear ones, so we have to modify (\ref{T2jet1}) in order to single out the explicit contributions arising from each of the two kinematical configurations. Let us then consider a hard 
parton with momentum $p^{\mu}$ in the $S_{\vec{n}_{T}}$  hemisphere which produces a hard collinear final state parton with momentum $k^{\mu}$ after emitting a soft gluon with momentum $q^{\mu}$ 
(Fig.~\ref{branching}). The parent parton is moving along $n^{\mu}$ so its hard momentum component $p\cdot \bar{n}$ is of order $\mathcal{O}(Q)$, while the remaining components $p\cdot n$ and 
$|p_{\perp}|$ are suppressed. The soft gluon momentum components tend to zero with the same scaling ($q\cdot n\sim q\cdot\bar{n}\sim|q_{\perp}|$). It is then easy to see that 
$k^{2} \simeq p^{2}-Qq\cdot n$. Plugging the expression for $p^{2}$ arising from the previous equation into (\ref{T2jet1}), we end up with the expression
\begin{align}
1-T =& \frac{k^{2}}{Q^{2}}+\frac{\bar{k}^{2}}{Q^{2}}+\frac{q\cdot n}{Q}+\frac{\bar{q}\cdot\bar{n}}{Q}+\mathcal{O}((1-T)^{2})\notag\\
\label{T2jet2}
=& \frac{k^{2}}{Q^{2}}+\frac{\bar{k}^{2}}{Q^{2}}+\frac{w}{Q}+\mathcal{O}((1-T)^{2}),
\end{align}
where $k$ ($\bar{k}$) is the total collinear momentum in the $S_{\vec{n}_{T}}$ ($\bar{S}_{\vec{n}_{T}}$) hemisphere
 and $q$ ($\bar{q}$) is the total soft momentum in the $S_{\vec{n}_{T}}$ ($\bar{S}_{\vec{n}_{T}}$) hemisphere.

An important simplification in the calculation is achieved if the recoil effects due to emissions of wide angle soft particles are neglected. Before neglecting them we have to make sure
that they do not give rise to any logarithmically enhanced terms. To see it, we rewrite the thrust as
\begin{align}
\label{T2jet3}
 1-T = \sum_{i}\frac{\omega_{i}}{Q}(1-{\rm cos}\theta_{i}),
\end{align}
where $\omega_{i}$ is the energy of the $i-$th final state particle and $\theta_{i}$ is the angle of its direction to the thrust axis. We now consider the wide angle soft contributions to $T$, but we neglect their effect on
the determination of the thrust axis $\vec{n}_{T}$. We call this fake thrust axis $\vec{n}_{{\rm fake}}$ and we define $\delta$ as the small angle of $\vec{n}_{{\rm fake}}$ to the physical axis $\vec{n}_{T}$.
In approximating $\vec{n}_{T}$ with $\vec{n}_{{\rm fake}}$ the angles $\theta_{i}$ are replaced by $\theta_{i}^{\prime}=\theta_{i}-\delta$.

We now consider the angle $\delta$ due to a single large angle soft emission. From simple kinematics we obtain $\delta\sim \omega_{s}/Q$, where $\omega_{s}$ is the soft particle energy. We observe
that $\delta$ is of the same order as the soft emission contribution to the thrust ({\it i.e.}$\sim\tau$). We want to estimate the effect of neglecting the recoil $\delta$  on the thrust itself using expression (\ref{T2jet3}). 
The effect of the approximation on collinear emissions ($\theta\simeq0$, $\omega_{c}\sim Q$) is $\Delta\tau \sim ({\omega_{c}}/{Q})\delta^{2}\sim\tau^{2}$, while for large angle soft emissions
($\theta\gg0$, $\omega_{s}\ll Q$) we find $\Delta\tau\sim ({\omega_{s}}/{Q})\delta\sim \tau^{2}$.

We see that in both cases the effect of the recoil amounts to a contribution to $T$ of relative order (at least) $\mathcal{O}(1-T)$, so it produces power-suppressed terms. We can then replace the thrust axis
in the dijet region with the direction of the hardest (jet-initiating) parton. From now on this approximation is understood.

\begin{figure}[!htp]
\begin{center}
\includegraphics[width=80mm]{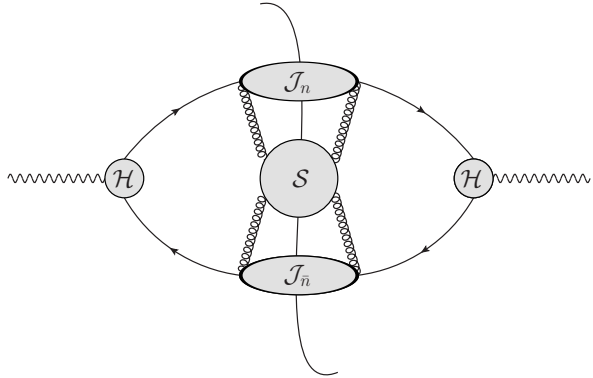}
\caption{Leading regions in dijet factorization.\label{fig:factorization}}
\end{center}
\end{figure}

Factorization properties of event-shapes have been widely studied in the 
literature~\cite{CollinsSoper,BerSterKucs1,schwartzMHnll}. Referring to Fig. \ref{fig:factorization} we recast the cross section (\ref{eq:Rfixed})
as
\begin{align}
\label{factorization1}
R_{T}(\tau)=\,\,H\left(\frac{Q}{\mu},\alpha_{s}(\mu)\right)&\int dk^{2}d\bar{k}^{2}\mathcal{J}(\frac{k}{\mu},\alpha_{s}(\mu))\mathcal{\bar{J}}(\frac{\bar{k}}{\mu},\alpha_{s}(\mu))\notag\\
\times&\int dw\mathcal{S}(\frac{w}{\mu},\alpha_{s}(\mu))\Theta(Q^{2}\tau-\bar{k}^{2}-k^{2}-wQ)+\mathcal{O}(\tau).
\end{align}
We use the integral representation of the $\Theta$-function
\begin{align}
 \Theta(Q^{2}\tau-\bar{k}^{2}-k^{2}-wQ)=\frac{1}{2\pi i}\int_{C} \frac{d\nu}{\nu}\,{\rm e}^{\nu\tau Q^{2}}{\rm e}^{-\nu k^{2}}{\rm e}^{-\nu \bar{k}^{2}}{\rm e}^{-\nu wQ},
\end{align}
and the Laplace transform to recast Eq.~(\ref{factorization1}) as
\begin{align}
R_{T}(\tau) =& H\left(\frac{Q}{\mu},\alpha_{s}(\mu)\right)\frac{1}{2\pi i}\int_{C} \frac{dN}{N}\,{\rm e}^{\tau N}
\tilde{J}^{2}\left(\sqrt{\frac{N_{0}}{N}}\frac{Q}{\mu},\alpha_{s}(\mu)\right)
\tilde{S}\left(\frac{N_{0}}{N}\frac{Q}{\mu},\alpha_{s}(\mu)\right)\notag\\
=& \frac{1}{2\pi i}\int_{C} \frac{dN}{N}\,{\rm e}^{\tau N} \tilde{\sigma}_{N}(Q^{2},\alpha_{s}),
\label{eq:RT}
\end{align}
where we set $N=\nu Q^{2}$ and $N_{0}=e^{-\gamma_{E}}$. For the sake of simplicity we defined $\tilde{\sigma}_{N}(Q^{2},\alpha_{s})$ as the Laplace-transformed cross section and we
omitted the term $\mathcal{O}(\tau)$.
The soft subprocess $\tilde{S}\left({N_{0}}/{N}{Q}/{\mu},\alpha_{s}(\mu)\right)$ describes the interaction between the two jets of hard collinear particles through soft gluon exchange. 
It can be therefore defined in a gauge invariant way as a correlator of Wilson lines
\begin{align}
\tilde{S}\left(\frac{N_{0}}{N}\frac{Q}{\mu},\alpha_{s}(\mu)\right) = \frac{Q}{N_{c}}\int d\tau_{s}{\rm e}^{-\tau_{s}N}\sum_{k_{eik}}\langle0|W^{\dagger}_{\bar{n}}(0)W^{\dagger}_{n}(0)|k_{eik}\rangle
\mathcal{J}_{cut}(\tau_{s}Q)\langle k_{eik}|W_{n}(0)W_{\bar{n}}(0)|0\rangle,
\label{eq:softsub}
\end{align}
 where we defined $\tau_{s}=w/Q$. $W_{n}$ and $W_{\bar{n}}$ are Wilson lines 
\begin{align}
W_{n}(y) = {\rm \textbf{P}}{\rm exp}\left(ig\int_{0}^{\infty}ds\,n\cdot A(ns+y)\right),
\end{align}
describing the eikonal interaction of soft gluons with the fast moving quarks along the directions $n^{\mu}$ and $\bar{n}^{\mu}$ respectively. $A(ns+y)$ in the previous expression denotes the gluon field
in QCD. The sum runs over the final states $|k_{eik}\rangle$ involving $k$ soft particles whose phase space is constrained according to the thrust measurement function $\mathcal{J}_{cut}(\tau Q^{2})$. 
Both soft and soft-collinear contributions are encoded into the soft subprocess.
The one-loop expression of the soft subprocess is known since a long time and we compute it with two-loop accuracy in the next section.

The collinear subprocess $\mathcal{J}$ ($\mathcal{\bar{J}}$) describes the decay of the jet-initiating hard quark (antiquark) into a jet of collinear particles moving along the $n^{\mu}$ ($\bar{n}^{\mu}$)
direction. It is therefore an inclusive quantity which can be found in many other relevant QCD processes such as deep inelastic scattering and heavy quarks decay \cite{Ster87,KorSter,SCETJet}.
Double-counting of soft-collinear and $\bar{n}$-collinear ($n$-collinear) contributions has to be avoided when computing the $n$-collinear ($\bar{n}$-collinear) jet subprocess. To this end different
regularization schemes can be found in the literature and they all provide the same results for the logarithmic structure of the jet subprocess.
For the purposes of the present paper the explicit expression of the collinear subprocess is not required. All we need is its non-logarithmic term at two-loop order, which has been 
computed in \cite{SCETJet} using dimensional regularization. In that work,
all pure virtual corrections to the collinear subprocess are given by scaleless integrals and thus they vanish 
in dimensional regularization. As we
will see in detail in Section \ref{sec:soft} this property holds true also for the soft subprocess and it ensures that the whole virtual contribution is encoded into the hard subprocess
 defined below, which can be identified with the squared of the constant part of the quark form factor. 
 
The short-distance hard function $H\left({Q}/{\mu},\alpha_{s}(\mu)\right)=|\mathcal{H}\left({Q}/{\mu},\alpha_{s}(\mu)\right)|^{2}$ takes into account the hard virtual corrections to 
the quark-antiquark production subprocess. It is free of large logarithms and it can be generally defined such that Eq. (\ref{factorization1}) reproduces the fixed-order cross section up to
power suppressed terms. Since the hard function contains only constants, one can avoid it by performing a $\log(R)$-matching to fixed-order as it will be shown in Section \ref{sec:matching}. 
We nevertheless take it
into account in order to compute the full constant part of the cross section with $\mathcal{O}(\alpha_{s}^{2})$ accuracy.

\section{Soft gluon emission at two-loop level}
\label{sec:soft}
Let us turn into the computation of the soft subprocess (\ref{eq:softsub}). 
Expanding the generating functional to the desired order we obtain the relevant set of cut Feynman diagrams. For real emissions, the integration runs over the soft gluons phase space
constrained by the thrust measurement function in the dijet limit.
The leading order contribution trivially reduces to $\delta(\tau_{s})$. 
In what follows we compute $S(\tau_{s},{Q}/{\mu},\alpha_{s}(\mu))$ to two-loop level.

\subsection{One-loop result}

\begin{figure}[!htp]  
\begin{center}
\includegraphics[width=7cm]{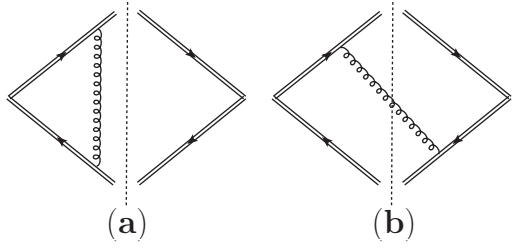}
\caption{NLO contribution to the soft subprocess.\label{singleemission}}
\end{center}
\end{figure}

At one-loop level the evaluation of the Wilson loop is straightforward. The contributing diagrams are listed in Figure \ref{singleemission}. 

The virtual soft correction (Fig.~\ref{singleemission}(a)) to the vertex identically vanishes in dimensional regularization since the integration over the loop momentum is given by the scaleless integral 
\begin{equation}
 S^{(1)}_{V}\left(\tau_{s}, \frac{Q}{\mu},\alpha_{s}(\mu)\right) = -n\cdot\bar{n}  g^{2} C_{F} \mu^{2\epsilon}\int\frac{d^{d}q}{(2 \pi)^{d}} \frac{\delta(\tau_{s})}{(q\cdot\bar{n}+i0^{+}) (q\cdot n-i0^{+})
(q^{2}+i0^{+})},
\end{equation}
where the phase space constraint reduces to $\delta(\tau_{s})$ since no real gluons have been emitted.\\
The real contribution involves the emission of a soft gluon off one of the two Wilson lines. The emitted gluon may go either into the $S_{\vec{n}_{T}}$ hemisphere or into the $\bar{S}_{\vec{n}_{T}}$
 one. The resulting phase space
measure is then
\begin{equation}
\label{NLOphasespace}
\frac{d^{d}q}{(2 \pi)^{d}} (2 \pi) \delta^{(+)}(q^{2}) (\delta(\tau_{s} Q-q\cdot n) \Theta(q\cdot\bar{n}-q\cdot n)+\delta(\tau_{s} Q-q\cdot \bar{n}) \Theta(q\cdot n-q\cdot \bar{n})),
\end{equation}
where the two $\Theta$-functions forbid the emitted gluon from going backwards, heading towards the opposite hemisphere.
The integrand is the eikonal factor corresponding to the single emission shown in Figure \ref{singleemission}(b) and its mirror conjugate diagram. The sum of the two then reads
\begin{align}
\label{NLOReal}
 S^{(1)}\left(\tau_{s}, \frac{Q}{\mu},\alpha_{s}(\mu)\right) = n\cdot \bar{n} g^{2} C_{F} Q \mu^{2 \epsilon} \int \frac{d^{d}q}{(2 \pi)^{d}} (2 \pi) \delta^{(+)}(q^{2})\notag\\
\times\frac{\delta(\tau_{s} Q-q\cdot n) \Theta(q\cdot\bar{n}-q\cdot n)+\delta(\tau_{s} Q-q\cdot \bar{n}) \Theta(q\cdot n-q\cdot \bar{n})}{(q\cdot n+i0^{+}) (q\cdot \bar{n}+i0^{+})}.
\end{align}
We first replace 
\begin{equation}
\label{measure}
d^{d}q = \frac{1}{2}d(q\cdot n)d(q\cdot \bar{n}) d^{d-2}q_{\perp}\,,
\end{equation}
and evaluate the integral over $q_{\perp}$. Since there is no explicit dependence on $q_{\perp}$ in the integrand function, it simply reduces to the replacement 
\begin{equation}
\label{replace} 
d^{d-2}q_{\perp}\delta^{(+)}(q^{2})\rightarrow \frac{\pi^{1-\epsilon}}{\Gamma(1-\epsilon)} (q\cdot n)^{-\epsilon}(q\cdot\bar{n})^{-\epsilon}\Theta(q\cdot\bar{n})\Theta(q\cdot n).
\end{equation}
It is then straightforward to show that the final result reads
\begin{equation}
\label{realoneloop}
 S^{(1)}\left(\tau_{s}, \frac{Q}{\mu},\alpha_{s}(\mu)\right)= C_{F} \frac{\alpha_{s}(\mu)}{\pi}\frac{e^{\epsilon \gamma_{E}}}{\epsilon \Gamma(1-\epsilon)} \left(\frac{Q}{\mu}\right)^{-2\epsilon}
2\left(\frac{1}{\tau_{s}}\right)^{1+2\epsilon},
\end{equation}
where we set $n\cdot \bar{n} = 2$ and we replaced the bare coupling $\alpha_{s}^{0} = {g^{2}}/{4\pi}$ with the renormalized one in the $\overline{{\rm MS}}$ scheme, {\it i.e.}
\begin{equation}
\label{msbar}
 \alpha^{0}_{s} \mu^{2 \epsilon}\rightarrow \alpha_{s}(\mu) \mu^{2\epsilon} \frac{e^{\epsilon \gamma_{E}}}{(4\pi)^{\epsilon}} \left(1-\frac{11 C_{A}-2 N_{F}}{6 \epsilon}
\big(\frac{\alpha_{s}(\mu)}{2\pi}\big)+\ldots \right),
\end{equation}
with $\gamma_{E} = 0.5772\ldots$ being the Euler-Mascheroni constant.
Using the relation
\begin{align}
 \left(\frac{1}{\tau_{s}}\right)^{1+2\epsilon} = -\frac{\delta(\tau_{s})}{2\epsilon}+\bigg[\frac{1}{\tau_{s}}\bigg]_{+}-2\epsilon\bigg[\frac{\log{\tau_{s}}}{\tau_{s}}\bigg]_{+}+\mathcal{O}(\epsilon^{2}),
\end{align}
we recast (\ref{realoneloop}) as
\begin{align}
 S^{(1)}\left(\tau_{s}, \frac{Q}{\mu},\alpha_{s}(\mu)\right)=C_{F} \frac{\alpha_{s}(\mu)}{\pi}
\bigg(-\frac{\delta(\tau_{s})}{\epsilon^{2}}+\frac{2}{\epsilon}\bigg[\frac{1}{\tau_{s}}\bigg]_{+}-\frac{2}{\epsilon}\log{\frac{\mu}{Q}}\delta(\tau_{s})+\notag\\
\label{oneloopsoft}
\delta(\tau_{s})\left(\frac{\pi^{2}}{12}-2\log^{2}{\frac{Q}{\mu}}\right)- 4\bigg[\frac{\log{\frac{Q}{\mu}}+\log{\tau_{s}}}{\tau_{s}}\bigg]_{+}+\mathcal{O}(\epsilon)\bigg) .
\end{align}
The above result in Laplace space reads
\begin{align}
\tilde{S}^{(1)}\left(\frac{N_{0}Q}{N\mu},\alpha_{s}(\mu)\right) = \int_{0}^{\infty}d\tau_{s}e^{-N\tau_{s}}S^{(1)}\left(\tau_{s}, \frac{Q}{\mu},\alpha_{s}(\mu)\right)\notag \\ =
C_{F}\frac{\alpha_{s}(\mu)}{\pi}\bigg(-\frac{1}{\epsilon^{2}}-\frac{2}{\epsilon}\log{\frac{N\mu}{N_{0}Q}}-\frac{\pi^{2}}{4}-2\log^{2}{\frac{N\mu}{N_{0}Q}}\bigg),
\end{align}
where $N_{0} = e^{-\gamma_{E}}$.

Note the absence of single logarithms in the $\mathcal{O}(\epsilon^{0})$ term meaning that the one loop soft subprocess receives logarithmic contributions only when the emitted gluon is both soft and
collinear to one of the eikonal legs. Wide angle soft emissions do not contribute at this order, but they become relevant at two-loop level.

\subsection{Two-loop result}
\label{soft function calculation}

\begin{figure}[!htp]  
\begin{center}
\includegraphics[width=10cm]{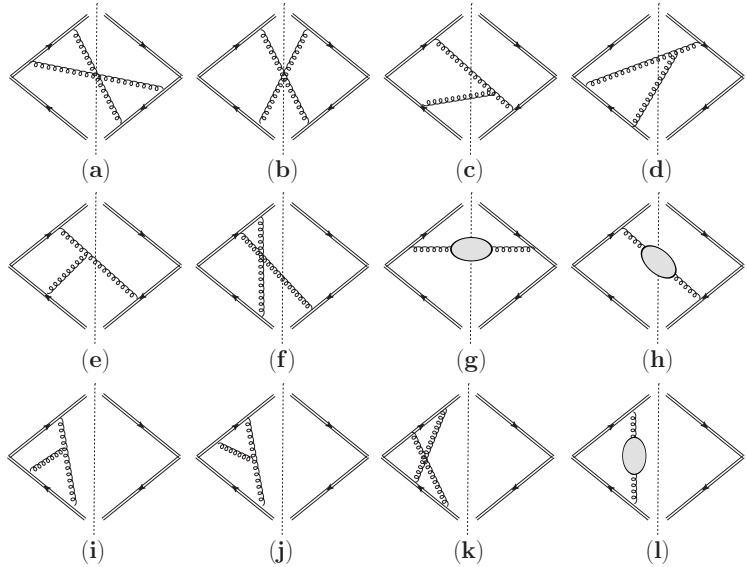}
\caption{NNLO contribution to the soft subprocess. Grey blobs stand for the sum of vacuum polarization bubbles due to fermions, gluons and ghosts.
 To complete the set one has to take into account mirror conjugate diagrams in addition. Pure abelian diagrams ({\it i.e.} proportional to $C_{F}^{2}$) are omitted.}
\label{doubleemission}
\end{center}
\end{figure}

\begin{figure}[!htp]  
\begin{center}
\includegraphics[width=15cm]{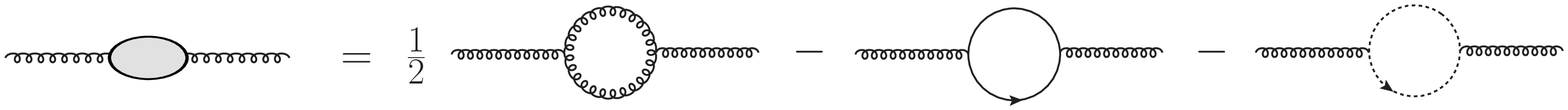}
\caption{Vacuum polarization diagrams in the Feynman gauge.}
\label{selfenergy}
\end{center}
\end{figure}

At order $\mathcal{O}(\alpha_{s}^{2})$ up to two real gluons are emitted. Note that diagrams with virtual dressing of eikonal lines vanish identically since the Wilson path lies on the light cone
({\it i.e.} $n^{2}=\bar{n}^{2}=0$).
 Furthermore, we only need to evaluate those diagrams contributing with maximal non-abelian ({\it i.e.} $C_{F}C_{A}$) and fermionic ({\it i.e.} $C_{F}T_{F}N_{F}$) terms, 
because of the non-abelian exponentiation theorem \cite{FT,NAET}. According to it we can write
\begin{align}
 \label{NAET}
\tilde{S}\left(\frac{N_{0}Q}{N\mu},\alpha_{s}(\mu)\right) = 1+\sum_{l=1}^{\infty}\tilde{S}^{(l)}\left(\frac{N_{0}Q}{N\mu},\alpha_{s}(\mu)\right) = {\rm
exp}\left(\sum_{l=1}^{\infty}\tilde{s}^{(l)}\left(\frac{N_{0}Q}{N\mu},\alpha_{s}(\mu)\right)\right),
\end{align}
where $\tilde{s}^{(2)}({N_{0}Q}/{(N\mu)},\alpha_{s}(\mu))$ involves $C_{F}C_{A}$ and $C_{F}T_{F}N_{F}$ contributions while $C_{F}^{2}$ terms arise from exponentiation of the 
$\mathcal{O}(\alpha_{s})$ result $\tilde{S}^{(1)}({N_{0}Q}/{(N\mu)},\alpha_{s}(\mu))$.
It follows  that the $\mathcal{O}(\alpha_{s}^{2})$ result is given by 
\begin{align}
\label{twoloopexp}
 \tilde{S}^{(2)}\left(\frac{N_{0}Q}{N\mu},\alpha_{s}(\mu)\right) = \frac{1}{2}\tilde{S}^{(1)}\left(\frac{N_{0}Q}{N\mu},\alpha_{s}(\mu)\right)^{2}+\tilde{s}^{(2)}\left(\frac{N_{0}Q}{N\mu},\alpha_{s}(\mu)\right),
\end{align}
where $\tilde{S}^{(1)}({N_{0}Q}/{(N\mu)},\alpha_{s}(\mu))$ was computed in the previous section.
The remaining set of (non-vanishing) diagrams contributing to $\tilde{s}^{(2)}({N_{0}Q}/{(N\mu)},\alpha_{s}(\mu))$ is shown in Figure \ref{doubleemission}, where the mirror conjugate diagrams are
omitted. The vacuum polarization blob includes fermions, gluons and ghosts as depicted in Figure \ref{selfenergy}, since the calculation will be carried out in the Feynman gauge. 

The double virtual contribution (Fig.~\ref{doubleemission}(i--l)) is made of scaleless integrals which identically vanish in dimensional regularization. \\
The first non-trivial contribution to consider is the one-loop virtual diagram with an extra real gluon, depicted in Fig.~\ref{doubleemission}(e,f). The sum of such diagrams and their mirror conjugate
 ones is gauge invariant. Using the known expression of the soft current at one loop \cite{OneLoopSoft}, we can write their contribution  as
\begin{align}
s^{(2)}_{(e),(f)}\left(\tau_{s}, \frac{Q}{\mu},\alpha_{s}(\mu)\right) = -4 \frac{g^{4}}{8\pi^{2}} (4 \pi \mu^{4})^{\epsilon} Q C_{F} C_{A}
\frac{1}{\epsilon^{2}}\frac{\Gamma^{4}(1-\epsilon)\Gamma^{3}(1+\epsilon)}{\Gamma^{2}(1-2\epsilon)\Gamma(1+2\epsilon)} \notag\\
\times\int\frac{d^{d}q}{(2\pi)^{d}}\frac{(2\pi)\delta^{(+)}(q^{2})(\delta(\tau_{s} Q-q\cdot n) \Theta(q\cdot\bar{n}-q\cdot n)+\delta(\tau_{s} Q-q\cdot \bar{n}) \Theta(q\cdot n-q\cdot
\bar{n}))}{((q\cdot n) (q\cdot \bar{n}))^{1+\epsilon}},
\end{align}
where the phase space constraint is the same as in the one loop case, since only one real soft gluon is emitted. Notice that the pole prescription ($\pm i0$) has been omitted since the poles are never
``touched'' during the integration because of the $\Theta$-functions in the numerator. From now on we write explicitly the factor ($\pm i0$) only when relevant.
Using the relation (\ref{measure}) and evaluating the straightforward integral left we obtain
\begin{align}
\label{RplusV}
s^{(2)}_{(e),(f)}\left(\tau_{s}, \frac{Q}{\mu},\alpha_{s}(\mu)\right) = -\frac{\alpha_{s}(\mu)^{2}}{\pi^{2}} \frac{C_{F}C_{A}}{2}\frac{e^{2\gamma_{E} \epsilon}}{\epsilon^{3}}
\frac{\Gamma^{3}(1-\epsilon)\Gamma^{3}(1+\epsilon)}{\Gamma^{2}(1-2\epsilon)\Gamma(1+2\epsilon)}\left(\frac{\mu}{Q}\right)^{4\epsilon} \left(\frac{1}{\tau_{s}}\right)^{1+4\epsilon},
\end{align}
where we expressed the bare coupling according to equation (\ref{msbar}). \\

Let us now consider the double real emission.  The phase space constraint is more  involved since two distinct gluons are emitted. Each of them can indeed go either into the $S_{\vec{n}_{T}}$
 or the $\bar{S}_{\vec{n}_{T}}$ hemisphere, leading to the following phase space cut
\begin{align}
 Q (2\pi)\delta^{(+)}(q^{2})&(2\pi)\delta^{(+)}(k^{2}) \mathcal{J}_{cut}(\tau_{s} Q) = Q (2\pi)\delta^{(+)}(q^{2})(2\pi)\delta^{(+)}(k^{2})\notag\\
\label{cut1}
\times&(\delta(\tau_{s} Q-q\cdot n-k\cdot n) \Theta(q\cdot \bar{n}-q\cdot n) \Theta(k\cdot \bar{n}-k\cdot n)\\
\label{cut2} 
&+\delta(\tau_{s} Q-q\cdot \bar{n}-k\cdot \bar{n}) \Theta(q\cdot n-q\cdot \bar{n}) \Theta(k\cdot n-k\cdot \bar{n})\\
\label{cut3}
&+\delta(\tau_{s} Q-q\cdot n-k\cdot \bar{n}) \Theta(q\cdot \bar{n}-q\cdot n)\Theta(k\cdot n-k\cdot \bar{n})\\
\label{cut4}
&+\delta(\tau_{s} Q-k\cdot n-q\cdot \bar{n}) \Theta(k\cdot \bar{n}-k\cdot n)\Theta(q\cdot n-q\cdot \bar{n})).
\end{align}
Such a constraint gives rise to non-trivial phase space integrals which are solvable using analytic techniques. We evaluate the integrals in {\tt MATHEMATICA}, partly using the 
 package {\tt HypExp} \cite{HypExp}.
An independent numerical check is also performed using the computer code {\tt SecDec} \cite{secdec}, a recent implementation of sector decomposition
based on the algorithms presented in \cite{sectordeco1,sectordeco2}. The numerical integration is then carried out using the Monte Carlo routines {\tt BASES} \cite{bases} and {\tt
VEGAS}
 \cite{vegas} included in \cite{secdec} with $10^{7}$ Monte Carlo events per coefficient. The two results agree within an uncertainty of $0.001\%$.

 We organise the calculation considering first the class of diagrams without any internal 
 gluons ({\it i.e.}~gluons which are not involved in the cut) shown in Fig.~\ref{doubleemission}(a,b) 
and mirror conjugate diagrams, then the class of diagrams with only one internal gluon 
(Fig.~\ref{doubleemission}(c,d) and mirror conjugate diagrams)
 and finally vacuum polarization diagrams, containing two internal gluons 
 (Fig.~\ref{doubleemission}(g,h) and mirror conjugate diagrams).

\subsubsection{Box-type diagrams}
We first consider the Box-type diagrams depicted in Fig.~\ref{doubleemission}(a,b). The second diagram (Fig.~\ref{doubleemission}(b)) has a simple structure and can be evaluated easily. Dropping
the abelian part of the colour factor  we are left with
\begin{align}
 s^{(2)}_{(b)}\left(\tau_{s}, \frac{Q}{\mu},\alpha_{s}(\mu)\right) = -\frac{C_{A} C_{F}}{2}	(n\cdot\bar{n})^{2} g^{4} \mu^{4 \epsilon}\notag\\
\times\int\frac{d^{d}q}{(2\pi)^{d}}\frac{d^{d}k}{(2\pi)^{d}}\frac{Q (2\pi)\delta^{(+)}(q^{2})(2\pi)\delta^{(+)}(k^{2}) \mathcal{J}_{cut}(\tau_{s} Q)}{(q\cdot \bar{n})(q\cdot n)(k\cdot n)(k\cdot
\bar{n})},
\end{align}
where $\mathcal{J}_{cut}(\tau_{s} Q)$ is the sum of terms in round brackets defined in 
Eqs.~(\ref{cut1}--\ref{cut4}).
The integrand function does not depend on the transverse component of the integrated momenta, so we can use Eqs.~(\ref{measure},\ref{replace}) getting
\begin{align}
 s^{(2)}_{(b)}\left(\tau_{s}, \frac{Q}{\mu},\alpha_{s}(\mu)\right) = -\frac{C_{A} C_{F}}{2}Q \frac{(n\cdot\bar{n})^{2}}{(2\pi)^{2d-2}} g^{4} \mu^{4
\epsilon}\frac{\pi^{2-2\epsilon}}{4\Gamma^{2}(1-\epsilon)}\hspace{4cm}\notag\\
\times\int d(q\cdot n)d(q\cdot \bar{n})d(k\cdot n)d(k\cdot \bar{n})\bigg(\frac{\delta(\tau_{s} Q-q\cdot n-k\cdot n) \Theta(q\cdot \bar{n}-q\cdot n) \Theta(k\cdot \bar{n}-k\cdot n)}{(q\cdot
\bar{n})^{1+\epsilon}(q\cdot n)^{1+\epsilon}(k\cdot n)^{1+\epsilon}(k\cdot \bar{n})^{1+\epsilon}}&+&\notag\\
\frac{\delta(\tau_{s} Q-q\cdot \bar{n}-k\cdot \bar{n}) \Theta(q\cdot n-q\cdot \bar{n}) \Theta(k\cdot n-k\cdot \bar{n})}{(q\cdot \bar{n})^{1+\epsilon}(q\cdot n)^{1+\epsilon}(k\cdot
n)^{1+\epsilon}(k\cdot \bar{n})^{1+\epsilon}}&+&\notag\\
\frac{\delta(\tau_{s} Q-q\cdot n-k\cdot \bar{n}) \Theta(q\cdot \bar{n}-q\cdot n)\Theta(k\cdot n-k\cdot \bar{n})}{(q\cdot \bar{n})^{1+\epsilon}(q\cdot n)^{1+\epsilon}(k\cdot n)^{1+\epsilon}(k\cdot
\bar{n})^{1+\epsilon}}&+&\notag\\
\label{explosion1}	
\frac{\delta(\tau_{s} Q-k\cdot n-q\cdot \bar{n}) \Theta(k\cdot \bar{n}-k\cdot n)\Theta(q\cdot n-q\cdot \bar{n})}{(q\cdot \bar{n})^{1+\epsilon}(q\cdot n)^{1+\epsilon}(k\cdot n)^{1+\epsilon}(k\cdot
\bar{n})^{1+\epsilon}}\bigg).
\end{align}
We now analyse each of the integrals in Eq.~(\ref{explosion1}). We first integrate out the $k\cdot n$ component in the first and fourth integral and the $k\cdot \bar{n}$ component in the second
and third integral by using the $\delta$-functions. We then make the following four changes of variables
\begin{align}
 \label{param1}
 k\cdot\bar{n}\rightarrow Q\tau_{s}u(1-t) && q\cdot\bar{n}\rightarrow Q\tau_{s} ts && q\cdot n\rightarrow Q\tau_{s} t,\\
 \label{param2}
 k\cdot n\rightarrow Q\tau_{s}u(1-t) && q\cdot n\rightarrow Q\tau_{s} ts && q\cdot\bar{n}\rightarrow Q\tau_{s} t,\\
 \label{param3}
 k\cdot n\rightarrow Q\tau_{s}u(1-t) && q\cdot \bar{n}\rightarrow Q\tau_{s} ts && q\cdot n\rightarrow Q\tau_{s} t,\\
 \label{param4}
 k\cdot\bar{n}\rightarrow Q\tau_{s}u(1-t) && q\cdot n\rightarrow Q\tau_{s} ts && q\cdot\bar{n}\rightarrow Q\tau_{s} t,
\end{align}
in the first, second, third and fourth integrals of Eq. (\ref{explosion1}) respectively, followed by the replacements $ s\rightarrow\frac{1}{s}\,,\, u\rightarrow\frac{1}{u}$ in each of them. We end up
with four identical integrals on a three dimensional unit cube. Summing them up we obtain the following expression
\begin{align}
  s^{(2)}_{(b)}\left(\tau_{s}, \frac{Q}{\mu},\alpha_{s}(\mu)\right) = 
  -\frac{C_{A} C_{F}}{2}Q \frac{(n\cdot\bar{n})^{2}}{(2\pi)^{2d-2}} g^{4} \mu^{4
\epsilon}\frac{\pi^{2-2\epsilon}}{\Gamma^{2}(1-\epsilon)}\notag\\
\times (Q\tau_{s})^{-1-4\epsilon} \int_{0}^{1} dt\,du\,ds (1-t)^{-1-2\epsilon}t^{-1-2\epsilon}u^{-1+\epsilon}s^{-1+\epsilon},
\end{align}
that can be easily evaluated in terms of Euler beta functions yielding
\begin{align}
 s^{(2)}_{(b)}\left(\tau_{s}, \frac{Q}{\mu},\alpha_{s}(\mu)\right) = -\frac{C_{A} C_{F}}{2}Q \frac{(n\cdot\bar{n})^{2}}{(2\pi)^{2d-2}} g^{4} \mu^{4
\epsilon}\frac{\pi^{2-2\epsilon}}{\Gamma^{2}(1-\epsilon)}(Q\tau_{s})^{-1-4\epsilon} \frac{1}{\epsilon^{2}}B(-2\epsilon,-2\epsilon).
\end{align}
After replacing the bare coupling with the renormalized one (Eq. (\ref{msbar})) and some algebra, we can recast the previous equation as follows
\begin{align}
\label{box1}
s^{(2)}_{(b)}\left(\tau_{s}, \frac{Q}{\mu},\alpha_{s}(\mu)\right) = -\frac{C_{A} C_{F}}{2}\frac{\alpha_{s}(\mu)^{2}}{\pi^{2}}\left(\frac{\mu}{Q}\right)^{4\epsilon} \left(\frac{1}{\tau_{s}}\right)^{1+4\epsilon}
\frac{e^{2\gamma_{E}\epsilon}B(-2\epsilon,-2\epsilon)}{\epsilon^{2}\Gamma^{2}(1-\epsilon)},
\end{align}
where we set $n\cdot\bar{n}=2$. 

We now consider the box diagram in Fig.~\ref{doubleemission}(a). We note that the mirror symmetrical diagram has the same expression provided we exchange $k\leftrightarrow q$, so it can be taken into
account by just including a factor of two. The full result reads
\begin{align}
 s^{(2)}_{(a)}\left(\tau_{s}, \frac{Q}{\mu},\alpha_{s}(\mu)\right) = -C_{A} C_{F}(n\cdot\bar{n})^{2} g^{4} \mu^{4 \epsilon}\notag\\
\times\int\frac{d^{d}q}{(2\pi)^{d}}\frac{d^{d}k}{(2\pi)^{d}}\frac{Q (2\pi)\delta^{(+)}(q^{2})(2\pi)\delta^{(+)}(k^{2}) \mathcal{J}_{cut}(\tau_{s} Q)}{((q+k)\cdot \bar{n})(q\cdot
n)(q\cdot\bar{n})((k+q)\cdot\bar{n})}.
\end{align}

\noindent Furthermore, exploiting the symmetry of the integrand under the transformation $$\{ k\cdot n\leftrightarrow k\cdot\bar{n},q\cdot n\leftrightarrow q\cdot\bar{n} \},$$ we see that the integrals arising
from the terms (\ref{cut1}) and (\ref{cut3}) equal those arising from (\ref{cut2}) and (\ref{cut4}) respectively. Using the parametrization shown in Eq.~(\ref{param1}--\ref{param4}), we are led to
the following expression
\begin{align}
\label{box2tilde}
& s^{(2)}_{(a)}\left(\tau_{s},\frac{Q}{\mu},\alpha_{s}(\mu)\right) = -\frac{C_{A} C_{F}}{2}\frac{\alpha_{s}(\mu)^{2}}{\pi^{2}}\left(\frac{\mu}{Q}\right)^{4\epsilon}
\left(\frac{1}{\tau_{s}}\right)^{1+4\epsilon}\frac{e^{2\gamma_{E}\epsilon}}{\Gamma^{2}(1-\epsilon)}\notag\\
\;&\times\bigg(\int_{0}^{1}dt\,du\,ds\frac{(1-t)^{-1-2\epsilon}t^{1-2\epsilon}s^{-1+\epsilon}u^{\epsilon}}{tu+s(1-t)}+\int_{0}^{1}dt\,du\,ds\frac{(1-t)^{-1-2\epsilon}t^{1-2\epsilon}s^{-1+\epsilon}u^{
\epsilon}}{(t+s(1-t))(1-t(1-u))}\bigg).
\end{align}
The two integrals in Eq.~(\ref{box2tilde}) can be easily evaluated with the desired accuracy, yielding
\begin{align}
\int_{0}^{1}dt\,du\,ds\frac{(1-t)^{-1-2\epsilon}t^{1-2\epsilon}s^{-1+\epsilon}u^{\epsilon}}{tu+s(1-t)} = \frac{\pi^{2}}{6}\frac{1}{\epsilon}+4
\zeta_{3}+\frac{\pi^{4}}{9}\epsilon+\mathcal{O}(\epsilon^{2}),
\end{align}
\begin{align}
\int_{0}^{1}dt\,du\,ds\frac{(1-t)^{-1-2\epsilon}t^{1-2\epsilon}s^{-1+\epsilon}u^{\epsilon}}{(t+s(1-t))(1-t(1-u))} = \frac{1}{2\epsilon^{3}}+\frac{\pi^{2}}{3}\frac{1}{\epsilon}+6
\zeta_{3}+\frac{\pi^{4}}{5}\epsilon+\mathcal{O}(\epsilon^{2}).
\end{align}
Plugging them back into Eq.~(\ref{box2tilde}) we find 
\begin{align}
 \label{box2}
s^{(2)}_{(a)}\left(\tau_{s}, \frac{Q}{\mu},\alpha_{s}(\mu)\right) = -\frac{C_{A} C_{F}}{2}\frac{\alpha_{s}(\mu)^{2}}{\pi^{2}}\left(\frac{\mu}{Q}\right)^{4\epsilon}
\left(\frac{1}{\tau_{s}}\right)^{1+4\epsilon}\frac{e^{2\gamma_{E}\epsilon}}{\Gamma^{2}(1-\epsilon)}\notag\\\times\bigg(\frac{1}{2\epsilon^{3}}+\frac{\pi^{2}}{2}\frac{1}{\epsilon}+10\zeta_{3}+\frac{14}{45}\pi^{4}
\epsilon+\mathcal{O}(\epsilon^{2})\bigg).
\end{align}
This completes the evaluation of Box-type diagrams contribution to the two loop soft subprocess.

\subsubsection{Non-abelian diagrams}
The class of non-abelian diagrams (Fig.~\ref{doubleemission}(c,d) and mirror symmetrical diagrams) is much more involved due to the presence of the three gluon vertex. 
The mirror conjugate diagrams in which the real gluon is connected to the opposite leg to the right of the cut are related to those depicted in Fig.~\ref{doubleemission}(c,d) 
by the transformation $$\{k\cdot n\leftrightarrow q\cdot\bar{n},k\cdot\bar{n}\leftrightarrow q\cdot n\}.$$ They can be taken into account by including a factor of two.

Thus we can write the whole non-abelian contribution as follows
\begin{align}
s^{(2)}_{(c),(d)}\left(\tau_{s}, \frac{Q}{\mu},\alpha_{s}(\mu)\right) =
C_{F}C_{A}\mu^{4\epsilon}g^{4}(n\cdot\bar{n})\int\frac{d^{d}q}{(2\pi)^{d-1}}\frac{d^{d}k}{(2\pi)^{d-1}}Q\frac{\delta^{(+)}(q^{2})\delta^{(+)}(k^{2})}{((q+k)^{2}-i0)}\notag\\
\times\mathcal{J}_{cut}(\tau_{s} Q)(\frac{2k\cdot n+q\cdot n}{(k\cdot n)(q\cdot\bar{n})((q+k)\cdot n)}+\frac{q\cdot n-k\cdot n}{((k+q)\cdot n)(k\cdot n)((k+q)\cdot \bar{n})}+\notag\\
\{k\cdot n\leftrightarrow q\cdot\bar{n},k\cdot\bar{n}\leftrightarrow q\cdot n\}).\hspace{7cm}
\end{align}
Looking at the expression of the phase space constraint $\mathcal{J}_{cut}$ we see that the transformation $$\{k\cdot n\leftrightarrow q\cdot\bar{n},k\cdot\bar{n}\leftrightarrow q\cdot n\}$$ has only the
effect of exchanging the terms (\ref{cut1}) and (\ref{cut3}) with (\ref{cut2}) and (\ref{cut4}) respectively, so the contribution due to the last term in round brackets amounts to multiply once again
by two.
Unlike the box-type case, the integrand function depends explicitly on the transverse components through the gluon propagator. We use (\ref{measure}) and perform the integral over the transverse
components as shown in Eq.~(\ref{soltransverse1}). Writing the result in terms of the parametrization (\ref{param1}--\ref{param4}) we end up with the following
expression
\begin{align}
s^{(2)}_{(c),(d)}\left(\tau_{s}, \frac{Q}{\mu},\alpha_{s}(\mu)\right) =
\frac{C_{F}C_{A}}{2}g^{4}\frac{(n\cdot\bar{n})}{(2\pi)^{2d-2}}\left(\frac{\mu}{Q}\right)^{4\epsilon}\left(\frac{1}{\tau_{s}}\right)^{1+4\epsilon}\frac{\pi^{3/2-2\epsilon}}{\Gamma(1-\epsilon)}4^{-\epsilon}
\frac{\Gamma(\frac{1}{2}-\epsilon)}{\Gamma(1-2\epsilon)}\notag\\
\times\int_{0}^{1}dt\,du\,ds\bigg((1-t)^{-1-2\epsilon}t^{-1-2\epsilon}s^{-1+\epsilon}u^{-1+\epsilon}(s+u)\frac{s(t-2)(t-1)+tu(1-t)}{s(1-t)+tu}\notag\\
\times\frac{{}_2F_{1}\left(1,\frac{1}{2}-\epsilon,1-2\epsilon,4\frac{\sqrt{su}}{(\sqrt{s}+\sqrt{u})^{2}}\right)}{(\sqrt{s}+\sqrt{u})^{2}}+(1-t)^{-1-2\epsilon}t^{-1-2\epsilon}s^{-1+\epsilon}u^{-1+\epsilon
}(1+su)\hspace{0.7cm}\notag\\
\label{nonabeltilde}
\times\frac{s(1-t)(2-t(2-u))+t(1-t(1-2u))}{(s(1-t)+t)(1-t(1-u))}\frac{{}_2F_{1}\left(1,\frac{1}{2}-\epsilon,1-2\epsilon,4\frac{\sqrt{su}}{(1+\sqrt{su})^{2}}\right)}{(1+\sqrt{su})^{2}}\bigg)\,,\hspace{0.55cm
}
\end{align}
where the first integral is made of the contributions due to the terms (\ref{cut1}) and (\ref{cut2}), while the ones due to (\ref{cut3}) and (\ref{cut4}) are encoded in the second integral.
Since $s\leq1$ and $u\leq1$ we can use Eq. (\ref{hyper1}) to perform the following replacements
\begin{align}
\label{id1}
{}_2F_{1}\left(1,\frac{1}{2}-\epsilon,1-2\epsilon,4\frac{\sqrt{su}}{(\sqrt{s}+\sqrt{u})^{2}}\right) &=
(\sqrt{s}+\sqrt{u})^{2}\bigg(\Theta(s-u)\frac{1}{s}\,{}_{2}F_{1}\big(1,1+\epsilon,1-\epsilon,\frac{u}{s}\big)+\notag\\
&+\Theta(u-s)\frac{1}{u}\,{}_{2}F_{1}\big(1,1+\epsilon,1-\epsilon,\frac{s}{u}\big)\bigg),\\	
\label{id2}
{}_2F_{1}\left(1,\frac{1}{2}-\epsilon,1-2\epsilon,4\frac{\sqrt{su}}{(1+\sqrt{su})^{2}}\right) &= (1+\sqrt{su})^{2} {}_{2}F_{1}\big(1,1+\epsilon,1-\epsilon,su\big),\hspace{2.3cm}
\end{align}
in the first and second integral of (\ref{nonabeltilde}) respectively.

Identity (\ref{id1}) splits the first integral in (\ref{nonabeltilde}) into two simpler integrals. We substitute $u\rightarrow zs$ in the first of such integrals and $s\rightarrow zu$ in the second
one in order to remap the integration range to the unit cube and we sum them up. After renaming the variables we finally obtain
\begin{align}
\label{nonabeltilde2}
s^{(2)}_{(c),(d)}\left(\tau_{s}, \frac{Q}{\mu},\alpha_{s}(\mu)\right) =
\frac{C_{F}C_{A}}{8}(n\cdot\bar{n})\frac{\alpha^{2}_{s}(\mu)}{\pi^{2}}\frac{4^{-\epsilon}e^{2\gamma_{E}\epsilon}}{\sqrt{\pi}}\frac{\Gamma(\frac{1}{2}-\epsilon)}{\Gamma(1-\epsilon)\Gamma(1-2\epsilon)}
\left(\frac{\mu}{Q}\right)^{4\epsilon}\left(\frac{1}{\tau_{s}}\right)^{1+4\epsilon}\notag\\
\times\int_{0}^{1}dt\,du\,dz\bigg((1-t)^{-1-2\epsilon}t^{-1-2\epsilon}u^{-1+2\epsilon}z^{-1+\epsilon}(1+z)\frac{4z+t(1-t)(3-z)(1-3z)}{(1-t(1-z))(z+t(1-z))}\hspace{1cm}\notag\\
\times{}_{2}F_{1}(1,1+\epsilon,1-\epsilon,z)+(1-t)^{-1-2\epsilon}t^{-1-2\epsilon}u^{-1+\epsilon}z^{-1+\epsilon}(1+uz)\hspace{4cm}\notag\\
\times\frac{z(1-t)(2-t(2-u))+t(1-t(1-2u))}{(z(1-t)+t)(1-t(1-u))}{}_{2}F_{1}\big(1,1+\epsilon,1-\epsilon,zu\big)\bigg).\hspace{3.4cm}	
\end{align}
where we used Eq.~(\ref{msbar}) for the coupling.
The two integrals in round brackets of Eq.~(\ref{nonabeltilde2}) are given by
\begin{align}
\int_{0}^{1}dt\,du&\,dz\bigg((1-t)^{-1-2\epsilon}t^{-1-2\epsilon}u^{-1+2\epsilon}z^{-1+\epsilon}(1+z)\frac{4z+t(1-t)(3-z)(1-3z)}{(1-t(1-z))(z+t(1-z))}\notag\\
&\times{}_{2}F_{1}(1,1+\epsilon,1-\epsilon,z)\bigg) =
\frac{1}{2\epsilon}\frac{\Gamma(1-\epsilon)}{\Gamma(1+\epsilon)\Gamma(-2\epsilon)}\bigg(-\frac{1}{\epsilon^{3}}-\frac{2}{\epsilon^{2}}-\frac{4+\pi^{2}}{\epsilon}-\notag\\
&-\frac{4}{3} \left(6+\pi ^2+6 \zeta_{3}\right)-\frac{1}{9} \left(144+24 \pi ^2+\pi ^4+180 \zeta_{3}\right)\epsilon+\mathcal{O}(\epsilon^{2})\bigg),\\
\int_{0}^{1}dt\,du&\,dz\bigg((1-t)^{-1-2\epsilon}t^{-1-2\epsilon}u^{-1+\epsilon}z^{-1+\epsilon}(1+uz)\frac{z(1-t)(2-t(2-u))+t(1-t(1-2u))}{(z(1-t)+t)(1-t(1-u))}\notag \\
&\times{}_{2}F_{1}\big(1,1+\epsilon,1-\epsilon,zu\big)\bigg) =
\frac{\Gamma(1-\epsilon)}{\Gamma(1+\epsilon)\Gamma(-2\epsilon)}\bigg(-\frac{\pi^{2}}{3}\frac{1}{\epsilon^{2}}-\frac{1}{\epsilon}\left(14\zeta_{3}-\frac{2}{3}\pi^{2}\right)-\notag\\
&-\frac{1}{15} \left(20 \pi ^2+9 \pi ^4-420 \zeta_{3}\right)+\mathcal{O}(\epsilon)\bigg).
\end{align}
Plugging them into (\ref{nonabeltilde2}) and setting $n\cdot\bar{n}=2$ we find
\begin{align}
 \label{nonabel}
s^{(2)}_{(c),(d)}\left(\tau_{s}, \frac{Q}{\mu},\alpha_{s}(\mu)\right) &=
\frac{C_{F}C_{A}}{4}\frac{\alpha^{2}_{s}(\mu)}{\pi^{2}}\left(\frac{\mu}{Q}\right)^{4\epsilon}\left(\frac{1}{\tau_{s}}\right)^{1+4\epsilon}\bigg(\frac{1}{\epsilon^{3}}+\frac{2}{\epsilon^{2}}+\frac{1}{
\epsilon}\left(4+\frac{7}{6}\pi^{2}\right)+\notag\\
&+\left(8-\pi ^2+\frac{100 \zeta_{3}}{3}\right)+\notag\\
&+\left(16+\frac{10 \pi ^2}{3}+\frac{199 \pi ^4}{360}-\frac{124 \zeta_{3}}{3}\right)\epsilon+\mathcal{O}(\epsilon^{2})\bigg).
\end{align}

\subsubsection{Vacuum polarization diagrams}
The last class of diagrams we have to take into account to complete the computation of the double real radiation contribution involves diagrams of the type shown in Fig.~\ref{doubleemission}(g,h).
Summing up diagrams (g) and (h) and their mirror symmetrical ones we end up with the following expression
\begin{align}
s^{(2)}_{(g),(h)}\left(\tau_{s}, \frac{Q}{\mu},\alpha_{s}(\mu)\right) &=
g^{4}\mu^{4\epsilon}\int\frac{d^{d}q}{(2\pi)^{d-1}}\frac{d^{d}k}{(2\pi)^{d-1}}Q\delta^{(+)}(q^{2})\delta^{(+)}(k^{2})\mathcal{J}_{cut}(Q\tau_{s})\notag\\
&\times\bigg(2\frac{((q\cdot\bar{n})(k\cdot n)-(k\cdot\bar{n})(q\cdot n))^{2}}{((k+q)\cdot\bar{n})^{2}((k+q)\cdot
n)^{2}}\frac{2(1-\epsilon)C_{A}C_{F}-4C_{F}T_{F}n_{F}}{((k+q)^{2}+i0)((k+q)^{2}-i0)}+\notag\\
\label{vacuumint}
&+8\frac{1}{((k+q)\cdot\bar{n})((k+q)\cdot n)}\frac{C_{F}T_{F}n_{F}-C_{A}C_{F}}{(k+q)^{2}+i0}\bigg),\hspace{3.1cm}
\end{align}
where the phase space constraint $\mathcal{J}_{cut}(Q\tau_s)$ is the usual measurement function defined in (\ref{cut1}-\ref{cut4}).
We name the two integrals appearing in (\ref{vacuumint}) $I_{(g),(h)}^{(2),(a)}$ and $I_{(g),(h)}^{(2),(b)}$ respectively and we evaluate them below.
We first use (\ref{measure}) getting
\begin{align}
I_{(g),(h)}^{(2),(a)}=&\frac{1}{4}\int\frac{d(q\cdot n)d(q\cdot \bar{n}) }{(2\pi)^{d-1}}\frac{d(q\cdot n)d(q\cdot \bar{n})}{(2\pi)^{d-1}}Q\mathcal{J}_{cut}(Q\tau_{s})\frac{((q\cdot\bar{n})(k\cdot
n)-(k\cdot\bar{n})(q\cdot n))^{2}}{((k+q)\cdot\bar{n})^{2}((k+q)\cdot n)^{2}}\notag\\
\label{int01}
\times&\int\frac{d^{d-2}q_{\perp}d^{d-2}q_{\perp}\delta^{(+)}(q^{2})\delta^{(+)}(k^{2})}{((k+q)^{2}+i0)((k+q)^{2}-i0)},\notag\\ \\
I_{(g),(h)}^{(2),(b)}=&\frac{1}{4}\int\frac{d(q\cdot n)d(q\cdot \bar{n}) }{(2\pi)^{d-1}}\frac{d(q\cdot n)d(q\cdot
\bar{n})}{(2\pi)^{d-1}}Q\mathcal{J}_{cut}(Q\tau_{s})\frac{1}{((k+q)\cdot\bar{n})((k+q)\cdot n)}\hspace{1cm}\notag\\
\label{int02}
\times&\int\frac{d^{d-2}q_{\perp}d^{d-2}q_{\perp}\delta^{(+)}(q^{2})\delta^{(+)}(k^{2})}{(k+q)^{2}+i0}.
\end{align}
From the symmetry of the previous expressions under the transformation $$\{k\cdot n\leftrightarrow k\cdot\bar{n},q\cdot n\leftrightarrow q\cdot\bar{n}\},$$ we see that the terms arising from the
constraints (\ref{cut2}) and (\ref{cut4}) are identical to those due to (\ref{cut1}) and (\ref{cut3}) respectively.  
The two internal integrals over the transverse components of the soft gluon momenta are evaluated in (\ref{soltransverse2}) and (\ref{soltransverse1}).
Following the same technique used for the non-abelian integrals and using the following two identities (obtained from Eq.~(\ref{hyper3})) 
\begin{align}
\label{id3}
{}_2F_{1}\left(2,\frac{1}{2}-\epsilon,1-2\epsilon,4\frac{\sqrt{su}}{(\sqrt{s}+\sqrt{u})^{2}}\right) &=
(\sqrt{s}+\sqrt{u})^{4}\bigg(\Theta(s-u)\frac{1}{s^{2}}\,{}_{2}F_{1}\left(2,1+\epsilon,1-\epsilon,\frac{u}{s}\right)+\notag\\
&+\Theta(u-s)\frac{1}{u^{2}}\,{}_{2}F_{1}\left(2,1+\epsilon,1-\epsilon,\frac{s}{u}\right)\bigg),\hspace{2.3cm}\\	
\label{id4}
{}_2F_{1}\left(2,\frac{1}{2}-\epsilon,1-2\epsilon,4\frac{\sqrt{su}}{(1+\sqrt{su})^{2}}\right) &= (1+\sqrt{su})^{4} {}_{2}F_{1}\left(2,1+\epsilon,1-\epsilon,su\right),
\end{align}
in addition to (\ref{id1},\ref{id2}), after renaming some of the variables we recast $I_{(g),(h)}^{(2),(a)}$   as
\begin{align}
&I_{(g),(h)}^{(2),(a)}=\frac{1}{2}\frac{\pi^{1-\epsilon}}{\Gamma(1-\epsilon)}\frac{\pi^{\frac{1}{2}-\epsilon}}{\Gamma(\frac{1}{2}-\epsilon)}\frac{4^{-\epsilon}}{(2\pi)^{2d-2}}\frac{\Gamma^{2}(\frac{1}
{2}-\epsilon)}{\Gamma(1-2\epsilon)}\left(\frac{1}{Q}\right)^{4\epsilon}\left(\frac{1}{\tau_{s}}\right)^{1+4\epsilon}\notag\\
&\times\int_{0}^{1}dt\,du\,dz\bigg(2(1-t)^{1-2\epsilon}t^{1-2\epsilon}u^{-1+2\epsilon}z^{\epsilon}\frac{(1-z)^{2}}{(1-t(1-z))^{2}}{}_{2}F_{1}\big(2,2+\epsilon,1-\epsilon,z\big)+\notag\\
&+(1-t)^{1-2\epsilon}t^{1-2\epsilon}z^{\epsilon}u^{\epsilon}\frac{(1-zu)^{2}}{(z(1-t)+t)^{2}(1-t(1-u))^{2}}{}_{2}F_{1}\big(2,2+\epsilon,1-\epsilon,zu\big)\bigg),
\end{align}
and $I_{(g),(h)}^{(2),(b)}$ as
\begin{align}
&I_{(g),(h)}^{(2),(b)}=\frac{1}{2}\frac{\pi^{1-\epsilon}}{\Gamma(1-\epsilon)}\frac{\pi^{\frac{1}{2}-\epsilon}}{\Gamma(\frac{1}{2}-\epsilon)}\frac{4^{-\epsilon}}{(2\pi)^{2d-2}}\frac{\Gamma^{2}(\frac{1}
{2}-\epsilon)}{\Gamma(1-2\epsilon)}\left(\frac{1}{Q}\right)^{4\epsilon}\left(\frac{1}{\tau_{s}}\right)^{1+4\epsilon}\notag\\
&\times\int_{0}^{1}dt\,du\,dz\bigg(2\frac{(1-t)^{-2\epsilon}t^{-2\epsilon}u^{-1+2\epsilon}z^{\epsilon}}{t+z(1-t)}{}_{2}F_{1}\big(1,1+\epsilon,1-\epsilon,z\big)+\notag\\
&+\frac{(1-t)^{-2\epsilon}t^{-2\epsilon}u^{\epsilon}z^{\epsilon}}{(t+z(1-t))(1-t(1-u))}{}_{2}F_{1}\big(1,1+\epsilon,1-\epsilon,zu\big)\bigg).
\end{align}
For the four integrals appearing above we obtain
\begin{align}
\int_{0}^{1}dt\,du\,dz&\frac{(1-t)^{-2\epsilon}t^{-2\epsilon}u^{-1+2\epsilon}z^{\epsilon}}{t+z(1-t)}{}_{2}F_{1}\big(1,1+\epsilon,1-\epsilon,z\big) 
=\frac{1}{2\epsilon}\frac{\Gamma(1-\epsilon)}{\Gamma(1+\epsilon)\Gamma(-2\epsilon)}\notag\\
&\times\bigg(\frac{1}{4\epsilon^{2}}+\frac{1}{2\epsilon}+\bigg(1+\frac{\pi ^2}{6}\bigg)+\bigg(2+\frac{\pi ^2}{3}+\frac{5 \zeta_{3}}{2}\bigg)\epsilon+\mathcal{O}(\epsilon^{2})\bigg),\\
\int_{0}^{1}dt\,du\,dz&\frac{(1-t)^{-2\epsilon}t^{-2\epsilon}u^{\epsilon}z^{\epsilon}}{(t+z(1-t))(1-t(1-u))}{}_{2}F_{1}\big(1,1+\epsilon,1-\epsilon,zu\big)
=-\frac{\Gamma(1-\epsilon)}{\Gamma(1+\epsilon)\Gamma(-2\epsilon)}\notag\\
&\times\bigg(\frac{\pi^{2}}{6}\frac{1}{\epsilon}+7\zeta_{3}-\frac{\pi^{2}}{3}+\mathcal{O}(\epsilon)\bigg),\\
\int_{0}^{1}dt\,du\,dz&(1-t)^{1-2\epsilon}t^{1-2\epsilon}u^{-1+2\epsilon}z^{\epsilon}\frac{(1-z)^{2}}{(1-t(1-z))^{2}}{}_{2}F_{1}\big(2,2+\epsilon,1-\epsilon,z\big) =\notag\\
 &-\frac{1}{2\epsilon}\frac{\Gamma(1-\epsilon)}{\Gamma(2+\epsilon)\Gamma(-1-2\epsilon)}\times\bigg(\frac{1}{12\epsilon^{2}}+\frac{5}{36\epsilon}+\frac{16}{27}+\frac{\pi^{2}}{18}+\notag\\
&+\left(\frac{59}{81}+\frac{5 \pi ^2}{54}+\frac{5 \zeta_{3}}{6}\right)\epsilon+\mathcal{O}(\epsilon^{2})\bigg),\\
\int_{0}^{1}dt\,du\,dz&(1-t)^{1-2\epsilon}t^{1-2\epsilon}z^{\epsilon}u^{\epsilon}\frac{(1-zu)^{2}}{(z(1-t)+t)^{2}(1-t(1-u))^{2}}{}_{2}F_{1}\big(2,2+\epsilon,1-\epsilon,zu\big) =\notag\\
 &\frac{\Gamma(1-\epsilon)}{\Gamma(2+\epsilon)\Gamma(-1-2\epsilon)}\times\bigg(\frac{1}{\epsilon}\left(\frac{1}{6}+\frac{\pi^{2}}{18}\right)-\frac{7}{9}-\frac{5 \pi ^2}{54}+\frac{7
\zeta_{3}}{3}+\mathcal{O}(\epsilon)\bigg).
\end{align}
We finally plug these expressions back into (\ref{vacuumint}) and we obtain the following result for the vacuum polarization diagrams
\begin{align}
s^{(2)}_{(g),(h)}&\left(\tau_{s}, \frac{Q}{\mu},\alpha_{s}(\mu)\right) = \frac{\alpha_{s}(\mu)}{\pi^{2}}\bigg(\frac{\mu}{Q}\bigg)^{4\epsilon}\bigg(\frac{1}{\tau_{s}}\bigg)^{1+4\epsilon}\notag\\
&\times\bigg(-C_{F}T_{F}n_{F}\bigg(\frac{1}{3\epsilon^{2}}+\frac{5}{9\epsilon}+\bigg(\frac{28}{27}-\frac{\pi^{2}}{6}\bigg)+\bigg(\frac{20}{81}+\frac{37 \pi ^2}{54}-\frac{62
\zeta_{3}}{9}\bigg)\epsilon+\mathcal{O}(\epsilon^{2})\bigg)+\notag\\
\label{vacuumpol}
&+C_{A}C_{F}\bigg(\frac{5}{12\epsilon^{2}}+\frac{31}{36\epsilon}+\bigg(\frac{47}{27}-\frac{5 \pi ^2}{24}\bigg)+\bigg(\frac{211}{81}+\frac{155 \pi ^2}{216}-\frac{155
\zeta_{3}}{18}\bigg)\epsilon+\mathcal{O}(\epsilon^{2})\bigg)\bigg),
\end{align}
where we used (\ref{msbar}) to replace the bare coupling with the renormalized one in the $\overline{{\rm MS}}$ scheme.
This completes the computation of the relevant contributions to the two-loop soft subprocess. As a further check of our calculation, we observe that summing up all the integrand functions contributing to
the two-loop soft subprocess we reproduce the known double-soft current derived in \cite{catanigrazzinidoublesoft}.

\subsection{Renormalization of the two-loop soft subprocess}
Before the subtraction of the overall divergences we need to handle the subdivergences. The coupling renormalization (\ref{msbar}) leads to a counter-term
\begin{align}
\label{counterterm}
s^{(2)}_{{\rm c.t.}}&\left(\tau_{s}, \frac{Q}{\mu},\alpha_{s}(\mu)\right)=  \frac{\alpha_{s}(\mu)}{\pi}\frac{\beta_{0}}{\epsilon}S^{(1)}\left(\tau_{s}, \frac{Q}{\mu},\alpha_{s}(\mu)\right) = -C_{F} \frac{\alpha_{s}^{2}(\mu)}{\pi^{2}}\frac{e^{\epsilon
\gamma_{E}}}{\epsilon^2 \Gamma(1-\epsilon)} \notag\\
&\times\left(\frac{Q}{\mu}\right)^{-2\epsilon}\left(\frac{1}{\tau_{s}}\right)^{1+2\epsilon}\bigg(\frac{11}{6}C_{A}-\frac{4}{6}T_{F}n_{F}\bigg),
\end{align}
where $S^{(1)}(\tau_{s}, {Q}/{\mu},\alpha_{s}(\mu))$ is the one-loop contribution (\ref{realoneloop}).

After performing the Laplace transform of Eqs.~(\ref{RplusV}),(\ref{box1}),(\ref{box2}),(\ref{nonabel}),(\ref{vacuumpol}),(\ref{counterterm}) by means of the relation
\begin{align}
\int_{0}^{\infty}d\tau_{s}e^{-N\tau_{s}}\bigg(\frac{Q}{\mu}\bigg)^{-k\epsilon}\tau_{s}^{-1-k\epsilon} = e^{-k\epsilon\gamma_{E}}\Gamma(-k\epsilon)\bigg(\frac{N\mu}{N_{0}Q}\bigg)^{k\epsilon},
\end{align}
and summing them up we obtain the following expression for the non-abelian part of the unrenormalized two loop soft subprocess $\tilde{s}^{(2)}({N_{0}Q}/{(N\mu)},\alpha_{s}(\mu))$
\begin{align}
\label{laplacesoft}
\tilde{s}^{(2)}\left(\frac{N_{0}Q}{N\mu},\alpha_{s}(\mu)\right) =
\frac{\alpha^{2}_{s}(\mu)}{\pi^{2}}\left(\tilde{s}^{(2)}_{3}\log^{3}{\frac{N\mu}{N_{0}Q}}+\tilde{s}^{(2)}_{2}\log^{2}{\frac{N\mu}{N_{0}Q}}+\tilde{s}^{(2)}_{1}\log{\frac{N\mu}{N_{0}Q}}+\tilde{s}^{(2)}_{
0}\right)+\mathcal{O}(\epsilon),
\end{align}
where we find
\begin{align}
\tilde{s}^{(2)}_{3} &= -\frac{11}{9}C_{A}C_{F}+\frac{4}{9}C_{F}T_{F}n_{F}, \qquad  \tilde{s}^{(2)}_{2} =-C_{A}C_{F}\bigg(\frac{67}{18}-\frac{\pi^{2}}{6}\bigg)+\frac{10}{9}C_{F}T_{F}n_{F},\\
\tilde{s}^{(2)}_{1} &= -C_{F}T_{F}n_{F}\bigg(\frac{1}{3 \epsilon^2}-\frac{5}{9\epsilon}-\frac{28}{27}-\frac{\pi^{2}}{9}\bigg)\notag\\
&-C_{A}C_{F}\bigg(-\frac{11}{12 \epsilon^2}+\frac{1}{\epsilon}\bigg(\frac{67}{36}-\frac{\pi^{2}}{12}\bigg)+\frac{11}{36}\pi^{2}+\frac{101}{27}-\frac{7}{2}\zeta_{3}\bigg), \\
\tilde{s}^{(2)}_{0} &= -C_{F}T_{F}n_{F}\bigg(\frac{1}{4\epsilon^{3}}-\frac{5}{36\epsilon^{2}}+\frac{1}{\epsilon}\bigg(-\frac{7}{27}+\frac{\pi^{2}}{72}\bigg)\notag\\
&-\frac{5}{81}-\frac{77 \pi ^2}{216}+\frac{13 \zeta_{3}}{18}\bigg)\notag\\
&+C_{A}C_{F}\bigg(\frac{11}{16\epsilon^{3}}-\frac{1}{\epsilon^{2}}\bigg(\frac{67}{144}-\frac{\pi^{2}}{48}\bigg)-\frac{1}{\epsilon}\bigg(\frac{101}{108}-\frac{11}{288}\pi^{2}-\frac{7}{8}\zeta_{3}
\bigg)\notag\\
&-\frac{535}{324}-\frac{871 \pi ^2}{864}+\frac{7 \pi ^4}{120}+\frac{143 \zeta_{3}}{72}\bigg).
\end{align}

Renormalization properties of Wilson loops have been studied in detail in \cite{wrge1,wrge2,wrge3}.  The Wilson path we considered has two cusps and light-cone segments leading to
additional light cone singularities. 
This leads us to the following evolution equation \cite{wrge1,wrge2}
\begin{align}
\label{RGEsoft}
\left(\mu\frac{\partial}{\partial\mu}+\beta(\alpha_{s})\frac{\partial}{\partial \alpha_{s}}\right)\log{\tilde{S}\left(\frac{N_{0}Q}{N\mu},\alpha_{s}(\mu)\right)} =
-2\Gamma_{\rm{cusp}}(\alpha_{s}(\mu))\log{\frac{N^{2}\mu^{2}}{N^{2}_{0}Q^{2}}}-2\Gamma_{\rm{soft}}(\alpha_{s}(\mu)),
\end{align}
where $\Gamma_{\rm{cusp}}(g)$ is the well-known universal cusp anomalous dimension while $\Gamma_{\rm{soft}}(g)$ is a path-dependent coefficient often called soft anomalous dimension. The factor 2 in
front  of the cusp anomalous dimension in the evolution equation (\ref{RGEsoft}) counts the number of cusps in the integration path.

The two quantities $\Gamma_{\rm{cusp}}(g)$ and $\Gamma_{\rm{soft}}(g)$ can be evaluated through $\mathcal{O}(\alpha_{s}^{2})$ considering the $\mathcal{O}(\alpha_{s})$ and
$\mathcal{O}(\alpha_{s}^{2})$ 
counter-terms 
\begin{align}
\label{oneloopct}
\delta_{c.t.}^{(1)} = &C_{F}\frac{\alpha_{s}(\mu)}{\pi}\bigg(\frac{1}{\epsilon^{2}}+\frac{2}{\epsilon}\log{\frac{N\mu}{N_{0}Q}}\bigg),\\
\delta_{c.t.}^{(2)} = &\frac{\alpha_{s}^{2}(\mu)}{\pi^{2}}\bigg(\bigg(C_{F}T_{F}n_{F}\bigg(\frac{1}{3 \epsilon^2}-\frac{5}{9\epsilon}\bigg)
 +C_{A}C_{F}\bigg(-\frac{11}{12 \epsilon^2}+\frac{1}{\epsilon}\bigg(\frac{67}{36}-\frac{\pi^{2}}{12}\bigg)\bigg)\bigg)\log{\frac{N\mu}{N_{0}Q}}\notag\\
&+C_{F}T_{F}n_{F}\bigg(\frac{1}{4\epsilon^{3}}-\frac{5}{36\epsilon^{2}}+\frac{1}{\epsilon}\bigg(-\frac{7}{27}+\frac{\pi^{2}}{72}\bigg)\bigg)\notag\\
&+C_{A}C_{F}\bigg(-\frac{11}{16\epsilon^{3}}+\frac{1}{\epsilon^{2}}\bigg(\frac{67}{144}-\frac{\pi^{2}}{48}\bigg)+
\frac{1}{\epsilon}\bigg(\frac{101}{108}-\frac{11}{288}\pi^{2}-\frac{7}{8}\zeta_{3}\bigg)\bigg)\bigg),
\end{align}
leading to the following results
\begin{align}
\label{gammacusp}
\Gamma_{\rm{cusp}}(\alpha_{s}) &=
\frac{\alpha_{s}}{\pi}C_{F}+\frac{\alpha_{s}^{2}}{\pi^{2}}C_{F}\bigg(C_{A}\bigg(\frac{67}{36}-\frac{\pi^{2}}{12}\bigg)-\frac{5}{9}T_{F}n_{F}\bigg)+\mathcal{O}(\alpha_{s}^{3}), \\
\label{gammasoft}
\Gamma_{\rm{soft}}(\alpha_{s}) &=
-\frac{\alpha_{s}^{2}}{\pi^{2}}C_{F}\bigg(T_{F}n_{F}\bigg(\frac{14}{27}-\frac{\pi^{2}}{36}\bigg)+C_{A}\bigg(-\frac{101}{54}+\frac{11}{144}\pi^{2}+\frac{7}{4}\zeta_{3}\bigg)\bigg)+\mathcal{O}(\alpha_{s
}^{3}).
\end{align}
The two-loop cusp anomalous dimension (\ref{gammacusp}) was computed in \cite{cuspkorch}, while the two-loop value of $\Gamma_{\rm{soft}}(g)$ was first deduced in \cite{Becher:2008cf} using
renormalization group invariance of the cross section but it was never obtained by a direct calculation.

Exploiting the non-abelian exponentiation theorem (\ref{NAET}) and (\ref{twoloopexp}),
 we derive a complete two-loop expression for the soft subprocess
\begin{align}
\tilde{S}\left(\frac{N_{0}Q}{N\mu},\alpha_{s}(\mu)\right) =& 1-C_{F}\frac{\alpha_{s}(\mu)}{\pi}\bigg(\frac{\pi^{2}}{4}+2\log^{2}{\frac{N\mu}{N_{0}Q}}\bigg)\notag\\
&+\frac{\alpha_{s}^{2}(\mu)}{\pi^{2}}\bigg(2C_{F}^2\log^{4}{\frac{N\mu}{N_{0}Q}}-\bigg(\frac{11}{9}C_{A}C_{F}-\frac{4}{9}C_{F}T_{F}n_{F}\bigg)\log^{3}{\frac{N\mu}{N_{0}Q}}\notag\\
&+\bigg(\frac{\pi^{2}}{2}C_{F}^{2}-C_{A}C_{F}\bigg(\frac{67}{18}-\frac{\pi^{2}}{6}\bigg)+\frac{10}{9}C_{F}T_{F}n_{F}\bigg)\log^{2}{\frac{N\mu}{N_{0}Q}}\notag\\
&-\bigg(C_{F}T_{F}n_{F}\bigg(-\frac{28}{27}-\frac{\pi^{2}}{9}\bigg)+C_{A}C_{F}\bigg(\frac{11}{36}\pi^{2}+\frac{101}{27}-\frac{7}{2}\zeta_{3}\bigg)\bigg)\log{\frac{N\mu}{N_{0}Q}}\bigg)\notag\\
\label{twoloopcomplete}
&+\tilde{S}_{0}^{(2)}+\mathcal{O}(\alpha_{s}^{3}),
\end{align}
where $\tilde{S}_{0}^{(2)}$ is the non-logarithmic piece at two-loop order for which we provide an analytic expression
\begin{align}
\tilde{S}_{0}^{(2)} =& \frac{\alpha_{s}^{2}(\mu)}{\pi^{2}}\bigg(\frac{\pi^{4}}{32}C_{F}^{2}+C_{F}T_{F}n_{F}\bigg(\frac{5}{81}+\frac{77 \pi ^2}{216}-\frac{13 \zeta_{3}}{18}\bigg)\notag\\
\label{c2soft}
&+C_{A}C_{F}\bigg(-\frac{535}{324}-\frac{871 \pi ^2}{864}+\frac{7 \pi ^4}{120}+\frac{143 \zeta_{3}}{72}\bigg)\bigg).
\end{align}

The constant part (\ref{c2soft}) has also been calculated as a specific case of the
two-loop soft hemisphere function in a work done in parallel with ours~\cite{schwartznew}, in full agreement with our result.  
Previously, 
it had been fitted by two different groups \cite{Becher:2008cf,Chien:2010kc,hoangkluth} using the Monte Carlo program {\tt EVENT2} \cite{Catani:1996jh}. 
Their results are reported below
\begin{align}
&\tilde{S}_{0}^{(2),\textrm{\tiny\cite{Becher:2008cf}}}=\frac{\alpha_{s}^{2}(\mu)}{(4\pi)^{2}}\big((58\pm2)C_{F}^{2}-(60\pm1)C_{F}C_{A}+(43\pm1)C_{F}T_{F}n_{F}\big),\\
&\tilde{S}_{0}^{(2),\textrm{\tiny\cite{Chien:2010kc}}}=\frac{\alpha_{s}^{2}(\mu)}{(4\pi)^{2}}\big(48.7045 C_{F}^{2}-(57.8)C_{F}C_{A}+(43.4)C_{F}T_{F}n_{F}\big),\\
&\tilde{S}_{0}^{(2),\textrm{\tiny\cite{hoangkluth}}}=\frac{\alpha_{s}^{2}(\mu)}{(4\pi)^{2}}\big(48.7045 C_{F}^{2}-(58.8\pm2.25)C_{F}C_{A}+(43.8\pm3.06)C_{F}T_{F}n_{F}\big),
\end{align}
while we obtain 
\begin{align}
\tilde{S}_{0}^{(2)} =\frac{\alpha_{s}^{2}(\mu)}{(4\pi)^{2}}\big(48.7045 C_{F}^{2}-(56.4989)C_{A}C_{F}+(43.3905)C_{F}T_{F}n_{F}\big),
\end{align}
which is partly consistent with \cite{hoangkluth,Chien:2010kc} but not with 
the earlier superseded numbers of \cite{Becher:2008cf}, with the exception of the $C_{F}T_{F}n_{F}$ term. Notice that the determination of the $\tilde{S}_{0}^{(2)}$
constant
is relevant to the matching of NNLL resummed cross section to the NNLO one since it is part of the $G_{31}$ coefficient as it will be shown in the next section.

We now solve the evolution equation for the soft subprocess (\ref{RGEsoft}), yielding
\begin{align}
\label{solsoft1}
\log{\frac{\tilde{S}\left(\frac{N_{0}Q}{N\mu_{R}},\alpha_{s}(\mu_{R})\right)}{\tilde{S}(1,\alpha_{s}(\frac{N_{0}Q}{N}))}} =-
\int^{\mu^{2}_{R}}_{\frac{N^{2}_{0}Q^{2}}{N^{2}}}\frac{dk^{2}}{2k^{2}}\big(2\Gamma_{\rm{cusp}}(\alpha_{s}(k^{2}))\log{\frac{N^{2}k^{2}}{N^{2}_{0}Q^{2}}}+2\Gamma_{\rm{soft}}(\alpha_{s}(k^{2}))\big),
\end{align} 
where $\mu_{R}$ is the renormalization scale of the process.
We can perform the substitution 
\begin{align}
\log{\frac{N^{2}k^{2}}{N^{2}_{0}Q^{2}}} = \int_{\frac{N^{2}_{0}Q^{2}}{N^{2}}}^{k^{2}}\frac{d\mu^{2}}{\mu^{2}},
\end{align}
and exchange the order of integration in the integral containing $\Gamma_{\rm{cusp}}$ getting
\begin{align}
\label{solsoft2}
\log{\frac{\tilde{S}\left(\frac{N_{0}Q}{N\mu_{R}},\alpha_{s}(\mu_{R})\right)}{\tilde{S}(1,\alpha_{s}(\frac{N_{0}Q}{N}))}} = -
\int_{\frac{N_{0}^{2}Q^{2}}{N^{2}}}^{\mu_{R}^{2}}\frac{d\mu^{2}}{\mu^{2}}\int_{\mu^{2}}^{\mu_{R}^{2}}\frac{dk^{2}}{k^{2}}\Gamma_{\rm{cusp}}(\alpha_{s}(k^{2}))-\int_{\frac{N_{0}^{2}Q^{2}}{N^{2}}}^{\mu_
{R}^{2}}\frac{dk^{2}}{k^{2}}\Gamma_{\rm{soft}}(\alpha_{s}(k^{2})).
\end{align}
We set $\mu_{R} = Q$ in order to minimize the logarithmic corrections coming from the hard function in (\ref{eq:RT}), moreover we replace ${\mu^{2}/}{Q^{2}} = u^{2}$ in the first integral of 
Eq.~(\ref{solsoft2}) and ${k^{2}}/{Q^{2}} = u^{2}$ in the second one and we finally obtain
\begin{align}
\label{solsoft}
\frac{\tilde{S}(\frac{N_{0}}{N},\alpha_{s}(Q))}{\tilde{S}(1,\alpha_{s}(\frac{N_{0}Q}{N}))} = {\rm
exp}\bigg(-2\int_{\frac{N_{0}}{N}}^{1}\frac{du}{u}\int_{u^{2}Q^{2}}^{Q^{2}}\frac{dk^{2}}{k^{2}}\Gamma_{\rm{cusp}}(\alpha_{s}(k^{2}))-2\int_{\frac{N_{0}}{N}}^{1}\frac{du}{u}\Gamma_{\rm{soft}}(\alpha_{s
}(u^{2}Q^{2}))  \bigg).
\end{align}

The thrust observable is symmetrical under the exchange of the two hemispheres, so the factor $2$ in the exponent accounts for the identical contributions due to both of them.

\section{Resummation of large logarithms}
\label{sec:resummation of large logs}
In the present section we derive a resummed expression for the cross section (\ref{eq:Rres}) in Laplace space starting from the renormalization  group (RG) evolution of  each of the subprocesses. The
effect of soft gluons has been taken into account in the previous section, but we still need to consider logarithmically enhanced terms due to hard gluons moving collinearly to one of the hard quark
legs. Such an effect is encoded in the jet subprocess which describes the decay of a hard quark into a jet of collinear particles. The same
subprocess can be found in other relevant QCD processes such as deep inelastic scattering and $B$-meson decay and it obeys the following evolution equation \cite{SCETJet,KorSter}:
\begin{align}
\label{rgejet}
\left(\mu\frac{\partial}{\partial\mu}+\beta(\alpha_{s})\frac{\partial}{\partial \alpha_{s}}\right)\log{\tilde{J}\left(\sqrt{\frac{N_{0}}{N}}\frac{Q}{\mu},\alpha_{s}(\mu)\right)} =
2\Gamma_{\rm{cusp}}(\alpha_{s}(\mu))\log{\frac{N\mu^{2}}{N_{0}Q^{2}}}-2\Gamma_{\rm{coll}}(\alpha_{s}(\mu)).
\end{align}
The collinear subprocess can be defined as a cut propagator of a massless quark in the axial gauge \cite{Ster87}. Indeed, the factorization used here is manifest in the axial gauge \cite{Ster87}.

Equation~(\ref{rgejet}) can be solved following the same technique used with the soft subprocess but now replacing ${\mu^{2}}/{Q^{2}} = u$ and ${k^{2}}/{Q^{2}} = u$ in the first and
second  integral in the exponent respectively. It leads to
\begin{align}
\label{soljet}
\frac{\tilde{J}\left(\frac{N_{0}}{N},\alpha_{s}(Q)\right)}{\tilde{J}\left(1,\alpha_{s}(\sqrt{\frac{N_{0}}{N}}Q)\right)} = {\rm
exp}\bigg(\int_{\frac{N_{0}}{N}}^{1}\frac{du}{u}\int_{uQ^{2}}^{Q^{2}}\frac{dk^{2}}{k^{2}}\Gamma_{\rm{cusp}}(\alpha_{s}(k^{2}))-\int_{\frac{N_{0}}{N}}^{1}\frac{du}{u}\Gamma_{\rm{coll}}(\alpha_{s}(uQ^{2
}))  \bigg).
\end{align}

We now combine (\ref{solsoft}) and (\ref{soljet}) together in the expression of the cross section
\begin{align}
\tilde{\sigma}_{N}(Q^{2},\alpha_{s}) &= H(1,\alpha_{s}(\mu=Q))\tilde{J}^{2}\left(\frac{N_{0}}{N},\alpha_{s}(Q)\right)\tilde{S}\left(\frac{N_{0}}{N},\alpha_{s}(Q)\right) = \notag\\
&=H(1,\alpha_{s}(Q))\tilde{J}^{2}\left(1,\alpha_{s}(\sqrt{\frac{N_{0}}{N}}Q)\right)\tilde{S}\left(1,\alpha_{s}(\frac{N_{0}Q}{N})\right)\notag\\
\label{cross1}
&\times{\rm
exp}\bigg(-2\int_{\frac{N_{0}}{N}}^{1}\frac{du}{u}\int_{u^{2}Q^{2}}^{uQ^{2}}\frac{dk^{2}}{k^{2}}\Gamma_{\rm{cusp}}(\alpha_{s}(k^{2})) \notag \\
&\hspace{2cm}-2\int_{\frac{N_{0}}{N}}^{1}\frac{du}{u}(\Gamma_{\rm{soft}}(\alpha_
{s}(u^{2}Q^{2})) +\Gamma_{\rm{coll}}(\alpha_{s}(uQ^{2})) ) \bigg),
\end{align}
where we explicitly set the renormalization scale $\mu = Q$.

Using the relation
\begin{align}
\Gamma_{\rm{soft}}(\alpha_{s}(u^{2}Q^{2})) = \Gamma_{\rm{soft}}(\alpha_{s}(uQ^{2}))-\int_{u^{2}Q^{2}}^{uQ^{2}}\frac{dk^{2}}{k^{2}}\beta(\alpha_{s}(k^{2}))\frac{\partial
\Gamma_{\rm{soft}}(\alpha_{s}(k^{2}))}{\partial\alpha_{s}},
\end{align}
we recast Eq.~(\ref{cross1}) as
\begin{align}
\tilde{\sigma}_{N}(Q^{2},\alpha_{s})=&H(1,\alpha_{s}(Q))\tilde{J}^{2}\left(1,\alpha_{s}(\sqrt{\frac{N_{0}}{N}}Q)\right)\tilde{S}\left(1,\alpha_{s}(\frac{N_{0}Q}{N})\right)\notag\\
\label{resummedcross}
&\times{\rm exp}\bigg\{-2\int_{\frac{N_{0}}{N}}^{1}\frac{du}{u}\bigg(\int_{u^{2}Q^{2}}^{uQ^{2}}\frac{dk^{2}}{k^{2}}\mathcal{A}_\Gamma(\alpha_{s}(k^{2}))+\mathcal{B}_\Gamma(\alpha_{s}(uQ^{2})) \bigg)\bigg\},
\end{align}
where we defined
\begin{align}
\mathcal{A}_\Gamma(\alpha_{s}) &= \Gamma_{\rm{cusp}}(\alpha_{s})-\beta(\alpha_{s})\frac{\partial \Gamma_{\rm{soft}}(\alpha_{s})}{\partial\alpha_{s}},\notag \\
\label{AB}
\mathcal{B}_\Gamma(\alpha_{s}) &= \Gamma_{\rm{soft}}(\alpha_{s}) +\Gamma_{\rm{coll}}(\alpha_{s}).
\end{align}
The two coefficients $\mathcal{A}_\Gamma(\alpha_{s})$ and $\mathcal{B}_\Gamma(\alpha_{s})$ can be computed in perturbative QCD. To this end we observe that the Altarelli-Parisi splitting function
$P_{qq}(\alpha_{s},z)$ fulfils the following limit~\cite{korchdis} 
as $z\rightarrow 1$
\begin{align}
\label{splittingfunction}
 P_{qq}(\alpha_{s},z) = 2\frac{\Gamma_{\rm{cusp}}(\alpha_{s})}{(1-z)_{+}}+2\mathcal{B}_\Gamma(\alpha_{s})\delta(1-z)+...,
\end{align}
where the dots stand for regular terms in the $z\rightarrow1$ limit. The asymptotic expression (\ref{splittingfunction}) is valid to all orders in perturbative QCD and it can be easily proven in the
context of deep inelastic scattering as shown in  \cite{bechneubdis}. The Mellin transform of the structure function $F_{2}(Q^{2},x)$ can be indeed factorized in the threshold limit $x\rightarrow1$ as
a product of a hard virtual function $H({Q}/{\mu},\alpha_{s}(\mu))$, a collinear jet function and a parton distribution function $\phi_{q}(N,\mu)$. Both the hard and collinear jet
functions are essentially the same ones as in the thrust case  (up to constants in the hard subprocess due to crossing). The collinear jet function evolution is described by Eq.~(\ref{rgejet}), while the hard function RG equation reads
\begin{align}
\label{rgehard}
\left(\mu\frac{\partial}{\partial\mu}+\beta(\alpha_{s})\frac{\partial}{\partial \alpha_{s}}\right)\log{H(\frac{Q}{\mu},\alpha_{s}(\mu))} =&
-2\Gamma_{\rm{cusp}}(\alpha_{s}(\mu))\log{\frac{\mu^{2}}{Q^{2}}}\notag\\
&+2\big(\Gamma_{\rm{soft}}(\alpha_{s}(\mu))+2\Gamma_{\rm{coll}}(\alpha_{s}(\mu))\big),
\end{align}
\noindent and the parton distribution function evolves according to the Altarelli-Parisi equation \cite{AP}
\begin{align}
\left(\mu\frac{\partial}{\partial\mu}+\beta(\alpha_{s})\frac{\partial}{\partial \alpha_{s}}\right)\log{\phi_{q}(N,\mu)}= \tilde{P}_{qq}(\alpha_{s},N),
\end{align}
where 
\begin{align}
\tilde{P}_{qq}(\alpha_{s},N)=-\int_{0}^{1}dz\,z^{N-1}P_{qq}(\alpha_{s},z)
\end{align}
\noindent is the Mellin transform of the splitting function using the conventions of \cite{NNLOAPS,NNLOAPNS}.
We now observe that for the structure function $F_{2}(Q^{2},x)$ to be RG invariant, we have to require that the anomalous dimensions of the hard, jet and parton distribution functions sum up to zero,
proving Eq.~(\ref{splittingfunction}).

As stated at the end of the previous section, since the thrust is symmetrical under the exchange of the two hemispheres, we can factorize the soft subprocess as a product of two independent ``hemisphere'' soft subprocesses.
It follows that the resummed cross section (\ref{resummedcross}) can be recast as a constant term multiplied by the evolution of two independent jets each of which is the product of the collinear
jet and the respective ``hemisphere'' soft subprocess. This is in analogy to the structure obtained at NLL using the coherent branching algorithm \cite{CTTW}. We compare the two expressions observing 
that the only difference between the two exponents is the term  $\beta(\alpha_{s}){\partial \Gamma_{\rm{soft}}(\alpha_{s})}/{\partial\alpha_{s}}$ which gives a non-vanishing contribution only
beyond NLL. It essentially accounts for large angle soft emissions whose effects do not contribute at NLL (it is easy to see that the first non-trivial term arises at $\mathcal{O}(\alpha_{s}^{3})$). 
A second interesting feature which shows up beyond the NLL approximation is the interplay between constant terms and logarithms due to the factor $\tilde{J}^{2}(1,\alpha_{s}(\sqrt{({N_{0}}/{N})}Q))\tilde{S}(1,\alpha_{s}({N_{0}Q}/{N}))$
that will be analyzed below.

The resummed cross section in the dijet limit (\ref{resummedcross}) takes the following form
\begin{align}
\tilde{\sigma}_{N}(Q^{2},\alpha_{s})=&\,\, \bigg(1+\sum_{k=1}^{\infty}\tilde{C}_{k}\bigg(\frac{\alpha_{s}}{2\pi}\bigg)^{k}\bigg)\Sigma_{N}(Q^{2},\alpha_{s}),\notag\\
\label{resummedform}
\Sigma_{N}(Q^{2},\alpha_{s}) =&\,\, {\rm
exp}\Bigg\{Lf_{1}\left(\frac{\alpha_{s}}{\pi}\beta_{0}L\right)+f_{2}\left(\frac{\alpha_{s}}{\pi}\beta_{0}L\right)+\frac{\alpha_{s}}{\pi}\beta_{0}f_{3}\left(\frac{\alpha_{s}}{\pi}\beta_{0}L\right)+\tilde{G}_{31}\bigg(\frac{\alpha_{s}}{2\pi
}\bigg)^{3}L\notag \\ &
+\mathcal{O}(\alpha_{s}^{4}L^{2})\Bigg\},
\end{align}
where here $L=\log N$.
The function $f_{1}(({\alpha_{s}}/{\pi})\beta_{0}L)$ resums all the leading logarithmic contributions $\alpha_{s}^{n}L^{n+1}$, $f_{2}(({\alpha_{s}}/{\pi})\beta_{0}L)$ resums the next to
leading terms $\alpha_{s}^{n}L^{n}$ and so on. We furthermore require that $f_{i}(0)=0$ so that at N$^{n}$LL we can write 
\begin{align}\label{eq:f_i_expansion}
f_{1}\left(\frac{\alpha_{s}}{\pi}\beta_{0}L\right) =& \sum_{k\geq 1}\tilde{G}_{k,k+1}\bigg(\frac{\alpha_{s}}{2\pi}\bigg)^{k}L^{k+1},\qquad &n=0;\\
f_{n+1}\left(\frac{\alpha_{s}}{\pi}\beta_{0}L\right) =& \sum_{k\geq n}\tilde{G}_{k,k+1-n}\bigg(\frac{\alpha_{s}}{2\pi}\bigg)^{k}L^{k+1-n}, \qquad &n\geq1.
\end{align}
With this notation we see that the term $\tilde{G}_{31}\alpha_{s}^{3}L$ is a N$^{3}$LL contribution due to the Taylor expansion of $f_{4}(\frac{\alpha_{s}}{\pi}\beta_{0}L)$. Nevertheless, such a
term is  relevant for the $R$-matching of the NNLL resummed cross section to the $\mathcal{O}(\alpha_{s}^{3})$ fixed-order result, which will be discussed below. 
After expanding the functions $\mathcal{A}_\Gamma(\alpha_{s})$ and 
$\mathcal{B}_\Gamma(\alpha_{s})$ as
\begin{align}
\label{ABseries}
\mathcal{A}_\Gamma(\alpha_{s}) = \sum_{k\geq1}A^{(k)}\bigg(\frac{\alpha_{s}}{\pi}\bigg)^{k}, \qquad \mathcal{B}_\Gamma(\alpha_{s}) = \sum_{k\geq1}B^{(k)}\bigg(\frac{\alpha_{s}}{\pi}\bigg)^{k},
\end{align}
in Eq.~(\ref{resummedcross}), we observe that $A^{(k)}$ gives rise to terms of order $\alpha_{s}^{n}L^{n+2-k}$ while $B^{(k)}$ contributes with terms of order $\alpha_{s}^{n}L^{n+1-k}$ with $n\geq k$.

The previous property ensures that the knowledge of $B^{(3)}$ is sufficient to compute $\tilde{G}_{31}$ and we do not need to know $A^{(4)}$ which has not been computed yet ($\Gamma_{\rm{soft}}$ is
known at three loops \cite{Becher:2008cf}, but the four loop value of $\Gamma_{\rm{cusp}}$ is still unknown). 

The coefficient $H(1,\alpha_{s}(Q))\tilde{J}^{2}(1,\alpha_{s}(\sqrt{({N_{0}}/{N})}Q))\tilde{S}(1,\alpha_{s}({N_{0}Q}/{N}))$ in Eq.~(\ref{resummedcross}) contains all the constant terms
and it can be expanded in perturbation theory. The function $H(1,\alpha_{s}(Q))$ is known at three loop order \cite{3Lform1,3Lform2} (this result has 
subsequently to be normalized to the total
hadronic cross section $\sigma$) and the two-loop non-logarithmic value of the collinear subprocess was computed in \cite{SCETJet}. The constant part of the two-loop soft subprocess was evaluated in
the previous section.
We see that the coupling is evaluated at different scales in each of the three functions. We use the expression for the running coupling (\ref{alpharunning}) to express them in terms of
$\alpha_{s}(Q)$ evaluated at the renormalization scale $\mu_{R}=Q$. The resulting expression has additional resummed logarithms sitting outside the exponent of Eq.~(\ref{resummedcross}) due to the
running of $\alpha_{s}$ and giving a well defined and finite contribution at large $N$. Such terms contribute from NNLL on and do not exponentiate naturally. Nevertheless, 
in order to bring the cross section to the form (\ref{resummedform}), we raise them to the exponent and we expand them to the desired order. 

 One finds that the one-loop constants of the collinear and soft subprocesses contribute to $f_{3}(({\alpha_{s}}/{\pi})\beta_{0}L)$ while the two-loop ones
contribute to $\tilde{G}_{31}$.

\noindent To evaluate the integrals in Eq.~(\ref{resummedcross}) we use the renormalization group equation (\ref{rgealpha}) to change the integration variable to $\alpha_{s}$. After imposing the normalization
condition  $f_{i}(0)=0$ we find
\begin{align}
\label{f1}
f_{1}(\lambda) =& -\frac{A^{(1)}}{\beta_{0}\lambda}[(1-2\lambda)\log(1-2\lambda)-2(1-\lambda)\log(1-\lambda)],\\
f_{2}(\lambda) =& -\frac{A^{(2)}}{\beta_{0}^{2}}[2\log(1-\lambda)-\log(1-2\lambda)]+2\frac{B^{(1)}}{\beta_{0}}\log(1-\lambda)\notag\\
&-\frac{A^{(1)}\beta_{1}}{\beta_{0}^{3}}[\log(1-2\lambda)+\frac{1}{2}\log^{2}(1-2\lambda)-\log(1-\lambda)(2+\log(1-\lambda)]\notag\\
\label{f2}
&-2\frac{A^{(1)}\gamma_{E}}{\beta_{0}}\log\frac{1-\lambda}{1-2\lambda},
\end{align}
{\allowdisplaybreaks
\begin{align}
f_{3}(\lambda) =&\,\,\,\frac{2c_{s}^{(1)}}{\beta_{0}}\frac{\lambda}{1-2\lambda}+\frac{2c_
{j}^{(1)}}{\beta_{0}}\frac{\lambda}{1-\lambda}-\frac{2B^{(2)}}{\beta_{0}^{2}}\frac{\lambda}{1-\lambda}-\frac{A^{(3)}}{\beta_{0}^{3}}\frac{\lambda^{2}}{(1-\lambda)(1-2\lambda)}\notag\\
&-\frac{2A^{(2)}\gamma_{E}}{\beta_{0}^{2}}\frac{\lambda}{(1-\lambda)(1-2\lambda)}+\frac{A^{(2)}\beta_{1}}{\beta_{0}^{4}}\frac{3\lambda^{2}+(1-\lambda)\log(1-2\lambda)-2(1-2\lambda)\log(1-\lambda)}{
(1-\lambda)(1-2\lambda)}\notag\\
&-2 \frac{B^{(1)}}{\beta_{0}}\gamma_{E}\frac{\lambda}{1-\lambda}+\frac{2B^{(1)}\beta_{1}}{\beta_{0}^{3}}\frac{\lambda+\log(1-\lambda)}{1-\lambda}\notag\\
&+\frac{A^{(1)}}{\beta_{0}}\frac{1}{(1-\lambda)(1-2\lambda)}\big[-\gamma_{E}^{2}\lambda(3-2\lambda)+\frac{2\gamma_{E}\beta_{1}}{\beta_{0}^{2}}[\lambda+(1-\lambda)\log(1-2\lambda)\notag\\
&-(1-2\lambda)\log(1-\lambda)]+\frac{\beta_{2}}{\beta_{0}^{3}}[-\lambda^{2}+(1-3\lambda+2\lambda^{2})(2\log(1-\lambda)-\log(1-2\lambda))]\big]\notag\\
&-\frac{A^{(1)}\beta_{1}^{2}}{\beta_{0}^{5}}\bigg[\frac{1-\lambda}{2(1-\lambda)(1-2\lambda)}\log(1-2\lambda)[4\lambda+\log(1-2\lambda)]\notag\\
\label{f3}
&-\frac{2}{2(1-\lambda)(1-2\lambda)}[\lambda^{2}-(1-2\lambda)\log(1-\lambda)(2\lambda+\log(1-\lambda))]\bigg],
\end{align}}
where $\lambda=\frac{\alpha_{s}(Q)}{\pi}\beta_{0}\log N$.

At NLL the previous functions reproduce the result obtained in \cite{CTTW}.
The constants $c_{j}^{(1)}$ and $c_{s}^{(1)}$ are the one-loop non-logarithmic terms of the collinear and the soft subprocesses respectively. They arise from the term
$\tilde{J}^{2}(1,\alpha_{s}(\sqrt{({N_{0}}/{N})}Q))\tilde{S}(1,\alpha_{s}({N_{0}Q}/{N}))$, as discussed above. The coefficients $A^{(i)}$ and $B^{(i)}$ can be determined as
shown in Eq.~(\ref{AB}) using the two-loop value of $\Gamma_{\rm{soft}}$ computed above and the three-loop splitting functions \cite{NNLOAPS,NNLOAPNS}. They are reported in Appendix A. 

We observe that the normalization condition $f_{i}(0)=0$ is automatically fulfilled by both $f_{1}(\lambda)$ and $f_{2}(\lambda)$ while it has to be imposed to obtain $f_{3}(\lambda)$ (\ref{f3}). This
could be considered as a signal of the breakdown of natural exponentiation beyond NLL.
Forcing such a constraint gives rise to a residual constant value which has to be taken out of the exponent and that contributes to the constants $\tilde{C}_{i}$. We will determine 
the value of such constants directly in thrust space in section \ref{Inversion of the integral transform}.

\subsection{Inversion of the integral transform}
\label{Inversion of the integral transform}
In the present section we perform the inverse Laplace transform of the resummed cross section. We recall the definition of the normalized cross section 
\begin{align}
\label{Rtau}
R_{T}(\tau)=\frac{1}{2\pi i}\int_{C}\frac{dN}{N}e^{N\tau}\tilde{\sigma}_{N}(Q^{2},\alpha_{s})+\mathcal{O}(\tau),
\end{align}
where $\tilde{\sigma}_{N}(Q^{2},\alpha_{s})$ is defined in Eq.~(\ref{resummedform}). The contour $C$ runs parallel to the imaginary axis to the right of all the singularities of the integrand
function.
A method to invert (\ref{Rtau}) was proposed in \cite{CTTW}. Keeping their notation we rewrite $\Sigma_{N}(Q^{2},\alpha_{s})$ in (\ref{resummedform}) as
\begin{align}
 \Sigma_{N}(Q^{2},\alpha_{s}) = e^{\mathcal{\tilde{F}}(\alpha_{s}(Q^{2}),\log N)},
\end{align}
where
\begin{align}
\mathcal{\tilde{F}}(\alpha_{s}(Q^{2}),L)
=& Lf_{1}\left(\frac{\alpha_{s}}{\pi}\beta_{0}L\right)+f_{2}\left(\frac{\alpha_{s}}{\pi}\beta_{0}L\right)+\frac{\alpha_{s}}{\pi}\beta_{0}f_{3}\left(\frac{\alpha_{s}}{\pi}\beta_{0}L\right)+\tilde{G}_{31}\bigg(\frac{\alpha_{s}}{2\pi}
\bigg)^{3}L\notag \\ &
+\mathcal{O}(\alpha_{s}^{4}L^{2}).
\end{align}

\noindent With this notation $\tilde{\sigma}_{N}(Q^{2},\alpha_{s})$ reads
\begin{align}
 \tilde{\sigma}_{N}(Q^{2},\alpha_{s}) = \bigg(1+\sum_{k=1}^{\infty}\tilde{C}_{k}\bigg(\frac{\alpha_{s}}{2\pi}\bigg)^{k}\bigg)e^{\mathcal{\tilde{F}}(\alpha_{s}(Q^{2}),\log N)}.
\end{align}
We now Taylor expand the exponent $\mathcal{\tilde{F}}$ with respect to $\log N$ around 
$\log N=\log({1}/{\tau})$ up to the desired order and we end up with the following expression for
$R_{T}(\tau)$
\begin{align}
R_{T}(\tau) =& \bigg(1+\sum_{k=1}^{\infty}\tilde{C}_{k}\bigg(\frac{\alpha_{s}}{2\pi}\bigg)^{k}\bigg)e^{\mathcal{\tilde{F}}(\alpha_{s}(Q^{2}),\log\frac{1}{\tau})}\frac{1}{2\pi
i}\int_{C}\frac{d\nu}{\nu}{\rm exp}\bigg[\nu+\mathcal{\tilde{F}}^{(1)}(\alpha_{s}(Q^{2}),\log\frac{1}{\tau})\log \nu\notag\\
\label{Rtauprov}
&+\frac{1}{2}\mathcal{\tilde{F}}^{(2)}(\alpha_{s}(Q^{2}),\log\frac{1}{\tau})\log^{2}\nu+\mathcal{O}(\alpha_{s}^{n}\log^{n-2}({1}/{\tau}))\bigg],
\end{align}
where $\nu = N\tau$ and
\begin{align}
 \mathcal{\tilde{F}}^{(n)}\left(\alpha_{s}(Q^{2}),\log\frac{1}{\tau}\right) = \frac{\partial^{n}}{\partial \log^{n}\frac{1}{\tau}}\mathcal{\tilde{F}}(\alpha_{s}(Q^{2}),\log({1}/{\tau})).
\end{align}
More precisely, we have
\begin{align}
\label{F1}
\mathcal{\tilde{F}}^{(1)}(\alpha_{s}(Q^{2}),\log({1}/{\tau})) =& \,\,\,f_{1}(\lambda)+\lambda
f_{1}^{\prime}(\lambda)+\frac{\alpha_{s}}{\pi}\beta_{0}f_{2}^{\prime}(\lambda)+\mathcal{O}(\alpha_{s}^{n}\log^{n-2}({1}/{\tau})),\\
\label{F2}
\mathcal{\tilde{F}}^{(2)}\left(\alpha_{s}(Q^{2}),\log({1}/{\tau})\right) =& \,\,\,2\frac{\alpha_{s}}{\pi}\beta_{0}f_{1}^{\prime}(\lambda)+\frac{\alpha_{s}}{\pi}\beta_{0}\lambda
f_{1}^{\prime\prime}(\lambda)+\mathcal{O}(\alpha_{s}^{n}\log^{n-2}({1}/{\tau})),
\end{align}
where, from now on, $\lambda = ({\alpha_{s}}/{\pi})\beta_{0}\log({1}/{\tau})$.
Without loss of generality we expand the factor 
\begin{align}
{\rm exp}\left[\frac{1}{2}\mathcal{\tilde{F}}^{(2)}(\alpha_{s}(Q^{2}),\log({1}/{\tau}))\log^{2}\nu\right]
\end{align}
in a Taylor series in $\mathcal{\tilde{F}}^{(2)}$ itself and we see that the series of terms $(\mathcal{\tilde{F}}^{(2)})^{k}$ with $k\geq2$ gives at most rise to terms of order
$\mathcal{O}(\alpha_{s}^{k}(\alpha_{s}\log({1}/{\tau}))^{n})$, so we drop them. Using the result \cite{CTTW}
\begin{align}
\label{eq:mastinv}
 \frac{1}{2\pi i}\int_{C}\frac{d\nu}{\nu}\log^{k}\nu\,{\rm exp}\left[\nu+\mathcal{\tilde{F}}^{(1)}\log \nu\right]=\frac{d^{k}}{d(\mathcal{\tilde{F}}^{(1)})^{k}}\frac{1}{\Gamma(1-\mathcal{\tilde{F}}^{(1)})},
\end{align}
we obtain
\begin{align}
 R_{T}(\tau) =&
\bigg(1+\sum_{k=1}^{\infty}\tilde{C}_{k}\bigg(\frac{\alpha_{s}}{2\pi}\bigg)^{k}\bigg)e^{\mathcal{\tilde{F}}(\alpha_{s}(Q^{2}),\log\frac{1}{\tau})}\bigg[\frac{1}{\Gamma(1-\gamma(\lambda))}+\frac{
\alpha_{s}}{\pi}\beta_{0}f_{2}^{\prime}(\lambda)\frac{\psi^{(0)}(1-\gamma(\lambda))}{\Gamma(1-\gamma(\lambda))}\notag\\
&+\frac{1}{2}\mathcal{\tilde{F}}^{(2)}\frac{d^{2}}{d\gamma^{2}(\lambda)}\frac{1}{\Gamma(1-\gamma(\lambda))}+\mathcal{O}(\alpha_{s}^{2}(\alpha_{s}\log({1}/{\tau}))^{n})\bigg],
\end{align}
where we neglected subleading terms and we defined $\gamma(\lambda)=f_{1}(\lambda)+\lambda f_{1}^{\prime}(\lambda)$. $\psi^{(m)}$ is the $(m+1)$-th derivative of the logarithm of the Euler 
$\Gamma$-function. 

After replacing the coefficients $\tilde{C}_{k}$ and $\tilde{G}_{31}$ in the previous expression with $C_{k}$ and $G_{31}$ computed as explained in Appendix \ref{g31andconstants}, we can recast $R_{T}(\tau)$ as
\begin{align}
 R_{T}(\tau) =&
\bigg(1+\sum_{k=1}^{\infty}C_{k}\bigg(\frac{\alpha_{s}}{2\pi}\bigg)^{k}\bigg)e^{\mathcal{F}(\alpha_{s}(Q^{2}),\log\frac{1}{\tau})}\frac{1}{\Gamma(1-\gamma(\lambda))}\bigg[1+\frac{\alpha_{s}}{\pi}
\beta_{0}f_{2}^{\prime}(\lambda)\psi^{(0)}(1-\gamma(\lambda))\notag\\
\label{finalresummed}
&+\frac{1}{2}\frac{\alpha_{s}}{\pi}\beta_{0}\gamma^{\prime}(\lambda)\Gamma(1-\gamma(\lambda))\frac{d^{2}}{d\gamma^{2}(\lambda)}\frac{1}{\Gamma(1-\gamma(\lambda))}+\frac{\alpha_{s}}{\pi}C_{F}
\left(\gamma_{E}\left(\frac{3}{2}-\gamma_{E}\right)-\frac{\pi^{2}}{6}\right)\bigg],
\end{align}
where now
\begin{align}
\mathcal{F}(\alpha_{s}(Q^{2}),L)
=&Lf_{1}\left(\frac{\alpha_{s}}{\pi}\beta_{0}L\right)+f_{2}\left(\frac{\alpha_{s}}{\pi}\beta_{0}L\right)+\frac{\alpha_{s}}{\pi}\beta_{0}f_{3}\left(\frac{\alpha_{s}}{\pi}\beta_{0}L\right)+G_{31}\bigg(\frac{\alpha_{s}}{2\pi}\bigg)^{3}
L \notag \\ &
+\mathcal{O}(\alpha_{s}^{4}L^{2}).
\end{align}
The expression (\ref{finalresummed}) for $R_{T}(\tau)$ resums all large logarithms through NNLL and it holds up to terms of order $\mathcal{O}(\alpha_{s}^{4}\log^{2}({1}/{\tau}))$.

The constant term ${\alpha_{s}}/{\pi}C_{F}\left(\gamma_{E}({3}/{2}-\gamma_{E}\right)-{\pi^{2}}/{6})$ is defined such that the term in square brackets is normalized to one as
$\lambda\rightarrow 0$ and ensures that all the constant terms are included in the $C_{k}$ coefficients which read
\begin{align}
\label{C1}
C_{1} =& C_{F}\bigg(-\frac{5}{2}+\frac{\pi^{2}}{3}\bigg),\\
C_{2} =&C_{F}^{2}\bigg(\frac{41}{8}-\frac{7}{8}\pi^{2}+\frac{5}{36}\pi^{4}-6\zeta_{3}\bigg)+C_{A}C_{F}\bigg(-\frac{491}{24}-\frac{53 \pi ^2}{108}+\frac{11 \pi ^4}{360}+\frac{104
\zeta_{3}}{3}\bigg)\notag\\
\label{C2}
&+C_{F}T_{F}n_{F}\bigg(\frac{35}{6}+\frac{7 \pi ^2}{27}-\frac{28 \zeta_{3}}{3}\bigg).
\end{align}
The coefficient $G_{31}$ is found to be
\begin{align}
G_{31} =& \,\,\,C_{F}^{2}n_{F} \left(-\frac{77}{8}-\frac{19 \pi ^2}{36}+\frac{8 \pi ^4}{45}-\frac{104 \zeta_{3}}{9}\right)+C_{A}C_{F}n_{F}\bigg(-\frac{1118}{27}+\frac{644 \pi ^2}{81}-\frac{17 \pi
^4}{135}-\frac{292 \zeta_{3}}{9}\bigg)\notag\\
&+C_{F}n_{F}^{2} \bigg(\frac{191}{54}-\frac{61 \pi ^2}{81}+\frac{32 \zeta_{3}}{9}\bigg)+C_{A}^{2}C_{F} \bigg(\frac{5951}{54}-\frac{6625 \pi ^2}{324}+\frac{383 \pi ^4}{540}+\frac{404 \zeta_{3}}{9}+10
\zeta_{5}\bigg)\notag\\
&+C_{A}C_{F}^{2} \bigg(\frac{23}{2}+\frac{161 \pi ^2}{72}-\frac{53 \pi ^4}{45}+\frac{452 \zeta_{3}}{9}+2 \pi ^2 \zeta_{3}+30 \zeta_{5}\bigg)\notag\\
\label{G31}
&+C_{F}^{3}\left(\frac{29}{8}+\frac{5 \pi ^2}{4}-\frac{8 \pi ^4}{15}+53 \zeta_{3}-\frac{44}{3} \pi ^2 \zeta_{3}+132 \zeta_{5}\right).
\end{align}

The value of the $\mathcal{O}(\alpha_{s}^{3})$ constant $C_{3}$ is unknown and we fit it using the Monte Carlo program {\tt EERAD3} \cite{GehrmannDeRidder:2007jk}. Details about the fitting
procedure are explained in the next section. Following the conventions of \cite{CTTW} the final resummed expression $R_{T}(\tau)$ (\ref{finalresummed}) can be written as
\begin{align}\label{eq:Rresummed}
R_{T}(\tau) = \bigg(1+\sum_{k=1}^{\infty}C_{k}\bigg(\frac{\alpha_{s}}{2\pi}\bigg)^{k}\bigg){\rm
exp}\left[\log\frac{1}{\tau}g_{1}(\lambda)+g_{2}(\lambda)+\frac{\alpha_{s}}{\pi}\beta_{0}g_{3}(\lambda)+\left(\frac{\alpha_{s}}{2\pi}\right)^{3}G_{31}\log\frac{1}{\tau}\right],
\end{align}
where 
\begin{align}\label{eq:g_is}
g_{1}(\lambda) =& \,\,\,f_{1}(\lambda),\notag\\
g_{2}(\lambda) =& \,\,\,f_{2}(\lambda)-\log\Gamma(1-f_{1}(\lambda)-\lambda f_{1}^{\prime}(\lambda)),\notag\\
g_{3}(\lambda) =& \,\,\,f_{3}(\lambda)+\left(f_{1}^{\prime}+\frac{1}{2}\lambda
f_{1}^{\prime\prime}(\lambda)\right)\left(\psi^{(0)}(1-\gamma(\lambda))^{2}-\psi^{(1)}(1-\gamma(\lambda))\right)+f_{2}^{\prime}(\lambda)\psi^{(0)}(1-\gamma(\lambda))\notag\\
&+\frac{C_{F}}{\beta_{0}}\left(\gamma_{E}\left(\frac{3}{2}-\gamma_{E}\right)-\frac{\pi^{2}}{6}\right),
\end{align}
notice that the functions $g_{i}(\lambda)$ do not generate any constants ({\it i.e.} $g_{i}(0)=0$).

\begin{figure}[tbh]
 \centering
 \includegraphics[width=0.7\textwidth]{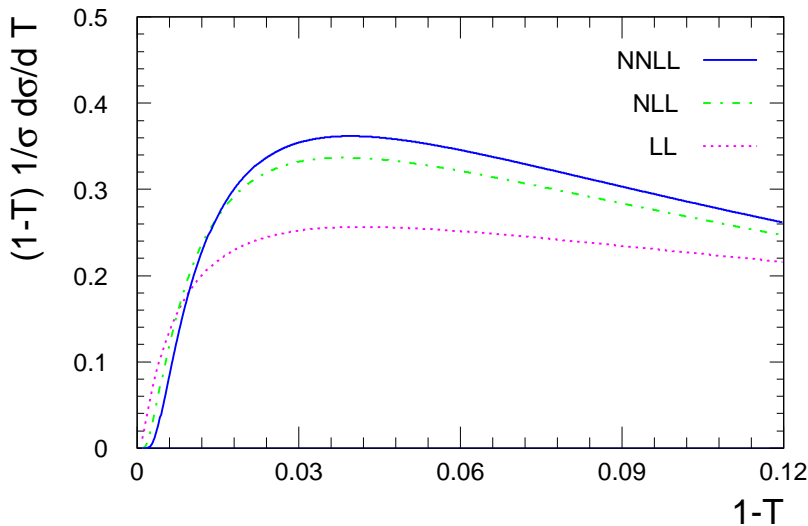}
 \caption{Comparison of the resummed result at different logarithmic orders around
the peak region.}
 \label{fig:resummation}
\end{figure}

The resummed expression at different logarithmic orders evaluated
around the peak region is shown in
Figure~\ref{fig:resummation}.

We observe that an exact inversion of Eq. (\ref{Rtau}) requires to choose the integration contour $C$ to the right of the Landau singularities present in
the resummed functions $f_{i}(\lambda)$. Following our prescription \cite{CTTW} for the inversion, we avoid such singularities by expanding around $\log N=\log\frac{1}{\tau}$ and 
then integrating by means of the residue theorem closing the contour in the left half-plane using Eq. (\ref{eq:mastinv}). In doing so, we are neglecting the contribution due to the residue at
the Landau pole, which gives rise to power suppressed terms \cite{Catani:1996yz}. The Landau singularity is then mapped onto the thrust space without contributing to the Laplace inversion.
By expressing the SCET result of~\cite{Becher:2008cf} in the form
(\ref{eq:Rresummed}) \cite{privBec},
we obtain full analytic agreement on $g_3(\lambda)$. This is a non-trivial
result, 
since in~\cite{Becher:2008cf} the scales in the resummation kernels are fixed at
the outset of 
the calculation, while in our analysis, they are integrated over in the 
Laplace inversion. This difference results in a different treatment
of the Landau singularity in the resummed expressions, which could produce a
power-suppressed
difference between the cross sections obtained in both approaches. The 
exact agreement of our $g_3(\lambda)$ with~\cite{Becher:2008cf} demonstrates that
there is no 
Landau-pole ambiguity between the two approaches. However, in \cite{Becher:2008cf} subleading terms arising from the scale fixing in $T$-space
and part of the constants are kept in the exponent resulting in a numerical difference when compared to our result.

\subsection{Determination of the $\mathcal{O}(\alpha_{s}^{3})$ constant $C_{3}$}
\label{sec:c3determination}
\begin{figure}[tb]
 \centering
 \includegraphics[scale=0.5]{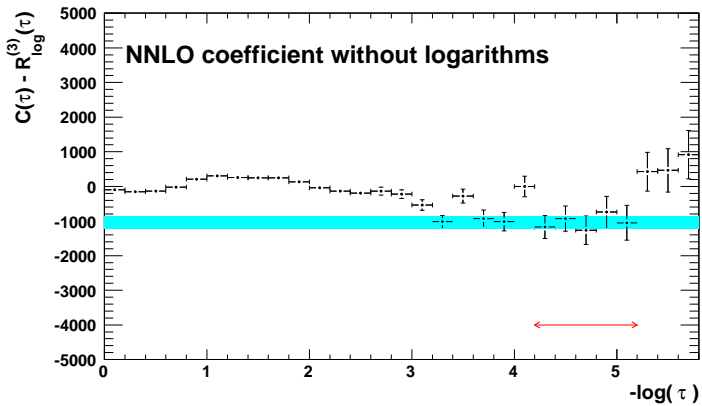}
 \caption{Fit of the $\mathcal{O}(\alpha_{s}^{3})$ coefficient $C_{3}$. The blue band shows the statistical
error of $C_{3}$ and the red arrow indicates the fitting interval.}
 \label{fig:c3fit}
\end{figure}

In order to match the resummed result to the NNLO cross section the $\mathcal{O}(\alpha_{s}^{3})$ constant
$C_{3}$ must be extracted from fixed-order data. We do it by subtracting the
logarithms from the fixed-order
$\mathcal{C}(\tau)$ coefficient obtained from {\tt EERAD3}. The logarithmic part
$R^{(3)}_{\rm log}(\tau)$ of the NNLO coefficient is decomposed into its different
color contributions
according to
\begin{align}
R^{(3)}_{\rm log}(\tau)=\left.R^{(3)}_{\rm log}\right|_{N^{2}} +  \left.R^{(3)}_{\rm
log}\right|_{N^{0}} + \left.R^{(3)}_{\rm log}\right|_{1/N^{2}} +
                        \left.R^{(3)}_{\rm log}\right|_{\nf\,N} + \left.R^{(3)}_{\rm log}\right|_{\nf/N}
+ \left.R^{(3)}_{\rm log}\right|_{\nf^{2}}\,.
\end{align}
{\tt EERAD3} is run with a technical cutoff $y_{0}=10^{-5}$ which
affects the thrust distribution below $\tau_{0}\sim \sqrt{y_{0}}$. This forbids us from probing
the far infrared region and we perform the fit for values of $\tau$ larger than $\tau_{0}$.
Numerical fixed order results are obtained with $6\times10^7$ points for the leading colour contribution and $10^7$ points for 
the subleading colour structures.
Because of the presence of large fluctuations in the Monte Carlo results, each color
contribution is fitted separately over an interval where the distribution is
stable and the different results
are combined to find the numerical value of $C_{3}$. 
The results of the fits and the
different fit intervals are given in Table~\ref{tab:c3intervals}. 
\begin{table}[tbh]
 \centering
 \begin{tabular}{|l|c|c|c|}
  \hline
  Color   & $(-\log\tau)_{\min}$ & $(-\log\tau)_{\max}$ & Fit result \\
  \hline
  $N^{2}  $  & $ 4.2 $  & $ 5.2 $ & $ 3541 \pm 51 $ \\
  $N^{0}  $  & $ 4.2 $  & $ 5.4 $ & $ -265 \pm 8  $ \\
  $1/N^{2}$  & $ 3.8 $  & $ 5.2 $ & $ -71  \pm 3  $ \\
  $N \nf  $  & $ 4.6 $  & $ 5.6 $ & $ -5078\pm 145$ \\
  $\nf/N  $  & $ 4.6 $  & $ 5.8 $ & $ 236  \pm 7  $ \\
  $\nf^2  $  & $ 4.2 $  & $ 5.2 $ & $ 95   \pm 120$ \\
  \hline
  \multicolumn{3}{|l|}{Sum of all colors} & $ -1543\pm 147$ \\
  \hline
  All colors & $ 4.2 $  & $ 5.2 $ & $ -1051\pm 178$ \\
  \hline
 \end{tabular}
  \caption{Intervals and results of the fits for $C_{3}$ for the different color
contributions.}\label{tab:c3intervals}
\end{table}
As an alternative approach we first sum all the color contributions to the $\mathcal{C}(\tau)$
 coefficient, then we subtract Eq.~(\ref{eq:R3}) and finally fit
$C_{3}$. The result of the
second approach is shown in Figure~\ref{fig:c3fit}.
We consider the difference between the two approaches as a systematic error and as
final result we obtain
\begin{equation}
C_{3}=-1050 \pm 180 (\textrm{stat.}) \pm 500 (\textrm{syst.})\,.
\end{equation}
Considering that there is  no statistical correlation between different bin errors, as a different possible estimate of the systematic 
uncertainty due to the sizeable fluctuation, we varied the fit range observing that it does not alter the result in any  
significant way outside the quoted systematic error margins. 
It is worth stressing that the numerical impact of $C_3$ on the distributions is negligible, as it will be shown in Section \ref{sec:results}, 
such that the large relative error range is tolerable for all practical purposes.
With the determination of the $\mathcal{O}(\alpha_{s}^{3})$ constant we now have all the needed ingredients
to perform the matching of the NNLL resummed result to the fixed NNLO.

\section{Matching of resummation to fixed-order calculations}
\label{sec:matching}
There are different matching schemes proposed in the literature, however mainly two
are used: the $R$-matching scheme and the $\log(R)$-matching
scheme~\cite{CTTW,Jones:2003yv}. The new results
presented in the previous sections allow us to compare for the first time the
predictions of the two schemes at NNLL+NNLO accuracy.
In the $R$-matching scheme the two expressions~(\ref{eq:Rfixed}) and~(\ref{eq:Rres})
are matched and logarithms appearing twice are subtracted. The explicit expression
for the matched integrated cross
section $R(\tau,Q)$ depends on both the logarithmic and fixed-order accuracy considered in the
matching. At NNLL+NNLO the following formula holds (for the sake of clarity we drop
any dependence on the
renormalization scale, which will be analyzed separately and write only $\asl$ as
arguments of the $g_{i}$ functions):
\begin{align}\label{eq:RmatchingNNLLNNLO}
R(\tau,Q)=&\left(1\,+\,C_{1}\asb+C_{2}\asb^{2}+C_{3}\asb^{3}\right)e^{\left(\Ltau\,g_{1}(\asl)+g_{2}(\asl)+\frac{\as}{\pi}\beta_{0}\,g_{3}(\asl)+\asb^{3}L\,G_{31}\right)}\nonumber\\
&+\asb    \left(\mathcal{A}(\tau)-R^{(1)}_{\rm log}\left(\tau\right)\right)
 +\asb^{2}\left(\mathcal{B}(\tau)-R^{(2)}_{\rm log}\left(\tau\right)\right)
 +\asb^{3}\left(\mathcal{C}(\tau)-R^{(3)}_{\rm log}\left(\tau\right)\right)\,.\notag\\
\end{align}
The terms in the second line correspond to the remainder functions
$d_{i}(\tau)$ defined above. It is however preferable to write it as difference
between the full fixed-order coefficient and
its logarithmic part since these are the functions which are known in practice.

The $\log(R)$-matching scheme~\cite{CTTW} is believed to be theoretically the most stable one
and for this reason it is generally preferred~\cite{Jones:2003yv}. In this case the
matching procedure is given by
\begin{equation}\label{eq:logRmatching}
\begin{split}
\log\left(R(\tau,\as)\right)\,=&L\,g_{1}(\asl)\,+\,g_{2}(\asl)+\asb g_{3}(\asl)\\
&{}+\,\asb\left(\mathcal{A}\left(\tau\right)-G_{11}L-G_{12}L^{2}\right)+{}\\
&{}+\,\asb^{2}\left(\mathcal{B}\left(\tau\right)-\frac{1}{2}\mathcal{A}_{1}^{2}\left(\tau\right)-G_{21}L-G_{22}L^{2}-G_{23}L^{3}\right){}\\
&{}+\,\asb^{3}\left(\mathcal{C}\left(\tau\right)-\mathcal{A}\left(\tau\right)\mathcal{B}\left(\tau\right)+\frac{1}{3}\mathcal{A}^{3}\left(\tau\right)-G_{32}L^{2}-G_{33}L^{3}-G_{34}L^{4}\right)\,.
\end{split}
\end{equation}

It is worth noting that the dependence on the $\mathcal{O}(\alpha_{s}^3)$ coefficients $C_{3}$ and $G_{31}$ disappears in this scheme.
The matching procedures presented above are valid over the whole phase space.
However, unlike to the fixed-order prediction~(\ref{eq:Rfixed}), in which every
coefficient vanishes in the kinematical limit such that
$R(\tau_{\rm max})=1$, in the limit $\tau\to \tau_{\rm max}$ the two
predictions~(\ref{eq:RmatchingNNLLNNLO}) and~(\ref{eq:logRmatching}) give a wrong ({\it i.e.} non-vanishing)
result. 
A fixed-order calculation takes into account only a finite
(in fact very few) number of final state
particles so the differential
cross section $\d\sigma/\d\tau$ obviously has to vanish at the kinematical limit $\tau=\tau_{{\rm max}}$.

For the NNLO fixed-order
prediction, the maximum number of
final state jets is five. Therefore the cross section should vanish at the kinematical limit for six partons. The limited predictive ability of the two matching schemes
in the multijet region can be solved
by modifying them slightly. This is done by imposing a kinematical constraint, which
assures the right prediction for $\tau\to \tau_{\rm max}$.

The constraints for the so-called modified $\log(R)$-matching scheme
are~\cite{Jones:2003yv}
\begin{equation}\label{eq:modmatchingconstr}
\log\left(R(\tau_{\rm
max},\as)\right)\,=\,0\qquad,\qquad\frac{1}{\sigma}\left.\frac{\d\sigma(\tau)}{\d
\tau}\right|_{\tau=\tau_{\rm max}} =\left.\frac{\d R(\tau)}{\d
\tau}\right|_{\tau=\tau_{\rm
max}}=\,0.
\end{equation}
In order to fulfil these two constraints, we follow the prescription proposed
in~\cite{Jones:2003yv} and redefine $L$ as follows:
\begin{equation}\label{eq:Ltilde}
L\,\longrightarrow\,\Lp\,=\,\frac{1}{p}\log\left(\left(\frac{1}{\tau}\right)^{p}-\left(\frac{1}{\tau_{\rm
max}}\right)^{p}+1\right)\,.
\end{equation}
The power $p$ is called ``degree of modification''. We choose $p=1$, as usual in
literature. It determines how fast the integrated cross section is damped at the
kinematical limit. The value of
$\tau_{\rm max}$ is given by symmetry arguments and at LO and NLO can be computed
exactly giving respectively $\tau_{\rm max,LO}=1/3$ and $\tau_{\rm
max,NLO}=1-1/\sqrt{3}$. At NNLO we can fix it
using the result given by {\tt EERAD3} at $\tau_{{\rm max}}=0.4275$.

For the $\log(R)$-matching scheme, the substitution~(\ref{eq:Ltilde}) is sufficient
to fulfil the constraints~(\ref{eq:modmatchingconstr}). In the $R$-matching scheme
one further modification
is needed
\begin{equation}\label{eq:modifiedGi1}
G_{i1}\to
G_{i1}(\tau)\,=\,G_{i1}\left[1-\left(\frac{\tau}{\tau_{\textrm{\tiny{max}}}}\right)^{p}\right],\quad\,i=1,2,3\,,
\end{equation}
leading to the following expression for the $R$-matching at NNLL+NNLO:
\begin{align}\label{eq:modRmatching}
R&\left(\tau\right)=\left(1+C_{1}\asb+C_{2}\asb^{2}+C_{3}\asb^{3}\right)\nonumber\\
&\times\exp\left[\Lp\,g_{1}(\aslp)+g_{2}(\aslp)+\frac{\as}{\pi}\beta_{0}g_{3}(\aslp)+\asb^{3}\Lp\,G_{31}(\tau)-\left(\frac{\tau}{\tau_{\textrm{\tiny{max}}}}\right)^{p}(G_{11}+G_{21})\right]\notag\\
&\qquad+\asb    \left(\mathcal{A}(\tau)-R^{(1)}_{\rm
log}\left(\tau\right)|_{L\to\Lp,G_{11}\to G_{11}(\tau)}\right)\notag\\
&\qquad+\asb^{2}\left(\mathcal{B}(\tau)-R^{(2)}_{\rm
log}\left(\tau\right)|_{L\to\Lp,G_{11}\to G_{11}(\tau),G_{21}\to
G_{21}(\tau)}\right)\notag\\
&\qquad+\asb^{3}\left(\mathcal{C}(\tau)-R^{(3)}_{\rm
log}\left(\tau\right)|_{L\to\Lp,G_{11}\to G_{11}(\tau),G_{21}\to
G_{21}(\tau),G_{31}\to G_{31}(\tau)}\right)\,.
\end{align}

The dependence on the renormalization scale was so far not considered. Every
term beyond the leading order acquires an explicit $\mu$-dependence, which for the fixed-order
coefficients is given
in~(\ref{eq:fixedordermudep}). For the resummation functions
$g_{i}\left(\asl\right)$ the renormalization scale dependence is given by
\begin{align}
g_{2}(\asl)\to\bar{g}_{2}\left(\asl,\mu^{2}\right)=&g_{2}(\asl)+\frac{\beta_{0}}{\pi}(\asl)^{2}g_{1}'(\asl)\log(x_{\mu}^{2})\\
g_{3}(\asl)\to\bar{g}_{3}\left(\asl,\mu^{2}\right)=&g_{3}(\asl)+\left[(\asl)g_{2}'(\asl)+\frac{\beta_{1}}{\pi\beta_{0}}(\asl)^{2}g_{1}'(\asl)\right]\log(x_{\mu}^{2})\notag\\
&\phantom{g_{3}(\asl)}+\left[\frac{\beta_{0}}{\pi}(\asl)^{2}g_{1}'(\asl)+\frac{\beta_{0}}{2\pi}(\asl)^{3}g_{1}''(\asl)\right]\log(x_{\mu}^{2})^{2}\,,\notag\\
\end{align}
where $g_{i}'(\asl)$ stands for the derivative of $g_{i}(\asl)$ with respect to $\alpha_{s}L$. Correspondingly the coefficients $G_{ij}$ and $C_{i}$ change as follows:
\begin{eqnarray}
G_{21}\rightarrow\,\bar{G}_{21}\left(\mu^{2}\right)&=&G_{21}\,+\,2\beta_{0}G_{11}\log(x_{\mu}^{2})\nonumber\\
G_{22}\rightarrow\,\bar{G}_{22}\left(\mu^{2}\right)&=&G_{22}\,+\,2\beta_{0}G_{12}\log(x_{\mu}^{2})\nonumber\\
G_{33}\rightarrow\,\bar{G}_{33}\left(\mu^{2}\right)&=&G_{33}\,+\,2\beta_{0} 2
G_{23}\log(x_{\mu}^{2})\nonumber\\
G_{31}\rightarrow\,\bar{G}_{31}\left(\mu^{2}\right)&=&G_{31}\,+\,\left(2\beta_{0}\log(x_{\mu}^{2})\right)^{2}G_{11}\,+\,2\log(x_{\mu}^{2})\left(2\beta_{0}G_{21}\,+\,2\beta_{1}G_{11}\right)\nonumber\\
G_{32}\rightarrow\,\bar{G}_{32}\left(\mu^{2}\right)&=&G_{32}\,+\,\left(2\beta_{0}\log(x_{\mu}^{2})\right)^{2}G_{12}\,+\,2\log(x_{\mu}^{2})\left(2\beta_{0}G_{22}\,+\,2\beta_{1}G_{12}\right)\nonumber\\
C_{2}\rightarrow\,\bar{C}_{2}\left(\mu^{2}\right)&=&C_{2}\,+\,2\beta_{0}C_{1}\log(x_{\mu}^{2})\nonumber\\
C_{3}\rightarrow\,\bar{C}_{3}\left(\mu^{2}\right)&=&C_{3}+2\left(2\beta_{0}C_{2}+2\beta_{1}C_{1}\right)\log(x_{\mu}^{2})+\left(2\beta_{0}\log(x_{\mu}^{2})\right)^{2}C_{1}.\label{
eq:Gijrenorm}
\end{eqnarray}

One further source of arbitrariness is the choice of the logarithm to be resummed.
In fact, it is not clear whether powers of $\as\log(1/\tau)$ or powers of
$\as\log(2/\tau)$ have to be resummed.
The origin of this arbitrariness has to do with how much of the non-logarithmic part
of the fixed-order prediction is exponentiated together with the logarithms. We can
express this arbitrariness by
introducing a new constant $x_{L}$, which rescales the logarithm to be
resummed~\cite{Jones:2003yv}:
\begin{equation}\label{eq:Lhat}
L\,\rightarrow\,\hat{L}=\frac{1}{p}\log\left[\left(\frac{1}{x_{L}\tau}\right)^{p}-\left(\frac{1}{x_{L}\tau_{\rm
max}}\right)^{p}+1\right].
\end{equation}
This rescaling modifies once more the resummed formulae and their expansion
coefficients. By requiring $\hat{R}(\tau)\,\stackrel{!}{=}\,R(\tau)\,,$ where
$\hat{R}(\tau)$ denotes the rescaled
integrated cross section according to (\ref{eq:Lhat}), we find the following replacements:
\begin{eqnarray}
\hat{C}_{1}(\tau)&=&
C_{1}+G_{11}\log(x_{L})+G_{12}\log(x_{L})^{2}\,,\label{eq:c1hattrafo}\\
\hat{C}_{2}(\tau)&=&
C_{2}+\left(G_{21}+C_{1}G_{11}\right)\log(x_{L})+\left(G_{22}+\frac{1}{2}G_{11}^{2}+C_{1}G_{12}\right)\log(x_{L})^{2}\nonumber\\
&&+\left(G_{23}+G_{12}G_{11}\right)\log(x_{L})^{3}+\frac{1}{2}G_{12}^{2}\log(x_{L})^{4}\,,\\
\hat{C}_{3}(\tau)&=&C_{3}+\left(G_{31}+C_{1}G_{21}+C_{2}G_{11}\right)\log(x_{L})\nonumber\\
&&+\left(G_{32}+C_{1}G_{22}+\frac{1}{2}C_{1}G_{11}^{2}+C_{2}G_{12}+G_{11}G_{21}\right)\log(x_{L})^{2}\nonumber\\
&&+\left(G_{33}+G_{11}G_{22}+G_{12}G_{21}+C_{1}G_{11}G_{12}+\frac{1}{6}G_{11}^{3}+C_{1}G_{23}\right)\log(x_{L})^{3}\nonumber\\
&&+\left(G_{34}+G_{12}G_{22}+\frac{1}{2}C_{1}G_{12}^{2}+G_{11}G_{23}+\frac{1}{2}G_{11}^{2}G_{12}\right)\log(x_{L})^{4}\nonumber\\
&&+\left(G_{12}G_{23}+\frac{1}{2}G_{12}^{2}G_{11}\right)\log(x_{L})^{5}\,+\,\frac{1}{6}G_{12}^{3}\log(x_{L})^{6}\,,
\end{eqnarray}
\begin{eqnarray}
G_{12}\,\rightarrow\,\hat{G}_{12}&=&G_{12}\,,\qquad\hat{G}_{23}\,=\,G_{23}\,,\qquad\hat{G}_{34}\,=\,G_{34}\,,\nonumber\\
G_{11}\,\rightarrow\,\hat{G}_{11}&=&G_{11}+2G_{12}\log(x_{L})\,,\nonumber\\
G_{22}\,\rightarrow\,\hat{G}_{22}&=&G_{22}+3G_{23}\log(x_{L})\,,\nonumber\\
G_{33}\,\rightarrow\,\hat{G}_{33}&=&G_{33}+4G_{34}\log(x_{L})\,,\nonumber\\
G_{21}\,\rightarrow\,\hat{G}_{21}&=&G_{21}+2G_{22}\log(x_{L})+3G_{23}\log(x_{L})^{2}\,,\nonumber\\
G_{32}\,\rightarrow\,\hat{G}_{32}&=&G_{32}+3G_{33}\log(x_{L})+6G_{34}\log(x_{L})^{2}\,,\nonumber\\
G_{31}\,\rightarrow\,\hat{G}_{31}&=&G_{31}+2G_{32}\log(x_{L})+3G_{33}\log(x_{L})^{2}+4G_{34}\log(x_{L})^{3}\,.\label{eq:Ghattrafo}
\end{eqnarray}
The corresponding changes of the $g_{i}$ functions are
\begin{align}
g_{1}(\asl)\rightarrow\,\hat{g}_{1}(\aslh)=&\,g_{1}(\aslh)\,,\notag\\
g_{2}(\asl)\rightarrow\,\hat{g}_{2}(\aslh)=&\,g_{2}(\aslh)+\left(g_{1}(\aslh)+(\aslh)g_{1}'(\aslh)\right)\log(x_{L})\,,\notag\\
g_{3}(\asl)\rightarrow\,\hat{g}_{3}(\aslh)=&\,g_{3}(\aslh)+\frac{\pi}{\beta_{0}}g_{2}'(\aslh)\log(x_{L})\notag\\
&\quad+\frac{\pi}{\beta_{0}}\left(g_{1}'(\aslh)+\frac{1}{2}(\aslh)g_{1}''(\aslh)\right)\log(x_
{ L } )^ { 2 } \,.
\end{align}
The transformations due to a variation of $x_{\mu}$ and $x_{L}$ are completely
general and hold for all possible event-shape observables which can be described
with this matching formalism.
Furthermore the order in which they are carried out is not important since they
commute.

\section{Results}
\label{sec:results}
Having set up the matching formalism in a way to access the theoretical
uncertainties, we can apply it to the case of thrust derived in
Section~\ref{sec:resummation of large logs} using the fixed
order result from {\tt EERAD3}~\cite{GehrmannDeRidder:2007hr,GehrmannDeRidder:2007bj}. 

\begin{figure}[tbh]
 \centering
 \includegraphics[width=0.7\textwidth]{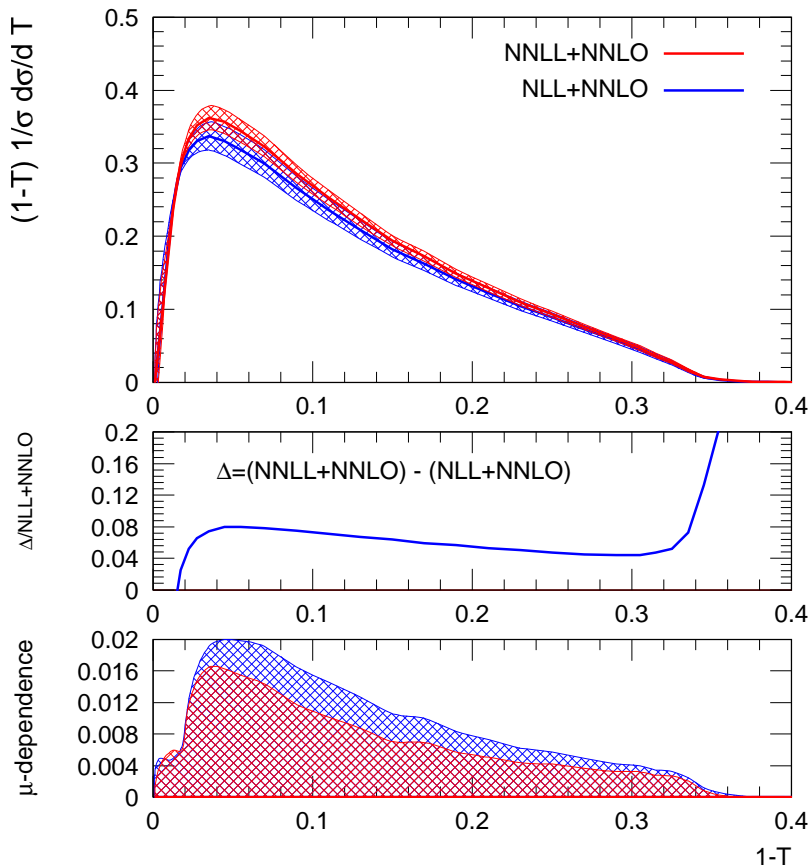}
 \caption{Comparison of the weighted cross section in the $\log(R)$-matching using
NNLL+NNLO and NLL+NNLO. The plot on the top shows the two distributions, with the
uncertainty band due to scale
dependence. The curve in the middle shows the difference between NNLL+NNLO and
NLL+NNLO normalized to the NLL+NNLO curve. The impact of the resummation at NNLL is
an increase in the distribution of
order 5-8\%. The lowest plot shows the absolute scale dependence of the two
curves.}
 \label{fig:logrmatching}
\end{figure}

\begin{figure}[tbh]
  \begin{minipage}[c]{.47\textwidth}
    \centering
    \includegraphics[width=1.0\textwidth]{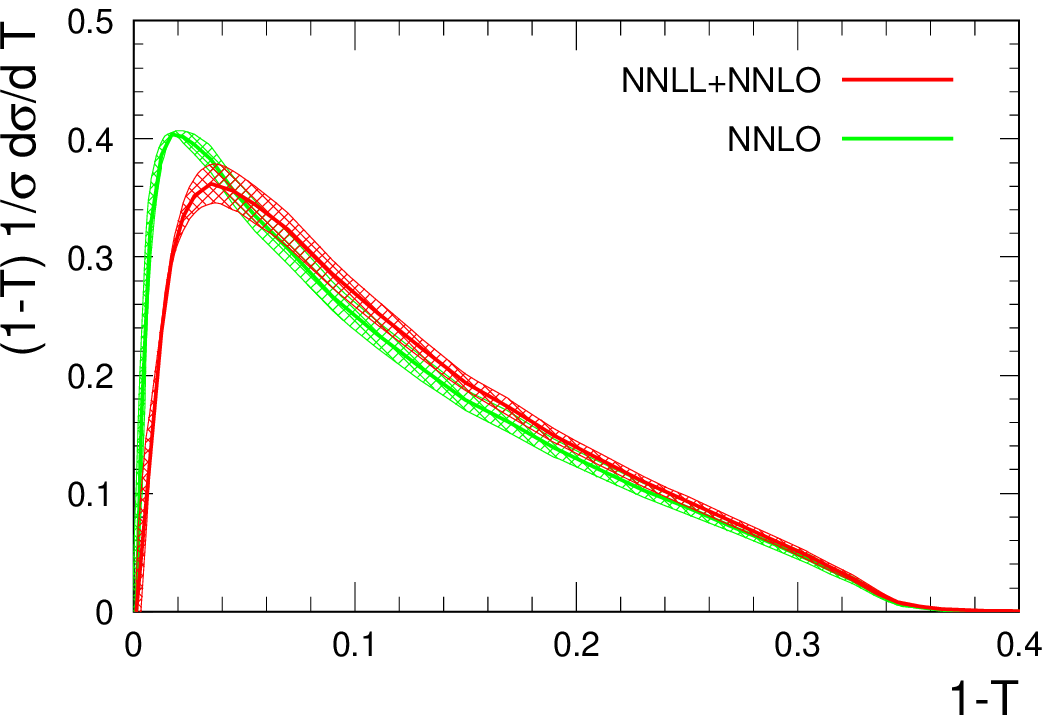}
    \caption{Comparison of the weighted cross section at NNLO with the matched
NNLL+NNLO predictions. The contribution of NNLL resummation is sizable over the
full thrust range.}
    \label{fig:nnllnnlo}
 \end{minipage}
 \hspace{5mm}
 \begin{minipage}[c]{.47\textwidth}
    \centering
    \includegraphics[width=1.0\textwidth]{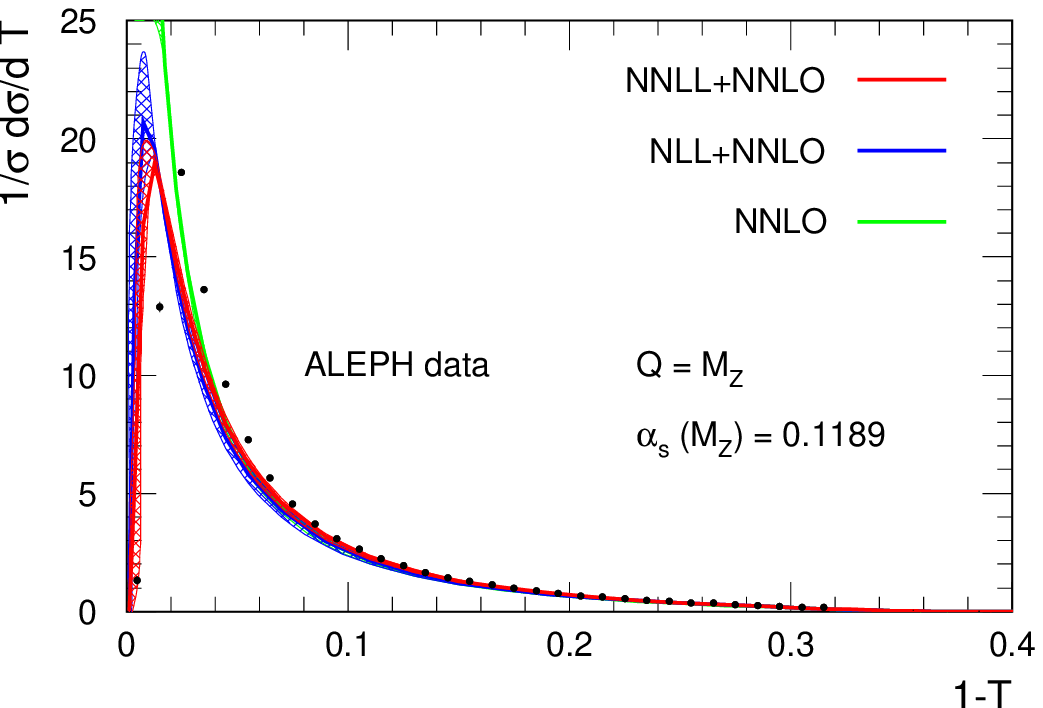}
    \caption{Comparison of the parton level fixed-order prediction at NNLO to NLL+NNLO
and NNLL+NNLO distributions. Experimental data are taken from the ALEPH collaboration~\cite{aleph}.}
    \label{fig:linplot}
  \end{minipage}
\end{figure}

In Figure~\ref{fig:logrmatching} we compare the weighted cross section of the new
matched NNLL+NNLO results with the old NLL+NNLO derived in~\cite{Gehrmann:2008kh}.
The modification
due to the resummation is sizable, leading to a $8\%$ increase of the distribution
around the peak region. The effect of the additional resummed subleading logarithms
becomes progressively less
important towards the multijet region, where the increase is nevertheless of about
$5\%$. It is interesting to note that the matching of NNLO with NNLL resummation
shifts the pure NNLO result also in
the multijet region (Figure~\ref{fig:nnllnnlo}). This was not the case for NLL+NNLO,
for which the impact of resummation in the region of large $\tau$ was negligible.
This is another sign of the
importance of the NNLL contribution. 

The renormalization scale dependence, which was observed to increase from pure NNLO
to NLL+NNLO~\cite{Gehrmann:2008kh,Dissertori:2009ik} because of a mismatch in the
cancellation of renormalization
scale logarithms, is obtained by varying $0.5<x_{\mu}<2$. It decreases at NNLL+NNLO
by 20\% in the peak region compared to NLL+NNLO. The magnitude of the scale
uncertainty varies between 4\%
in the 3-jet region and 5\% around the peak. 

In Figure~\ref{fig:linplot} we compare the unweighted parton-level cross section at
NNLO with the matched NLL+NNLO and NNLL+NNLO cross sections and with experimental
hadron-level data from the ALEPH
experiment~\cite{aleph}. We note that there is a visible shift of the theoretical
prediction towards the experimental data, which are best described by the newly
computed NNLL+NNLO distribution.
Around the peak region, where non-perturbative hadronization corrections are
large, the parton level prediction fails to describe the data. Hadronization
effects can account for this 
discrepancy. We will address this issue in a future publication. 

\begin{figure}[h]
 \centering
 \includegraphics[width=0.48\textwidth]{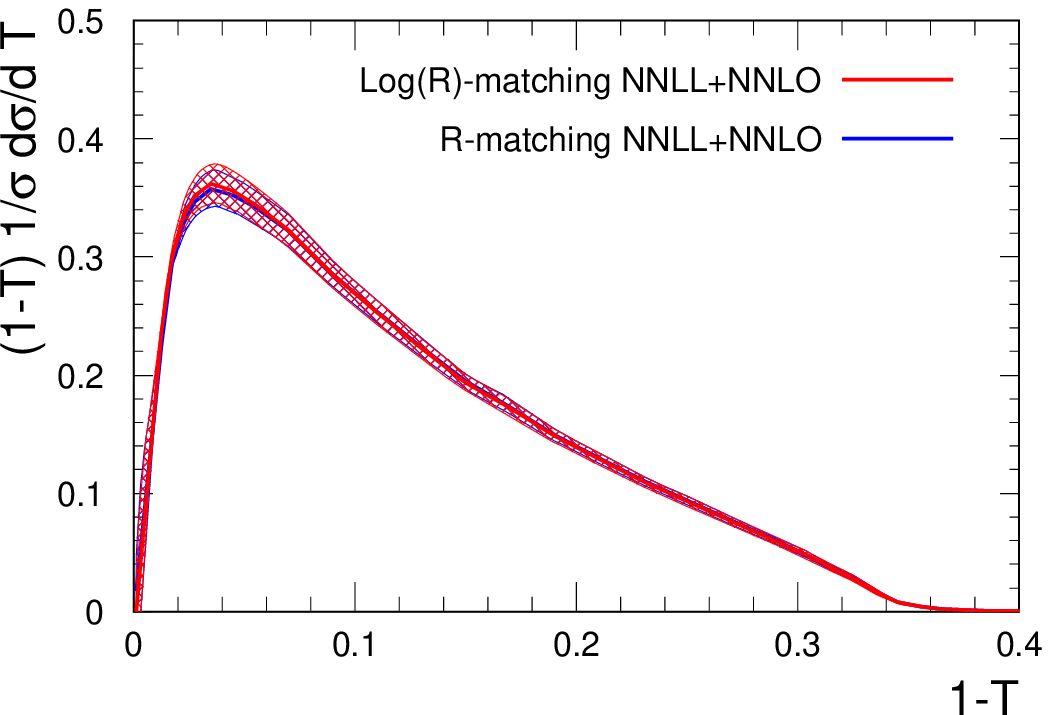}
 \quad
 \includegraphics[width=0.47\textwidth]{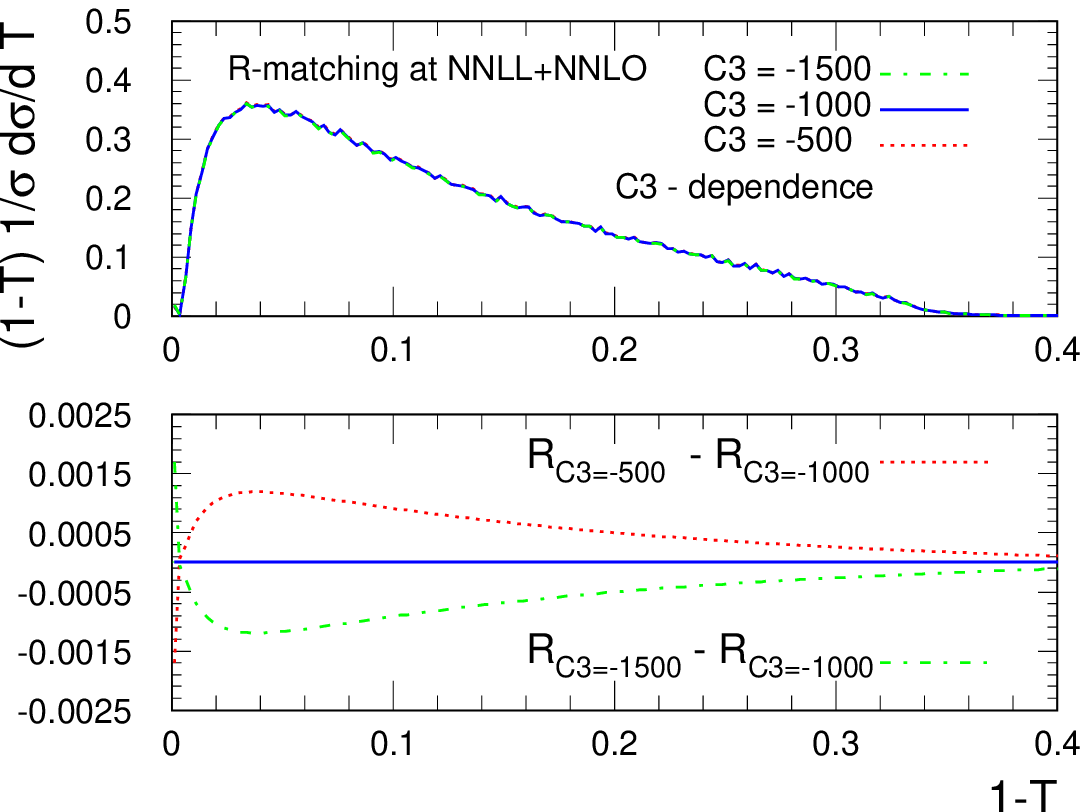}
 \caption{The left plot shows the comparison of the $R$-matching scheme and the
$\log(R)$-matching scheme results. The width of the curve shows the uncertainty
related to the scale variation. The two
matching schemes agree very well over the full thrust range. The right plot shows
the impact of the variation of the $\mathcal{O}(\alpha_{s}^{3})$ constant $C_3$ on the distribution in the
R-matching scheme. The difference
is at the per mille level.}\label{fig:rmatching}
\end{figure}

The computation of the two-loop constant $C_{2}$ presented in
Section~\ref{sec:soft} and the fit of the $\mathcal{O}(\alpha^{3}_{s})$ constant $C_{3}$ of
Section~\ref{sec:c3determination} allow us to perform for the first time the matching in the 
$R$-scheme using NNLL accuracy and
fixed NNLO results. On the left plot in Figure~\ref{fig:rmatching} we compare the $R$-matching
and the $\log(R)$-matching
scheme predictions at NNLL+NNLO. The difference between the two matching prescriptions is
very small and lies well below the scale uncertainty. This implies a very good
stability of the theoretical
predictions under variation of the matching scheme. Because of the big uncertainty in the
value of the constant $C_{3}$, we vary it within its error to investigate the
phenomenological impact on the cross
section. The results are shown in the right plot of Figure~\ref{fig:rmatching}. The
upper plot shows the different distributions obtained by setting $C_{3}=
-500,\,-1000,\,-1500$ respectively. The
three curves are almost indistinguishable and the tiny fluctuations in the
distributions are due to the fluctuations of the NNLO result. In the lower plot we
take the distribution with $C_{3}=-1000$
as reference and plot the difference between the reference cross section and those
obtained with $C_{3}=-500$ and $C_{3}=-1500$ respectively. The difference is less than $1.5\permil$ and it is
therefore completely negligible compared the other theoretical uncertainties.

\begin{figure}[h]
 \centering
 \includegraphics[width=0.48\textwidth]{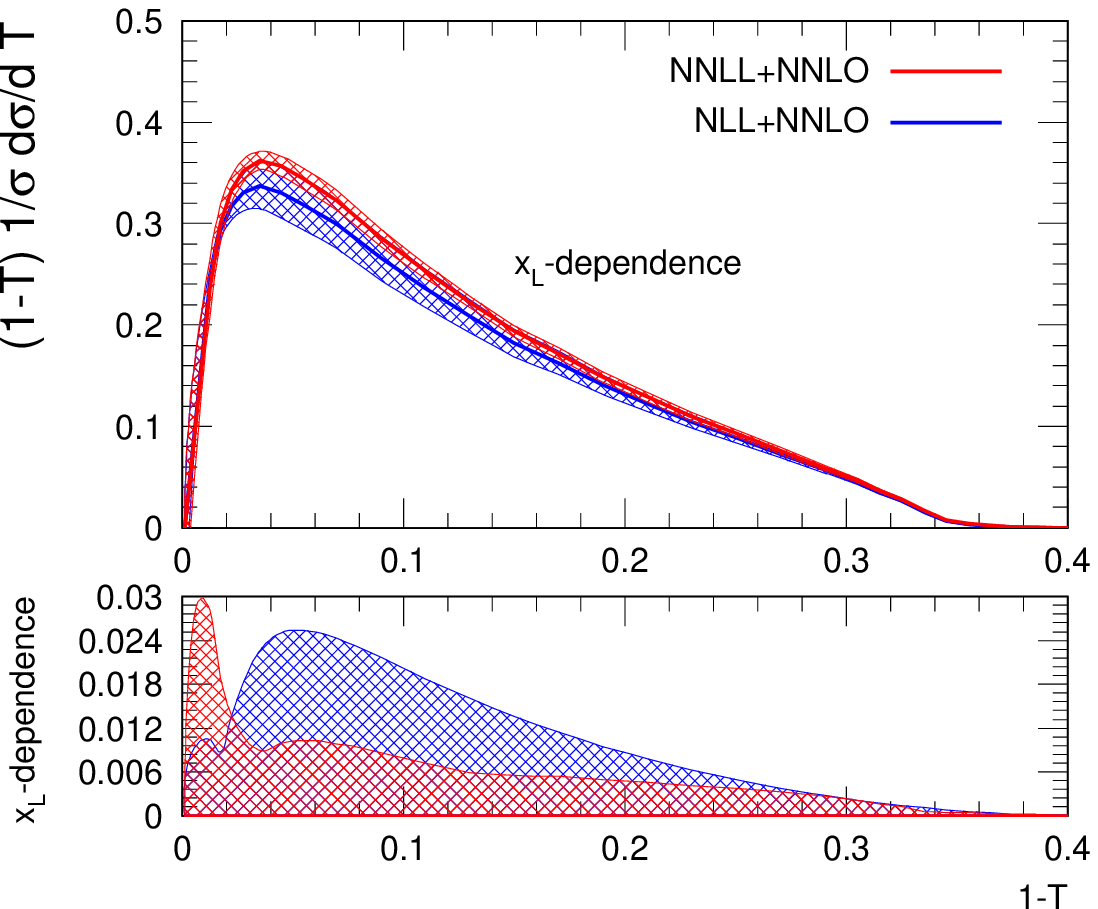}
 \quad
 \includegraphics[width=0.48\textwidth]{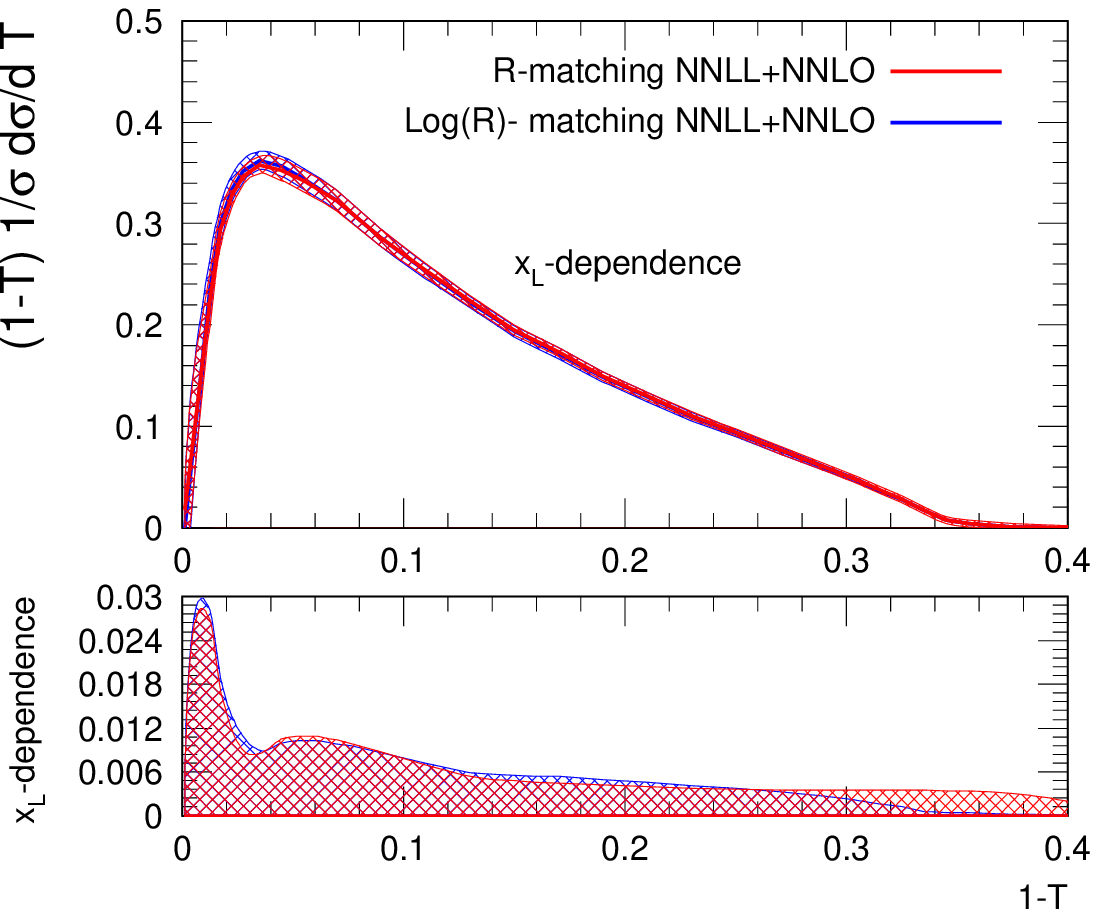}
 \caption{Dependence on the resummed logarithms, determined by varying the parameter
$x_{L}$. The left plot shows the change in the $x_{L}$ dependence between NLL+NNLO
and NNLL+NNLO. The upper plot
shows the distributions with the corresponding uncertainty band, in the lower plot
we compare only the uncertainties. In the right plots the $x_{L}$ dependence using
the two different matching
schemes is shown.}
 \label{fig:matchingxL}
\end{figure}

As discussed in the previous section, another source of theoretical uncertainty is the
choice of the logarithms to be resummed. We can estimate this uncertainty by varying
the parameter $x_{L}$. In
Ref.~\cite{Jones:2003yv} several prescriptions are given on how to set the correct
variation range for $x_{L}$ for different observables. For the sake of simplicity
and since we are not performing a
fit of the strong coupling constant, we choose to vary $x_{L}$
within the canonical interval $0.5<x_{L}<2$, 
similarly to what is chosen to quantify the
renormalization scale uncertainty. This
choice is also close to the nominal range of variation proposed
in~\cite{Jones:2003yv}. The impact of this variation is shown in
Figure~\ref{fig:matchingxL}. The left plots show a comparison of the
$x_{L}$-dependence between NLL+NNLO and NNLL+NNLO predictions. The lower plot
allows to quantify the reduction of the uncertainty due to a variation $x_{L}$. 
Apart from the far infrared
region, it is observed to decrease by $50\%$ in the peak region. The scale-dependence reduction is smaller towards the
multijet region, where the contribution of the logarithmic part becomes less
important. The resummation uncertainty at
NNLL+NNLO varies between $2\%$ and $3\%$. In the right plots the same comparison is
made at NNLL+NNLO using the $R$-matching and 
$\log(R)$-matching schemes. We observe a similar $x_{L}$-dependence in
both schemes.

\section{Conclusions and outlook}
\label{sec:conc} 
In this paper, we computed the resummed thrust cross section in $e^+e^-$ annihilation
to NNLL accuracy in Laplace space and we provided a compact expression for it in momentum space. 
We analytically derived the two-loop corrections to the soft subprocess for thrust, for which
only two numerical results~\cite{Chien:2010kc,hoangkluth} 
fitted to the subtracted fixed-order expressions were 
available previously. We find reasonable
agreement with them within their uncertainty. 
Our NNLL result for the thrust distribution confirms an earlier calculation carried out in 
the SCET framework~\cite{Becher:2008cf} in a different calculational approach. We 
find full analytic agreement between the two results, including the absence of power-suppressed 
terms which would be beyond the logarithmic accuracy. 

The NNLL corrections were matched to the fixed-order NNLO prediction for the 
thrust distribution in the $R$-matching and the $\log(R)$-matching schemes. By inspecting 
renormalization-scale logarithms in the resummed and fixed-order expressions, the 
appropriate combinations are NLL+NLO and NNLL+NNLO, while the NLL+NNLO 
combination leads to a mismatch in these logarithms, resulting in an enhanced scale-dependence. 

In SCET-based analyses \cite{Becher:2008cf,Abbate:2010xh,Chien:2010kc} the appropriate combinations are
usually determined by requiring that the order of $\alpha_{s}$ of the hard, jet and soft functions matches 
the fixed order expansion. This results in a different identification of matched distributions.
We observe that the NNLL contribution is sizable even outside the 
dijet region and 
that the inclusion of the NNLL corrections further stabilizes the perturbative prediction 
throughout the full kinematical range.

A comparison with experimental data has to include 
hadronization effects, which are most pronounced in the two-jet region. In the past, these 
were often obtained using leading-logarithmic parton shower Monte Carlo programs, which 
turn out to be clearly insufficient~\cite{Dissertori:2009ik} in view of the precision 
now attained by the perturbative description. Systematic approaches to 
hadronization within the dispersive model~\cite{alpha0,jaquier} or 
by using the shape function formalism~\cite{Abbate:2010xh} 
are offering  a more reliable description. We will address this issue in future work.

\section*{Acknowledgments}
We wish to thank L.~Magnea for interesting discussions on factorization issues, L.~Yang for stimulating comments on several aspects of the calculation and G.~Heinrich for help concerning {\tt SecDec}.
Furthermore we acknowledge several fruitful discussions with A.~Banfi, T.~Becher, E.~Gardi, M.~Grazzini and D.\ Ma\^{\i}tre. GL would like to thank the University of Z\"urich for its hospitality. TG would like to thank the Kavli Institute for 
Theoretical Physics (KITP) at UC Santa Barbara for hospitality while this 
work was completed. 
This research is supported in part by
the Swiss National Science Foundation (SNF) under contract
200020-126691, by the UK  STFC,  by the European Commission through the 
``LHCPhenoNet" Initial Training Network PITN-GA-2010-264564 and 
by the National Science Foundation under grant NSF PHY05-51164.

\appendix
\section{Constants and anomalous dimensions}
\label{sec:constants}
The renormalization group equation for the QCD coupling constant reads
\begin{align}
\label{rgealpha}
\frac{d\alpha_{s}(\mu)}{d\log\mu^{2}}=-\alpha_{s}(\mu)\bigg(\frac{\alpha_{s}(\mu)}{\pi}\beta_{0}+\frac{\alpha_{s}^{2}(\mu)}{\pi^{2}}\beta_{1}+\bigg(\frac{\alpha_{s}^{2}(\mu)}{\pi^{2}}\bigg)^2\beta_{2}+
\bigg(\frac{\alpha_{s}^{2}(\mu)}{\pi^{2}}\bigg)^3\beta_{3}+\ldots \bigg),
\end{align}
where the coefficients of the $\beta(\alpha_{s})$ functions are 
\begin{align}
\beta_{0} =& \frac{11}{12}C_{A}-\frac{1}{3}T_{F}n_{F},\notag\\
\beta_{1} =& \frac{17}{24}C_{A}^{2}-\frac{5}{12}C_{A}T_{F}n_{F}-\frac{1}{4}C_{F}T_{F}n_{F},\notag\\
\beta_{2} =& \frac{325}{3456}n_{F}^{2}-\frac{5033}{1152}n_{F}+\frac{2857}{128},\notag\\
\beta_{3} =&
\frac{1093}{186624}n_{F}^{3}+\bigg(\frac{50065}{41472}+\frac{809}{2592}\zeta_{3}\bigg)n_{F}^{2}-\bigg(\frac{1078361}{41472}+\frac{1627}{1728}\zeta_{3}\bigg)n_{F}+\frac{891}{64}\zeta_{3}+\frac{149753}{
256}.
\end{align}
Equation (\ref{rgealpha}) can be solved in perturbation theory and it gives the following resummed expression for the strong coupling
\begin{align}
\alpha_{s}(\mu) &= \frac{\alpha_{s}(\mu_{R})}{1+2\lambda}\bigg(1-\frac{\alpha_{s}(\mu_{R})}{\pi(1+2\lambda)}\frac{\beta_{1}}{\beta_{0}}\log(1+2\lambda)\notag\\
&+\frac{\alpha^{2}
_{s}(\mu_{R})}{\pi^2(1+2\lambda)^{2}}\bigg(\frac{\beta_{1}^{2}}{\beta_{0}^{2}}\big(\log(1+2\lambda)(\log(1+2\lambda)-1)+2\lambda\big)-\frac{\beta_{2}}{\beta_{0}}2\lambda\bigg)\notag\\
&+\frac{\alpha^{3}_{s}(\mu_{R})}{\pi^3(1+2\lambda)^{3}}\bigg(\frac{\beta_{1}^{3}}{\beta_{0}^{3}}\big(-4\lambda\log(1+2\lambda)+\frac{5}{2}\log^{2}
(1+2\lambda)-\log^{3}(1+2\lambda)-2\lambda^{2}\big)\notag\\
&-\frac{\beta_{3}}{\beta_{0}}2\lambda(1+\lambda)+\frac{\beta_{1}\beta_{2}}{\beta_{0}^{2}}
\big(2(1+2\lambda)\log(1+2\lambda)-3\log(1+2\lambda)+2\lambda(1+2\lambda)\big)\bigg)\bigg)\notag\\
\label{alpharunning}
&+...,
\end{align}
where here $\lambda=({\alpha_{s}(\mu_{R})}/{\pi})\beta_{0}\log ({\mu}/{\mu_{R}})$.
The coefficients $A^{(i)}$ and $B^{(i)}$ used in the resummed cross section (\ref{resummedcross}) can be computed using $\Gamma_{\rm{soft}}$ at two-loop order and the three-loop Altarelli-Parisi
splitting functions computed in \cite{NNLOAPS,NNLOAPNS}. Using the notation (\ref{ABseries}) we find
\begin{align}
A^{(1)} =& \,\,C_{F},\notag\\
A^{(2)} =& C_{F}\bigg(C_{A}\bigg(\frac{67}{36}-\frac{\pi^{2}}{12}\bigg)-\frac{5}{9}T_{F}n_{F}\bigg),\notag\\
A^{(3)} =& \,\,C_{F} n_{F}^{2} \left(\frac{25}{324}-\frac{\pi^{2}}{216}\right)+C_{A}C_{F}n_{F}\left(-\frac{2051}{1296}+\frac{7\pi^{2}}{72}\right)+C_{F}^{2} n_{F}
\left(-\frac{55}{96}+\frac{\zeta_{3}}{2}\right)\notag\\
&+C_{A}^{2} C_{F} \left(\frac{15503}{2592}-\frac{389\pi^{2}}{864}+\frac{11\pi^{4}}{720}-\frac{11\zeta_{3}}{4}\right),\notag\\
B^{(1)} =& -\frac{3}{4}C_{F},\notag\\
B^{(2)} =& \,\,C_{F}n_{F} \left(\frac{1}{48}+\frac{\pi^{2}}{36}\right)+C_{F}^{2}\left(-\frac{3}{32}+\frac{\pi^{2}}{8}-\frac{3 \zeta_{3}}{2}\right)+C_{A}
C_{F}\left(-\frac{17}{96}-\frac{11\pi^{2}}{72}+\frac{3\zeta_{3}}{4}\right),
\end{align}
\begin{align}
B^{(3)} =& C_{A}C_{F}n_{F}\bigg(-\frac{5}{16}+\frac{167 \pi ^2}{1296}-\frac{\pi ^4}{2880}-\frac{25 \zeta_{3}}{72}\bigg)+C_{F}n_{F}^{2}\bigg(\frac{17}{576}-\frac{5 \pi
^2}{648}+\frac{\zeta_{3}}{36}\bigg)\notag\\
&+C_{F}^{2}n_{F} \bigg(\frac{23}{64}-\frac{5 \pi ^2}{288}-\frac{29 \pi ^4}{4320}+\frac{17 \zeta_{3}}{24}\bigg)+C_{A}C_{F}^{2} \bigg(-\frac{151}{256}+\frac{205 \pi ^2}{576}+\frac{247 \pi
^4}{8640}-\frac{211 \zeta_{3}}{48}\notag\\
&-\frac{1}{24} \pi ^2 \zeta_{3}-\frac{15 \zeta_{5}}{8}\bigg)+C_{A}^{2}C_{F} \bigg(\frac{1657}{2304}-\frac{281 \pi ^2}{648}+\frac{\pi ^4}{1152}+\frac{97 \zeta_{3}}{36}-\frac{5
\zeta_{5}}{8}\bigg)\notag\\
&+C_{F}^{3} \bigg(-\frac{29}{128}-\frac{3 \pi ^2}{64}-\frac{\pi ^4}{40}-\frac{17 \zeta_{3}}{16}+\frac{1}{12} \pi ^2 \zeta_{3}+\frac{15 \zeta_{5}}{4}\bigg).
\end{align}
The functions $H(1,\alpha_{s}(Q))$, $\tilde{J}(1,\alpha_{s}(\sqrt{({N_{0}}/{N})}Q))$ and $\tilde{S}(1,\alpha_{s}({N_{0}Q}/{N}))$ can be expanded in a power series in the coupling as 
\begin{align}
H(1,\alpha_{s}(Q))=1+\sum_{i\geq1}&c_{h}^{(i)}\frac{\alpha_{s}^{i}(Q)}{\pi^{i}},\qquad
\tilde{J}\left(1,\alpha_{s}(\sqrt{({N_{0}}/{N})}Q)\right)=1+\sum_{i\geq1}c_{j}^{(i)}\frac{\alpha_{s}^{i}(\sqrt{({N_{0}}/{N}})Q)}{\pi^{i}},\notag\\
& \tilde{S}(1,\alpha_{s}({N_{0}Q}/{N}))=1+\sum_{i\geq1}c_{s}^{(i)}\frac{\alpha_{s}^{i}({N_{0}Q}/{N})}{\pi^{i}}.
\end{align}
The coefficients $c_{h}^{(i)}$ can be evaluated using the on-shell quark form factor \cite{3Lform1,3Lform2}, normalized to the total hadronic cross section
\begin{align}
c_{h}^{(1)} =& \,\,C_{F}\left(-\frac{19}{4}+\frac{7\pi^{2}}{12}\right),\notag\\
c_{h}^{(2)} =& \,\,C_{F}^{2}\left(\frac{745}{64}-\frac{13\pi^{2}}{6}+\frac{67\pi^{4}}{480}-\frac{15\zeta_{3}}{4}\right)+C_{F}n_{F}T_{F}\left(\frac{5867}{1296}-\frac{91\pi^{2}}{216}-\frac{17
\zeta_{3}}{18}\right)\notag\\
&+C_{A}C_{F}\left(-\frac{71083}{5184}+\frac{1061\pi^{2}}{864}-\frac{\pi^{4}}{90}+\frac{511\zeta_{3}}{72}\right).
\end{align}
The two loop non-logarithmic term of the collinear subprocess was computed in \cite{SCETJet}, resulting in
\begin{align}
c_{j}^{(1)} =& \,\,C_{F}\left(\frac{7}{4}-\frac{\pi^{2}}{6}\right),\notag\\
c_{j}^{(2)} =& \,\,C_{F}T_{F}n_{F}\left(-\frac{4057}{2592}+\frac{13\pi^{2}}{144}\right)+C_{F}C_{A}\left(\frac{53129}{10368}-\frac{155\pi^{2}}{576}-\frac{37\pi^{4}}{2880}-\frac{9
\zeta_{3}}{8}\right)\notag\\
&+C_{F}^{2}\left(\frac{205}{128}-\frac{97\pi^{2}}{192}+\frac{61\pi^{4}}{1440}-\frac{3\zeta_{3}}{8}\right),
\end{align}
and the non-logarithmic part of the two loop soft subprocess was computed in section 2 (\ref{twoloopcomplete}), (\ref{c2soft})
\begin{align}
c_{s}^{(1)} =& -C_{F}\frac{\pi^{2}}{4},\notag\\
c_{s}^{(2)} =& \,\,\frac{\pi^{4}}{32}C_{F}^{2}+C_{F}T_{F}n_{F}\bigg(\frac{5}{81}+\frac{77 \pi ^2}{216}-\frac{13 \zeta_{3}}{18}\bigg)+C_{A}C_{F}\bigg(-\frac{535}{324}-\frac{871 \pi ^2}{864}+\frac{7 \pi
^4}{120}+\frac{143 \zeta_{3}}{72}\bigg).
\end{align}

\section{Computation of the $G_{31}$ coefficient and of the constant terms}
\label{g31andconstants}
The starting point to obtain Eq.~(\ref{finalresummed}) is the cross section in Laplace space (\ref{resummedcross}) in which we include the coefficients $A^{(i)}$ and $B^{(i)}$ with $i\leq3$. In Section \ref{sec:resummation of large logs} 
we show that this is sufficient to compute the N$^{2}$LL function $f_{3}(\lambda)$ 
and the N$^{3}$LL coefficient $\tilde{G}_{31}$. The latter combines with terms arising from the inverse Laplace transform to produce $G_{31}$. With the normalization 
chosen for the functions $f_{i}(\lambda)$ (Eqs.~(\ref{f1}--\ref{f3})) the constant terms in 
Eq.~(\ref{resummedform}) are defined as 
\begin{align}
H(\frac{Q}{\mu},\alpha_{s}(\mu)){\rm exp}\left(\frac{\alpha_{s}}{2\pi}\tilde{G}_{10}+\frac{\alpha_{s}^{2}}{(2\pi)^{2}}\tilde{G}_{20}+\mathcal{O}(\alpha_{s}^{3})\right)=
1+\frac{\alpha_{s}}{2\pi}\tilde{C}_{1}+\left(\frac{\alpha_{s}}{2\pi}\right)^{2}\tilde{C}_{2}+\mathcal{O}(\alpha_{s}^{3}),
\end{align}
where $\tilde{G}_{10}$ and $\tilde{G}_{20}$ are the constant terms left in the exponent after computing $f_{3}(\lambda)$ and $\tilde{G}_{31}$. \\
In the Laplace inversion we have to add the N$^{3}$LL term $\frac{1}{6}\mathcal{\tilde{F}}^{(3)}(\alpha_{s}(Q^{2}),\log\frac{1}{\tau})\log^{3}\nu$ in square brackets 
of Eq. (\ref{Rtauprov}) and include the next subleading term in the definition of $\mathcal{\tilde{F}}^{(1)}$ and $\mathcal{\tilde{F}}^{(2)}$ getting
\begin{align}
\mathcal{F}^{(1)}(\alpha_{s}(Q^{2}),\log\frac{1}{\tau}) =& \,\,\,f_{1}(\lambda)+\lambda f_{1}^{\prime}(\lambda)+\frac{\alpha_{s}}{\pi}\beta_{0}f_{2}^{\prime}(\lambda)+\left(\frac{\alpha_{s}}{\pi}\beta_{0}\right)^{2}f_{3}^{\prime}(\lambda)+\mathcal{O}(\alpha_{s}^{n}\log^{n-3}\frac{1}{\tau}),\notag\\
\mathcal{F}^{(2)}(\alpha_{s}(Q^{2}),\log\frac{1}{\tau}) =& \,\,\,2\frac{\alpha_{s}}{\pi}\beta_{0}f_{1}^{\prime}(\lambda)+\frac{\alpha_{s}}{\pi}\beta_{0}\lambda f_{1}^{\prime\prime}(\lambda)+\left(\frac{\alpha_{s}}{\pi}\beta_{0}\right)^{2}f_{2}^{\prime\prime}(\lambda)+\mathcal{O}(\alpha_{s}^{n}\log^{n-3}\frac{1}{\tau}),\notag\\
\mathcal{F}^{(3)}(\alpha_{s}(Q^{2}),\log\frac{1}{\tau}) =& \,\,\,3\left(\frac{\alpha_{s}}{\pi}\beta_{0}\right)^{2}f_{1}^{\prime\prime}(\lambda)+\left(\frac{\alpha_{s}}{\pi}\beta_{0}\right)^{2}\lambda f_{1}^{\prime\prime\prime}(\lambda)+\mathcal{O}(\alpha_{s}^{n}\log^{n-3}\frac{1}{\tau}).
\end{align}
 It is important to notice that the terms $f_{2}^{\prime\prime}(\lambda)$ , $f_{3}^{\prime}(\lambda)$ and the whole $\mathcal{F}^{(3)}(\alpha_{s}(Q^{2}),\log\frac{1}{\tau})$ 
contribute at most with logarithmic order $\mathcal{O}(\alpha_{s}^{n}\log^{n-2}\frac{1}{\tau})$  ({\it i.e.} N$^{3}$LL) and we do
 not need them for a N$^{2}$LL order resummation. Nevertheless, they are relevant for the computation of the coefficient $G_{31}$, 
so we keep them. We now solve the integrals as shown in Section \ref{Inversion of the integral transform} neglecting subleading (N$^{4}$LL) 
terms and we expand the final result through order $\mathcal{O}(\alpha_{s}^{3}\log\frac{1}{\tau})$ obtaining $C_{1}$, $C_{2}$ explicitly as 
reported in Eq.~(\ref{C1},\ref{C2}). To determine $G_{31}$ we observe that the coefficient of $\alpha_{s}^{3}\log\frac{1}{\tau}$ in the previous expansion
is $G_{31}+C_{1}G_{21}+C_{2}G_{11}$, where $G_{21}$ and $G_{11}$ are known. The result is reported in Eq.~(\ref{G31}).

\section{Evaluation of the integrals over soft gluons transverse momenta}
In the present section we show how to evaluate the integrals over the transverse component of soft gluon momenta used in the text.
Let us consider the integral
\begin{align}
\label{transverse1}
 \int d^{d-2}q_{\perp}d^{d-2}k_{\perp} \frac{\delta^{(+)}(q^{2})\delta^{(+)}(k^{2})}{(q+k)^{2}-i0}.
\end{align}
Using the relations
\begin{align}
k^{2} = (k\cdot n)(k\cdot\bar{n})-k^{2}_{\perp},  && d^{d-2}k_{\perp} = \frac{\pi^{1-\epsilon}}{\Gamma(1-\epsilon)}(k^{2}_{\perp})^{-\epsilon}dk^{2}_{\perp},\\
q^{2} = (q\cdot n)(q\cdot\bar{n})-q^{2}_{\perp},  && d^{d-2}q_{\perp} = \frac{\pi^{\frac{1}{2}-\epsilon}}{\Gamma(\frac{1}{2}-\epsilon)}(q^{2}_{\perp})^{-\epsilon}dq^{2}_{\perp} {\rm
sin}^{-2\epsilon}(\theta)d\theta,
\end{align}
where $\theta$ is the angle between the $d-2$-dimensional euclidean vectors $q_{\perp}$ and $k_{\perp}$, we can recast (\ref{transverse1}) as follows

\begin{align}
\frac{\pi^{1-\epsilon}}{\Gamma(1-\epsilon)}\frac{\pi^{\frac{1}{2}-\epsilon}}{\Gamma(\frac{1}{2}-\epsilon)}\int dk^{2}_{\perp}dq^{2}_{\perp}
(k^{2}_{\perp})^{-\epsilon}(q^{2}_{\perp})^{-\epsilon}\frac{\delta^{(+)}((k\cdot n)(k\cdot\bar{n})-k^{2}_{\perp})\delta^{(+)}((q\cdot
n)(q\cdot\bar{n})-q^{2}_{\perp})}{-2|k_{\perp}||q_{\perp}|}\notag\\
\label{transverse12}
\times \frac{({\rm sin}^{2}(\theta))^{-\epsilon} d\theta}{{\rm cos}(\theta)-\frac{(k\cdot n)(q\cdot\bar{n})+(q\cdot n)(k\cdot\bar{n})}{2|k_{\perp}||q_{\perp}|}+i0}.\hspace{6cm}
\end{align}
The integrals over $q^{2}_{\perp}$ and $k^{2}_{\perp}$ can be easily evaluated using the two $\delta$ functions while the angular integral needs some attention.
We set $\frac{(k\cdot n)(q\cdot\bar{n})+(q\cdot n)(k\cdot\bar{n})}{2|k_{\perp}||q_{\perp}|} = \mathcal{K}$ and we consider the angular part of Eq. (\ref{transverse12})
\begin{align*}
\int_{0}^{\pi}\frac{({\rm sin}^{2}(\theta))^{-\epsilon}}{{\rm cos}(\theta)-\mathcal{K}+i0}d\theta,
\end{align*}
where it is straightforward to show that $\mathcal{K}\geq1$.	
The previous integral can be evaluated by setting ${\rm cos}(\theta) \rightarrow 2t-1$ and using the integral representation of the Hypergeometric function
\begin{align}
 \int_{0}^{\pi}\frac{({\rm sin}^{2}(\theta))^{-\epsilon}}{{\rm cos}(\theta)-\mathcal{K}+i0}d\theta = -4^{-\epsilon}\frac{1}{1+\mathcal{K}-i0}\int_{0}^{1} dt
\frac{(1-t)^{-\frac{1}{2}-\epsilon}t^{-\frac{1}{2}-\epsilon}}{1-\frac{2}{1+\mathcal{K}-i0}t}=\hspace{0.3cm}\notag\\
=-4^{-\epsilon}\frac{\Gamma^{2}(\frac{1}{2}-\epsilon)}{\Gamma(1-2\epsilon)}\frac{{}_{2}F_{1}(1,\frac{1}{2}-\epsilon,1-2\epsilon,\frac{2}{1+\mathcal{K}-i0})}{1+\mathcal{K}-i0}.
\end{align}\\
This leads to the solution of (\ref{transverse12}) 
\begin{align}
\label{soltransverse1}
\int d^{d-2}q_{\perp}d^{d-2}k_{\perp} &\frac{\delta^{(+)}(q^{2})\delta^{(+)}(k^{2})}{(q+k)^{2}-i0} =
\frac{\pi^{1-\epsilon}}{\Gamma(1-\epsilon)}\frac{\pi^{\frac{1}{2}-\epsilon}}{\Gamma(\frac{1}{2}-\epsilon)}4^{-\epsilon}\frac{\Gamma^{2}(\frac{1}{2}-\epsilon)}{\Gamma(1-2\epsilon)}\notag\\
\times(q\cdot n)^{-\epsilon}&(q\cdot\bar{n})^{-\epsilon}(k\cdot n)^{-\epsilon}(k\cdot\bar{n})^{-\epsilon}\frac{{}_{2}F_{1}\bigg(1,\frac{1}{2}-\epsilon,1-2\epsilon,4\frac{\sqrt{(q\cdot
n)(q\cdot\bar{n})(k\cdot n)(k\cdot\bar{n})}}{\big(\sqrt{(q\cdot n)(k\cdot\bar{n})}+\sqrt{(q\cdot\bar{n})(k\cdot n)}\big)^{2}}\bigg)}{\big(\sqrt{(q\cdot n)(k\cdot\bar{n})}+\sqrt{(q\cdot\bar{n})(k\cdot
n)}\big)^{2}}\hspace{-0cm},\notag\\
\end{align}
used in the text.
The second relevant integral is 
\begin{align}
\int d^{d-2}q_{\perp}d^{d-2}k_{\perp} \frac{\delta^{(+)}(q^{2})\delta^{(+)}(k^{2})}{((q+k)^{2}+i0)((q+k)^{2}-i0)},
\end{align}
appearing in the computation of the vacuum polarization diagrams. Such an integral can be evaluated easily using the previous reult (\ref{soltransverse1}).
We can indeed write it as
\begin{align}
-\frac{1}{2}\frac{1}{i0}\bigg(\int d^{d-2}q_{\perp}d^{d-2}k_{\perp} \frac{\delta^{(+)}(q^{2})\delta^{(+)}(k^{2})}{(q+k)^{2}+i0}- \int d^{d-2}q_{\perp}d^{d-2}k_{\perp}
\frac{\delta^{(+)}(q^{2})\delta^{(+)}(k^{2})}{(q+k)^{2}-i0}\bigg),
\end{align}
so it amounts to use (\ref{soltransverse1}) with two different pole prescriptions. It is then straightforward to show that
\begin{align}
\label{soltransverse2}
\int d^{d-2}q_{\perp}d^{d-2}k_{\perp} \frac{\delta^{(+)}(q^{2})\delta^{(+)}(k^{2})}{((q+k)^{2}+i0)((q+k)^{2}-i0)} =
\frac{\pi^{1-\epsilon}}{\Gamma(1-\epsilon)}\frac{\pi^{\frac{1}{2}-\epsilon}}{\Gamma(\frac{1}{2}-\epsilon)}4^{-\epsilon}\frac{\Gamma^{2}(\frac{1}{2}-\epsilon)}{\Gamma(1-2\epsilon)}\hspace{2cm}\notag\\
\times(q\cdot n)^{-\epsilon}(q\cdot\bar{n})^{-\epsilon}(k\cdot n)^{-\epsilon}(k\cdot\bar{n})^{-\epsilon}\frac{{}_{2}F_{1}\bigg(2,\frac{1}{2}-\epsilon,1-2\epsilon,4\frac{\sqrt{(q\cdot
n)(q\cdot\bar{n})(k\cdot n)(k\cdot\bar{n})}}{\big(\sqrt{(q\cdot n)(k\cdot\bar{n})}+\sqrt{(q\cdot\bar{n})(k\cdot n)}\big)^{2}}\bigg)}{\big(\sqrt{(q\cdot n)(k\cdot\bar{n})}+\sqrt{(q\cdot\bar{n})(k\cdot
n)}\big)^{4}}.\hspace{1cm}
\end{align}

In the last part of the present section we report some identities for the hypergeometric functions appearing in the calculation:

\begin{align}
\label{hyper1}
{}_{2}F_{1}\big(1,\frac{1}{2}-\epsilon,1-2\epsilon,4\frac{z}{(1+z)^{2}}\big) = (1+z)^{2} {}_{2}F_{1}\big(1,1+\epsilon,1-\epsilon,z^{2}\big), && |z|\leq1;\\
\label{hyper2}
{}_2F_{1}(1,1+\epsilon,1-\epsilon,z) = \frac{\Gamma(1-\epsilon)}{\Gamma(1+\epsilon)\Gamma(-2\epsilon)}\int_{0}^{1}dy\,\frac{y^{\epsilon}(1-y)^{-1-2\epsilon}}{1-yz}, && \epsilon<0;\\
\label{hyper3}
{}_{2}F_{1}\left(2,\frac{1}{2}-\epsilon,1-2\epsilon,4\frac{z}{(1+z)^{2}}\right) = (1+z)^{4} {}_{2}F_{1}\big(2,2+\epsilon,1-\epsilon,z^{2}\big), && |z|\leq1;\\
\label{hyper4}
{}_2F_{1}(2,2	+\epsilon,1-\epsilon,z) = \frac{\Gamma(1-\epsilon)}{\Gamma(1+\epsilon)\Gamma(-2\epsilon)}\frac{1}{1-z}\int_{0}^{1}dy\,y^{\epsilon}(1-y)^{-1-2\epsilon}\frac{1+zy}{(1-yz)^{2}}, &&
\epsilon<0.
\end{align}
To prove (\ref{hyper4}) we consider the following relation
\begin{align}
{}_2F_{1}\left(2,2+\epsilon,1-\epsilon,z\right) = \frac{1}{1-z}\left({}_2F_{1}(2,1+\epsilon,1-\epsilon,z)+z\frac{1+\epsilon}{1-\epsilon}{}_2F_{1}(2,2+\epsilon,2-\epsilon,z)\right),
\end{align}
and then we use the integral reppresentation of the hypergeometric functions in the right hand side of the previous equation getting
\begin{align}
{}_2F_{1}(2,2+\epsilon,1-\epsilon,z) = \frac{1}{1-z}\bigg(\frac{\Gamma(1-\epsilon)}{\Gamma(1+\epsilon)\Gamma(-2\epsilon)}\int_{0}^{1}dy\,\frac{y^{\epsilon}(1-y)^{-1-2\epsilon}}{(1-yz)^{2}}+\notag\\
\frac{1+\epsilon}{1-\epsilon}\frac{\Gamma(2-\epsilon)}{\Gamma(2+\epsilon)\Gamma(-2\epsilon)}z\int_{0}^{1}dy\,\frac{y^{1+\epsilon}(1-y)^{-1-2\epsilon}}{(1-yz)^{2}}\bigg).
\end{align}
Using the relation
\begin{align}
\frac{1+\epsilon}{1-\epsilon}\frac{\Gamma(2-\epsilon)}{\Gamma(2+\epsilon)\Gamma(-2\epsilon)} = \frac{\Gamma(1-\epsilon)}{\Gamma(1+\epsilon)\Gamma(-2\epsilon)},
\end{align}
we prove (\ref{hyper4}).

\addcontentsline{toc}{section}{
\end{document}
\textbf{References}} 


\begin{thebibliography}{99} 

\bibitem{aleph}
  D.~Buskulic {\it et al.}  [ALEPH Collaboration],
  Z.\ Phys.\  C {\bf 73} (1997) 409;
  A.~Heister {\it et al.}  [ALEPH Collaboration],
  Eur.\ Phys.\ J.\  C {\bf 35} (2004) 457.

\bibitem{opal}
  P.D.~Acton {\it et al.}  [OPAL Collaboration],
  Z.\ Phys.\  C {\bf 59} (1993) 1;
  G.~Alexander {\it et al.}  [OPAL Collaboration],
  Z.\ Phys.\  C {\bf 72} (1996) 191;
  K.~Ackerstaff {\it et al.}  [OPAL Collaboration],
  Z.\ Phys.\  C {\bf 75} (1997) 193;
  G.~Abbiendi {\it et al.}  [OPAL Collaboration],
  Eur.\ Phys.\ J.\  C {\bf 16} (2000) 185
  [hep-ex/0002012];
  G.~Abbiendi {\it et al.}  [OPAL Collaboration],
  Eur.\ Phys.\ J.\  C {\bf 40} (2005) 287
  [hep-ex/0503051].
  G.~Abbiendi {\it et al.}  [OPAL Collaboration],
  Eur.\ Phys.\ J.\  C {\bf 53} (2008) 21.

 \bibitem{l3}
  M.~Acciarri {\it et al.}  [L3 Collaboration],
  Phys.\ Lett.\  B {\bf 371} (1996) 137;
  M.~Acciarri {\it et al.}  [L3 Collaboration],
  Phys.\ Lett.\  B {\bf 404} (1997) 390;
  M.~Acciarri {\it et al.}  [L3 Collaboration],
  Phys.\ Lett.\  B {\bf 444} (1998) 569;
  P.~Achard {\it et al.}  [L3 Collaboration],
  Phys.\ Lett.\  B {\bf 536} (2002) 217
  [hep-ex/0206052];
  P.~Achard {\it et al.}  [L3 Collaboration],
  Phys.\ Rept.\  {\bf 399} (2004) 71
  [hep-ex/0406049].

\bibitem{delphi}
  P.~Abreu {\it et al.}  [DELPHI Collaboration],
  Phys.\ Lett.\  B {\bf 456} (1999) 322;
  J.~AbdaLLh {\it et al.}  [DELPHI Collaboration],
  Eur.\ Phys.\ J.\  C {\bf 29} (2003) 285
  [hep-ex/0307048];
  J.~AbdaLLh {\it et al.}  [DELPHI Collaboration],
  Eur.\ Phys.\ J.\  C {\bf 37} (2004) 1
  [hep-ex/0406011].

\bibitem{sld}
 K.~Abe {\it et al.}  [SLD Collaboration],
  Phys.\ Rev.\  D {\bf 51} (1995) 962
  [hep-ex/9501003].

\bibitem{jade}
 P.~Pfeifenschneider {\it et al.}  [JADE collaboration],
  Eur.\ Phys.\ J.\  C {\bf 17} (2000) 19
  [hep-ex/0001055].

\bibitem{thrust}
  S.~Brandt, C.~Peyrou, R.~Sosnowski and A.~Wroblewski,
  Phys.\ Lett.\  {\bf 12} (1964) 57;\\
  E.~Farhi,
  Phys.\ Rev.\ Lett.\  {\bf 39} (1977) 1587.


\bibitem{Ellis:1980wv}
  R.~K.~Ellis, D.~A.~Ross and A.~E.~Terrano,
  Nucl.\ Phys.\  B {\bf 178} (1981) 421.
\bibitem{Ellis:1980nc}
  R.~K.~Ellis, D.~A.~Ross and A.~E.~Terrano,
  Phys.\ Rev.\ Lett.\  {\bf 45} (1980) 1226.
\bibitem{Kunszt:1980vt}
  Z.~Kunszt,
  Phys.\ Lett.\  B {\bf 99} (1981) 429.
\bibitem{Vermaseren:1980qz}
  J.~A.~M.~Vermaseren, K.~J.~F.~Gaemers and S.~J.~Oldham,
  Nucl.\ Phys.\  B {\bf 187} (1981) 301.
\bibitem{Fabricius:1981sx}
  K.~Fabricius, I.~Schmitt, G.~Kramer and G.~Schierholz,
  Z.\ Phys.\  C {\bf 11} (1981) 315.
\bibitem{Kunszt:1989km}
  Z.~Kunszt and P.~Nason, 
  {\it QCD at LEP}, CERN Yellow Report 89-08 (1989), p.373.
\bibitem{Giele:1991vf}
  W.~T.~Giele and E.~W.~N.~Glover,
  Phys.\ Rev.\  D {\bf 46} (1992) 1980.

\bibitem{Catani:1996jh}
  S.~Catani, M.~H.~Seymour,
  Phys.\ Lett.\  {\bf B378} (1996)  287.
  [hep-ph/9602277];
S.~Catani and M.~H.~Seymour,
  Nucl.\ Phys.\  B {\bf 485} (1997) 291
  [hep-ph/9605323].


\bibitem{CTTW}
  S.~Catani, L.~Trentadue, G.~Turnock, B.~R.~Webber,
  Nucl.\ Phys.\  {\bf B407} (1993)  3.


\bibitem{broadenings}
 Y.L.~Dokshitzer, A.~Lucenti, G.~Marchesini and G.P.~Salam,
  JHEP {\bf 9801} (1998) 011
  [hep-ph/9801324].

\bibitem{y3}
  A.~Banfi, G.P.~Salam and G.~Zanderighi,
  JHEP {\bf 0201} (2002) 018
  [hep-ph/0112156].

\bibitem{Jones:2003yv}
  R.~W.~L.~Jones, M.~Ford, G.~P.~Salam, H.~Stenzel, D.~Wicke,
  JHEP {\bf 0312} (2003)  007.
  [hep-ph/0312016].



\bibitem{ourant}
 A.~Gehrmann-De Ridder, T.~Gehrmann and E.~W.~N.~Glover,
  JHEP {\bf 0509} (2005) 056
  [hep-ph/0505111].

\bibitem{GehrmannDeRidder:2007jk}
  A.~Gehrmann-De Ridder, T.~Gehrmann, E.~W.~N.~Glover, G.~Heinrich,
  JHEP {\bf 0711} (2007)  058
  [arXiv:0710.0346].
\bibitem{GehrmannDeRidder:2007hr}
  A.~Gehrmann-De Ridder, T.~Gehrmann, E.~W.~N.~Glover, G.~Heinrich,
  JHEP {\bf 0712} (2007)  094
  [arXiv:0711.4711].
\bibitem{GehrmannDeRidder:2007bj}
  A.~Gehrmann-De Ridder, T.~Gehrmann, E.~W.~N.~Glover, G.~Heinrich,
  Phys.\ Rev.\ Lett.\  {\bf 99} (2007)  132002
  [arXiv:0707.1285].
\bibitem{Weinzierl:2009nz}
  S.~Weinzierl,
  JHEP {\bf 0907} (2009)  009
  [arXiv:0904.1145].
\bibitem{Weinzierl:2009ms}
  S.~Weinzierl,
  JHEP {\bf 0906} (2009)  041
  [arXiv:0904.1077].


\bibitem{weinzierljetnew}
 S.~Weinzierl,
  Eur.\ Phys.\ J.\  C {\bf 71} (2011) 1565
  [arXiv:1011.6247].

\bibitem{Denner:2009gx}
  A.~Denner, S.~Dittmaier, T.~Gehrmann, C.~Kurz,
  Phys.\ Lett.\  {\bf B679} (2009)  219
  [arXiv:0906.0372].
\bibitem{Denner:2010ia}
  A.~Denner, S.~Dittmaier, T.~Gehrmann, C.~Kurz,
  Nucl.\ Phys.\  {\bf B836} (2010)  37
  [arXiv:1003.0986].

\bibitem{Dissertori:2007xa}
  G.~Dissertori, A.~Gehrmann-De Ridder, T.~Gehrmann, E.~W.~N.~Glover, G.~Heinrich, H.~Stenzel,
  JHEP {\bf 0802} (2008)  040.
  [arXiv:0712.0327].

\bibitem{asjet}
 G.~Dissertori, {\it et al.},
  Phys.\ Rev.\ Lett.\  {\bf 104} (2010) 072002
  [arXiv:0910.4283].
  
\bibitem{jaquier}
 T.~Gehrmann, M.~Jaquier and G.~Luisoni,
  Eur.\ Phys.\ J.\  C {\bf 67} (2010) 57
  [arXiv:0911.2422].

\bibitem{Gehrmann:2008kh}
  T.~Gehrmann, G.~Luisoni, H.~Stenzel,
  Phys.\ Lett.\  {\bf B664} (2008)  265
  [arXiv:0803.0695].
  
\bibitem{davisonwebber}
 R.~A.~Davison and B.~R.~Webber,
  Eur.\ Phys.\ J.\  C {\bf 59} (2009) 13
  [arXiv:0809.3326].

\bibitem{jadeas}
  S.~Bethke, S.~Kluth, C.~Pahl and J.~Schieck  [JADE Collaboration],
  Eur.\ Phys.\ J.\  C {\bf 64} (2009) 351
  [arXiv:0810.1389].


\bibitem{Dissertori:2009ik}
  G.~Dissertori, A.~Gehrmann-De Ridder, T.~Gehrmann, E.~W.~N.~Glover, G.~Heinrich, G.~Luisoni, H.~Stenzel,
  JHEP {\bf 0908} (2009)  036.
  [arXiv:0906.3436].


\bibitem{opalas}
  G.~Abbiendi {\it et al.}  [OPAL Collaboration],
  [arXiv:1101.1470].

\bibitem{Ster87} 
 G.~F.~Sterman,
  Nucl.\ Phys.\  B {\bf 281} (1987) 310.

\bibitem{eec}
 D.~de Florian and M.~Grazzini,
  Nucl.\ Phys.\  B {\bf 704} (2005) 387
  [hep-ph/0407241].


\bibitem{scet}
C.W.~Bauer, D.~Pirjol and I.W.~Stewart,
  Phys.\ Rev.\  D {\bf 65} (2002) 054022
  [hep-ph/0109045];\\
C.W.~Bauer, S.~Fleming, D.~Pirjol, I.Z.~Rothstein and I.W.~Stewart,
  Phys.\ Rev.\  D {\bf 66} (2002) 014017
  [hep-ph/0202088];\\
  M.~Beneke, A.~P.~Chapovsky, M.~Diehl and T.~Feldmann,
  Nucl.\ Phys.\  B {\bf 643} (2002) 431
  [hep-ph/0206152].

\bibitem{schwartzMHnll}
 M.~D.~Schwartz,
  Phys.\ Rev.\  D {\bf 77} (2008) 014026
  [arXiv:0709.2709].

\bibitem{hoangtop}
S.~Fleming, A.~H.~Hoang, S.~Mantry and I.~W.~Stewart,
  Phys.\ Rev.\  D {\bf 77} (2008) 074010
  [hep-ph/0703207];
  Phys.\ Rev.\  D {\bf 77} (2008) 114003
  [arXiv:0711.2079].


  

\bibitem{Becher:2008cf}
  T.~Becher, M.D.~Schwartz,
  JHEP {\bf 0807} (2008)  034
  [arXiv:0803.0342].
\bibitem{Abbate:2010xh}
  R.~Abbate, M.~Fickinger, A.~H.~Hoang, V.~Mateu, I.~W.~Stewart,
  Phys.\ Rev.\  {\bf D83} (2011)  074021
  [arXiv:1006.3080].
\bibitem{Chien:2010kc}
  Y.T.~Chien, M.D.~Schwartz,
  JHEP {\bf 1008} (2010)  058
  [arXiv:1005.1644].
  
\bibitem{Chiu:2011qc}
  J.~-y.~Chiu, A.~Jain, D.~Neill, I.~Z.~Rothstein,
  [arXiv:1104.0881].


\bibitem{Becher:2011pf}
  T.~Becher, G.~Bell, M.~Neubert,
  [arXiv:1104.4108].

\bibitem{3Lform1} 
 S.~Moch, J.~A.~M.~Vermaseren and A.~Vogt,
  JHEP {\bf 0508} (2005) 049
  [hep-ph/0507039].



\bibitem{3Lform2} 
  P.~A.~Baikov, K.~G.~Chetyrkin, A.~V.~Smirnov, V.~A.~Smirnov and M.~Steinhauser,
  Phys.\ Rev.\ Lett.\  {\bf 102} (2009) 212002
  [arXiv:0902.3519];\\
 R.~N.~Lee, A.~V.~Smirnov and V.~A.~Smirnov,
  JHEP {\bf 1004} (2010) 020
  [arXiv:1001.2887];\\
T.~Gehrmann, E.~W.~N.~Glover, T.~Huber, N.~Ikizlerli and C.~Studerus,
  JHEP {\bf 1006} (2010) 094
  [arXiv:1004.3653].




\bibitem{SCETJet} 
 T.~Becher and M.~Neubert,
  Phys.\ Lett.\  B {\bf 637} (2006) 251
  [hep-ph/0603140].


  \bibitem{bechneubdis} 
 T.~Becher, M.~Neubert and B.~D.~Pecjak,
  JHEP {\bf 0701} (2007) 076
  [hep-ph/0607228].


\bibitem{hoangkluth}
 A.~H.~Hoang and S.~Kluth,
  [arXiv:0806.3852].


\bibitem{schwartznew}
 R.~Kelley, R.~M.~Schabinger, M.~D.~Schwartz and H.~X.~Zhu,
  [arXiv:1105.3676].


\bibitem{stewartnew}
A.~Hornig, C.~Lee, I.~W.~Stewart, J.~R.~Walsh and S.~Zuberi,
  [arXiv:1105.4628].

\bibitem{mantrynew}
 Y.~Li, S.~Mantry and F.~Petriello,
  [arXiv:1105.5171].


\bibitem{Baikov:2010iw}
  P.A.~Baikov, K.G.~Chetyrkin and J.H.~K\"uhn,
  Phys.\ Rev.\ Lett.\  {\bf 101} (2008) 012002
  [arXiv:0801.1821].

\bibitem{CollinsSoper}
J.C Collins, D.E. Soper, G. Sterman, in {\it Perturbative Quantum Chromodynamics},\\ed. A.H. Mueller (World Scientific Singapore, 1989).

\bibitem{BerSterKucs1}
  C.~F.~Berger, T.~Kucs, G.~F.~Sterman,
  Phys.\ Rev.\  {\bf D68} (2003)  014012.
  [hep-ph/0303051];
  Int.\ J.\ Mod.\ Phys.\  {\bf A18} (2003)  4159.
  [hep-ph/0212343].

\bibitem{KorSter}
 G.~P.~Korchemsky and G.~F.~Sterman,
  Phys.\ Lett.\  B {\bf 340} (1994) 96
  [hep-ph/9407344].

 \bibitem{FT} 
 J.~Frenkel and J.~C.~Taylor,
  Nucl.\ Phys.\  B {\bf 246} (1984) 231.
\bibitem{NAET} 
  J.~G.~M.~Gatheral,
  Phys.\ Lett.\  B {\bf 133} (1983) 90.

\bibitem{OneLoopSoft} 
S.~Catani and M.~Grazzini,
  Nucl.\ Phys.\  B {\bf 591} (2000) 435
  [hep-ph/0007142].

\bibitem{HypExp}
T.~Huber and D.~Ma\^{\i}tre,
  Comput.\ Phys.\ Commun.\  {\bf 175} (2006) 122
  [hep-ph/0507094];
 Comput.\ Phys.\ Commun.\  {\bf 178} (2008)  755.
 [arXiv:0708.2443].
 
\bibitem{secdec} 
 J.~Carter and G.~Heinrich,
  Comput.\ Phys.\ Commun.\  {\bf 182} (2011) 1566
  [arXiv:1011.5493].
\bibitem{sectordeco1} 
 G.~Heinrich,
  Int.\ J.\ Mod.\ Phys.\  A {\bf 23} (2008) 1457
  [arXiv:0803.4177].
 \bibitem{sectordeco2} 
T.~Binoth and G.~Heinrich,
  Nucl.\ Phys.\  B {\bf 585} (2000) 741
  [hep-ph/0004013].
\bibitem{bases} 
S.~Kawabata,
  Comput.\ Phys.\ Commun.\  {\bf 88} (1995) 309.
\bibitem{vegas} 
   G.~P.~Lepage,
  J.\ Comput.\ Phys.\  {\bf 27} (1978) 192;\\
 T.~Hahn,
  Comput.\ Phys.\ Commun.\  {\bf 168} (2005) 78
  [hep-ph/0404043].
\bibitem{catanigrazzinidoublesoft} 
S.~Catani and M.~Grazzini,
  Nucl.\ Phys.\  B {\bf 570} (2000) 287
  [hep-ph/9908523].





\bibitem{wrge1} 
 G.~P.~Korchemsky and G.~Marchesini,
  Phys.\ Lett.\  B {\bf 313} (1993) 433.
\bibitem{wrge2}
G.~P.~Korchemsky and G.~Marchesini,
  Nucl.\ Phys.\  B {\bf 406} (1993) 225
  [hep-ph/9210281].
 
\bibitem{wrge3} 
I.~A.~Korchemskaya and G.~P.~Korchemsky,
  Phys.\ Lett.\  B {\bf 287} (1992) 169.
\bibitem{cuspkorch} 
 G.~P.~Korchemsky and A.~V.~Radyushkin,
  Nucl.\ Phys.\  B {\bf 283} (1987) 342.

\bibitem{korchdis}
 G.~P.~Korchemsky,  
Mod.\ Phys.\ Lett.\  A {\bf 4} (1989) 1257.


\bibitem{AP} 
 G.~Altarelli and G.~Parisi,
  Nucl.\ Phys.\  B {\bf 126} (1977) 298.
\bibitem{NNLOAPS} 
 A.~Vogt, S.~Moch and J.~A.~M.~Vermaseren,
  Nucl.\ Phys.\  B {\bf 691} (2004) 129
  [hep-ph/0404111].

\bibitem{NNLOAPNS} 
 S.~Moch, J.~A.~M.~Vermaseren and A.~Vogt,
  Nucl.\ Phys.\  B {\bf 688} (2004) 101
  [hep-ph/0403192].

\bibitem{Catani:1996yz}
  S.~Catani, M.~L.~Mangano, P.~Nason, L.~Trentadue,
  Nucl.\ Phys.\  {\bf B478} (1996)  273
  [hep-ph/9604351].



\bibitem{privBec} T. Becher, {\it private communication}.  


\bibitem{alpha0}
 Y.~L.~Dokshitzer, G.~Marchesini and B.~R.~Webber,
  Nucl.\ Phys.\  B {\bf 469} (1996) 93
  [hep-ph/9512336].


 \end{thebibliography}
